\documentclass[10pt]{article}
\pdfoutput=1
\usepackage{amsmath}
\usepackage{amssymb}
\usepackage{graphicx}

\newcommand{\be}{\begin{equation}}
\newcommand{\ee}{\end{equation}}

\newcommand{\p}{\partial}

\newcommand{\cH}{{\cal H}}

\newcommand{\me}{{\rm me}}

\newcommand{\el}{{\rm el}}
\newcommand{\te}{{\rm te}}
\newcommand{\ben}{{\rm be}}

\newcommand{\tr}{{\rm tr}}
\newcommand{\ci}{{\rm ci}}

\newcommand{\half}{\textstyle { \frac{1}{2}}\,}

\date{}

\title{Membrane adhesion and domain formation}
\author{Thomas R.\ Weikl and Reinhard Lipowsky\\[0.4cm]
  \small Max Planck Institute of Colloids and Interfaces \\[-0.05cm] \small Department of
   Theory and Bio-Systems \\[-0.05cm] \small 14424 Potsdam, Germany}

\begin{document}
\maketitle
\tableofcontents

\newpage

\section{Introduction}

 The molecular structure of biological and biomimetic 
membranes is provided by  bilayers of amphiphilic molecules 
such as lipids and proteins.  Lipid bilayers are rather  thin with 
a thickness between 4 and 5 nm.  At physiological temperatures, 
these  membranes are {\em fluid} in the sense that they have no 
in-plane shear modulus. On the nanometer scale, the molecules have
no fixed   neighborhoods since two adjacent molecules can easily exchange 
their position within the membrane.  This  local exchange leads to   
rapid lateral diffusion of the molecules along  the membrane. 
In this way, a typical lipid molecule is displaced by about one micrometer in one 
second. Therefore, biomembranes\footnote{Here and below,  the term `biomembrane' is used  as an
abbreviation for `biological or  biomimetic membrane'. It is always  assumed
implicitly that biomembranes have two characteristic features: (i) they contain  
several molecular  components and (ii) they are in a fluid phase. }
represent multicomponent  liquids in two dimensions. 

\subsection{Membrane adhesion via sticker molecules} 

Now, consider  such a multicomponent membrane that forms a giant vesicle 
which adheres to another membrane or, more generally, to  a substrate surface. 
Within the contact area of the vesicle, the membrane  molecules can still  
diffuse, at least to some extent, which implies that the different molecular
components may  attain many different {\em spatial patterns}. In particular, some of 
these components may form {\em  clusters} or extended {\em domains} within the 
contact area.  In this article, we will  review  recent work  on these patterns and their 
formation processes. 

We will focus on situations in which the membrane contains certain types of 
membrane-anchored molecules,  so-called stickers and repellers. 
[1-6]
Sticker molecules mediate  attractive 
interactions between the membrane and the adjacent surface whereas repeller molecules
act as repulsive spacers between these two surfaces.  Within the contact area, 
sticker molecules have a strong tendency to form clusters or intramembrane domains. 
In fact, there are several distinct mechanism for these pattern formation processes
as explained in the main part of this review.  

There are many different types of  stickers which can vary greatly 
in their size.  The smallest stickers are presumably charged head
groups of lipids which are attracted towards an oppositely charged 
surface. Likewise,  lipids may have large sticky head groups 
containing, e.g., polysaccharides.  Much larger stickers  mediate
the specific adhesion of biological membranes which governs  both 
cell-cell adhesion and the  signalling between cells of the immune system. 
\cite{alberts94,spri90}
Cell adhesion molecules, which govern the binding of two cells, and
 receptor molecules, which are expressed on the surface of immune cells, 
are usually  relatively stiff rodlike molecules which are anchored 
in the bilayer membrane and  have a linear extension of 10 - 30 nm. 

Presumably the first theoretical models for the adhesion of membranes via
mobile stickers were introduced  by Bell {\em et al.}  in Refs.~\cite{bell78,bell84} as
reviewed in \cite{bell88} and by Evans in Refs. \cite{evans85a,evans85b}. 
In these models,  the membrane is divided up into
a bound segment, which represents the  contact  area, and an unbound segment
which acts as a  reservoir for the mobile sticker molecules which can diffuse
in and out of the contact area. The repellers are taken to be more or less
immobile and the separation of the two interacting surfaces  is taken
to be constant within the contact area.  If the repeller molecules are longer than
the sticker molecules, 
the competing action of these two types of molecules should lead to a modulation of
this  separation:  sticker--rich membrane segments should have a
relatively small separation whereas   repeller--rich  segments should have a
relatively large one as discussed in the context of
gap junctions \cite{braun84,braun87,bruinsma94}.

Since the membrane-anchored stickers 
gain energy when they enter  the contact area and bind to the second surface, 
these molecules will be `recruited' by this surface  and, thus, will be enriched
within the contact area. Therefore, the  bound and unbound  segments of 
multicomponent membranes will usually  differ in their  molecular compositions
 \cite{lipo195}.\footnote{
This difference in composition has apparently been overlooked
by Bell  since he says    on page 242 of Ref.  \cite{bell88} "that at equilibrium
the number of receptors per unit area will be the same in the contact area 
and outside of it." } 
In the following,  we will focus on the bound membrane segment, i.e., on the 
contact area of the multicomponent membrane, and view its  unbound segment as a
reservoir for the  sticker and repeller molecules. It is then convenient to 
use a grand-canonical description and to replace 
this reservoir by chemical potentials for the sticker and repeller molecules. 
In this way, one arrives at  lattice gas models on flexible surfaces. 
\cite{lipo130,weikl00,weikl01} These lattice gas models provide a general 
theoretical framework by which one  can study the interplay of
membrane adhesion and domain formation in a systematic manner. 

The  problem of adhesion-induced domain formation has also been addressed
 using somewhat 
different theoretical models in Refs. \cite{komura00,komura03,bruinsma00,chen03}. 
In addition, this process  has also been observed experimentally
both in biomimetic and in biological membranes. The next two subsections
contain a short summary of these observations. 

\subsection{Stickers and repellers in biomimetic membranes}

Adhesion-induced  lateral phase separation into domains with small and large  membrane
separations  has been found to occur in several  biomimetic systems. The formation of
blisters has been observed in membranes containing cationic lipids in contact with a
negatively charged surface \cite{nardi97}, and between membranes containing both
negatively and positively charged lipids \cite{wong01}.  The coexistence of tightly and weakly
bound membrane segments  has been found for membranes with biotinylated lipids  bound to
another biotinylated surface via streptavidin \cite{albersdorfer97}, 
membranes with homophilic
csA-receptors from the slime mold {\it Dictyostelium discoideum} \cite{kloboucek99},
and   membranes containing specific ligands of integrin molecules adsorbed on a substrate
\cite{guttenberg01}. Attractive  membrane-mediated interactions between bound
csA-receptors of adhering  vesicles have been  inferred from membrane tension jumps
induced by the  micropipet aspiration technique \cite{maier01}. In addition to the
receptors,  the membranes studied in
\cite{albersdorfer97,kloboucek99,guttenberg01,maier01} also contain  repulsive
lipopolymers to prevent non-specific adhesion.  

\subsection{Specific adhesion of biological membranes}

The adhesion of cells plays a key role in important biological processes such as 
tissue development and immune response. The highly selective interactions
leading to cell  adhesion are mediated by a variety of specific receptors which are
embedded in the cell membranes.  Two prominent examples
for  domain formation within the contact area of biomembranes are provided   by
focal contacts \cite{alberts94,bala01}, which are formed during
cell adhesion at  substrate surfaces,
and  by the so-called `immunological synapses' which are formed during the adhesion of
helper T cells and antigen-presenting cells as part of the immune response. 
Within the contact area of these two cells, several groups have 
recently observed distinct domain patterns of membrane-anchored receptors and ligands.  

The antigen-presenting cells  (APCs)  display foreign peptide fragments on their surfaces. 
These
peptide fragments are presented by MHC molecules on the APC surfaces, and recognized by the
highly specific T cell receptors (TCR). At the cell-cell contact zone, the bound receptor-ligand
pairs are arranged in characteristic supramolecular patterns
\cite{monks98,grakoui99,krummel00,potter01,khlee02}, for reviews see
\cite{vanderMerwe00, dustin00,delon00,bromley01,dustin01,wulfing02}. The final,
`mature'  pattern of an adhering T cell consists of a central domain in which the TCRs are
bound to the MHC-peptides (MHCp), surrounded by a ring-shaped domain in which the
integrin receptors LFA-1 of the T cell are bound to their ligands ICAM-1 of the APC.
Intriguingly, the characteristic intermediate pattern formed earlier during T cell adhesion is
inverted, with a TCR/MHCp ring surrounding a central LFA-1/ICAM-1 domain in the contact
zone \cite{grakoui99,johnson00,khlee02}. This pattern inversion has been first observed
for T cells adhering to a supported lipid bilayer with embedded MHCp and ICAM-1,
\cite{grakoui99,johnson00}, more recently also in a cell-cell system \cite{khlee02}. A
significantly different type of pattern evolution has been found for immature T cells or
thymozytes, which form multifocal synapses with several nearly circular clusters of
TCR/MHC-peptide complexes in the contact zone \cite{hailman02,richie02}. 

\subsection{Outline of review}

This  review article is organized as follows. In section \ref{S.MembraneModels}, 
we   describe the theoretical 
framework used to study the interplay of membrane adhesion and domain 
formation. We first review the behavior of  interacting membranes 
that have a  homogeneous or uniform composition. We then describe 
general lattice gas models for  multicomponent membranes with anchored
sticker and repeller molecules. 
Section \ref{S.TheoreticalMethods} summarizes the different
theoretical methods used to elucidate the membrane behavior.  The 
remaining sections \ref{S.EntropicMechanisms} -- \ref{S.Dynamics}
represent the main part of this review and 
discuss  the interplay of  adhesion and domain formation for several membrane
systems. 

In section \ref{S.EntropicMechanisms}, we consider multicomponent membranes 
with one species of sticker molecules and describe several entropic mechanisms 
that enhance or induce the  formation of sticker-rich domains within the contact
area.  
These mechanisms reflect the large configuration space which the membranes and 
the sticker molecules can explore because of their mobility.  In order to 
elucidate these mechanisms, 
one must distinguish two types of molecular interactions: (i) {\em trans}-interactions 
between the sticker molecules and the second membrane or substrate surface; and
(ii) {\em cis}-interactions between two stickers which are 
anchored to the same membrane. 
If the cis-interactions between the stickers are {\em repulsive} and short-ranged, 
the thermally excited shape fluctuations of the membrane induce small clusters of
stickers but are not able to initiate the formation of 
sticker-rich  domains within a sticker-poor membrane matrix. In fact, in the 
latter case, sticker-mediated adhesion occurs only if the sticker concentration 
exceeds a certain threshold value. 
Sticker-rich and sticker-poor domains are formed if the stickers experience
{\em attractive} cis-interactions and these attractive interactions
are  effectively enhanced by the thermally excited shape 
fluctuations of the membranes. Furthermore, purely entropic mechanisms for the 
formation of sticker-rich domains
arise if the  stickers are  large compared to the smallest wavelength of the 
bending modes or  if the stickers  increase the local bending rigidity of the membrane. 

Another class of mechanisms for adhesion-induced domain formation is discussed
in section \ref{S.BarrierMechanisms} where we consider  the adhesion of 
multicomponent membranes  which contain both sticker and repeller molecules.  
If the length of  the repellers, say $l_r$,  exceeds
the length of the stickers, say $l_s$, the size mismatch between these two species of
molecules favors the formation and growth of sticker-rich domains.  This can be 
understood in terms of effective membrane potentials with a potential well 
arising from the stickers and a potential barrier arising from the repellers. 
 Phase separation into sticker-rich and sticker-poor (or repeller-rich) domains
occurs if the potential barrier is sufficiently high.  Similar barrier mechanisms
are also effective if the membranes contains two species of stickers, long and short
ones. 

Finally, section \ref{S.Dynamics} addresses the time evolution of  domains in 
multicomponent membranes which contain (i) stickers and repellers,  (ii) short and
long stickers, or (iii) short  stickers, long stickers as well as repellers. In all cases, 
the effective membrane potential exhibits a  potential barrier which implies that  the initial
dynamics of the domain formation represents a nucleation process.  Thus, 
in the presence of repellers, adhesion is governed by the nucleation of 
sticker  clusters or  islands. The diffusion and coalescence of these clusters
leads to  the formation of  distinct domain patterns  at intermediate times. 
This provides a simple and generic mechanism for the observed time evolution 
within the immunological synapse between helper T cells and antigen-presenting
cells. 

\pagebreak

\section{Modelling of membranes}
\label{S.MembraneModels}

In this section, we discuss the theoretical models that will be used in order to 
describe and understand the interplay between membrane adhesion and 
domain formation. This interplay arises from two degrees of freedom: the 
elastic deformations of  the membranes and the spatial 
patterns of membrane-anchored molecules.  

As mentioned in the introduction, we will focus on the contact area of  
adhering vesicles or cells and view their  unbound segments as 
reservoirs for the  sticker and repeller molecules. It is then convenient to 
use a grand-canonical description and to describe these
 reservoirs by chemical potentials   for the membrane-anchored 
molecules. In general,  the membranes may contain several species  of 
such molecules which will be distinguished by the index
$k$ with $k= 1, \dots K$. The membrane concentration of  species $k$ is then 
determined  by the chemical potential $\mu_k$. 

Because of the flexibility of the membranes and the lateral mobility of the 
anchored molecules, the membranes can attain many microscopic 
states within the contact area. 
Each of these states can be characterized by its configurational 
energy or effective Hamiltonian, $\cH$.\footnote{Here and below, 
the term  `effective
Hamiltonian' is equivalent to the term `configurational energy' which is standard practise in 
 statistical mechanics even though the configurations are described in terms of
classical, i.e., commuting variables and, thus, do not involve any 
quantum-mechanical degrees of freedom. 
The Hamiltonians used in this article are   `effective' in the sense
that they do not describe all molecular details of the biomembranes 
but  focus on the {\em relevant} degrees of freedom. }  
At temperature $T$, the statistical weight of a certain configuration, i.e., 
the probability to observe this configuration,  is then  proportional to  
the Boltzmann factor $\sim \exp(- \cH / k_B T)$ with Boltzmann constant
$k_B$. 

The  elastic deformations of an adhering membrane will be described 
by the separation field $l$. The spatial patterns of  anchored molecules
will be represented 
by  the composition variables $n$. These composition variables are
defined with respect to an underlying lattice of membrane patches. In this 
way, the well-known theoretical framework of lattice gas models 
is extended to flexible surfaces.  \cite{lipo130,weikl00,weikl01}

In general, the membrane-anchored molecules may experience a variety 
of intermolecular forces. We will distinguish between {\em cis}- and 
{\em trans}-interactions of these molecules. \cite{lipo130} Since two
molecules  that are anchored to the same membrane cannot occupy the 
same membrane patch, these molecules always experience
hardcore {\em cis}-interactions which are repulsive and short-ranged. 
In addition, these molecules may stick to each another, which corresponds
to short-ranged attractive cis-interactions, or they may carry 
electric charges, which can lead to long-ranged 
 cis-interactions. In addition, the membrane-anchored molecules
mediate the  trans-interactions between the  membrane and  the second surface. By
definition,  stickers mediate attractive trans-interactions whereas repellers mediate
repulsive ones. 

The  lattice gas models considered here have two rather useful features: (i) The hardcore
cis-interaction  between the anchored molecules, which leads to their 
mutual exclusion within the membrane,  is automatically incorporated; and 
(ii) If this hardcore interaction is the dominant cis-interaction, one can perform the
partial summation  over the composition variables in the partition function. As 
a result, one obtains  effective membrane models that depend only on the 
separation field $l$.

\subsection{Homogeneous or uniform membranes}

\subsubsection{Membrane configurations and effective Hamiltonian}
\label{S.EffectiveHamiltonian}

Our theoretical description starts with 
a  homogeneous or uniform membrane in contact with  another, planar surface
with Cartesian coordinates  $x \equiv  (x_1,x_2)$. 
This membrane will be viewed as a thin elastic sheet  that exhibits an average 
orientation parallel to this planar surface. If we ignore overhangs, 
the membrane shape can be parametrized by the separation field 
$ l(x)$  which describes  the local separation of the membrane from the planar  
reference state with  $l(x) \equiv 0$. 

The elastic deformations of the membrane are governed by two parameters: 
(i) the membrane tension $ \sigma$, which is conjugate to the total 
membrane area,  and (ii) the bending rigidity $ \kappa$ that governs the 
bending energy of the membrane. In addition, the membrane experiences 
direct interactions with the second surface that will be described by the interaction
energy per unit area or  {\em effective
membrane potential} $V_\me(l)$. 
The membrane configurations are then governed  by the effective
Hamiltonian  \cite{lipo33}
\be
\cH \{l \} = \cH_\el \{l \} + \cH_{\rm in} \{l \}
\label{HHomogen}
\ee
which consists of the elastic term $\cH_\el \{l \}$ and the interaction term 
$\cH_{\rm in} \{l \}$. The elastic term has the form 
\be
\cH_\el \{l \} = \int d^2 x \, \, \left[
 \half  \sigma  \,  (\nabla l)^2 + \half \kappa \,  (\nabla^2 l)^2 \right]
 \label{HElastic}
\ee
where the  $\sigma$--term is proportional to the excess area of the deformed membrane
and the $\kappa$--term is proportional to the squared mean curvature of the membrane, 
see Appendix A. The interaction term represents an integral over the effective membrane
potential as given by 
\be
\cH_{\rm in} \{l \} =   \int d^2 x \, \,  V_\me (l) . 
 \label{HInHomogen}
\ee

The effective Hamiltonian as given by (\ref{HHomogen}) -- (\ref{HInHomogen}) 
also applies to two interacting
membranes which are characterized by  bending rigidities $\kappa_1$ and $\kappa_2$ and 
membrane tensions $\sigma_1$ and $\sigma_2$, respectively.  In this case, the separation
field
$l$  describes the local distance between the two membranes which can both exhibit bending 
deformations and, thus, attain a nonplanar shape, see Appendix A. For two interacting
membranes, the paramaters
$\sigma$ and $\kappa$ in the elastic part  (\ref{HElastic}) now represent the effective
tension \cite{lipo69}
\be
\sigma  \equiv  \sigma_1 \sigma_2 / ( \sigma_1 + \sigma_2 )
\label{EffectiveTension}
\ee
and the effective bending rigidity \cite{lipo50}
\be
\kappa  \equiv  \kappa_1 \kappa_2 / (\kappa_1 + \kappa_2 ).
\label{EffectiveRigidity}
\ee

The interaction potential  between the two surfaces always contains a repulsive {\em  
hardwall potential}
which  ensures that the two surfaces cannot penetrate each other and that 
the separation field $l$ satisfies $l \ge 0$. 
This hardwall potential can be implemented in two ways: (i) It may be included into 
the definition of the effective membrane potential via $V_\me (l)
\equiv
\infty$  for $l < 0$; or (ii) it may be embodied by restricting the $l$--integration
 in the partition
function to  positive values. In the following, we will  use the second
implementation and, thus,  define the partition function via
\be
{\cal Z} = \int_{l > 0}  {\cal D} \{l \}  \, \exp[ - \cH \{ l \}/ k_B T]  . 
\ee

\subsubsection{Classification of effective membrane potentials} 

The effective membrane potential $V_\me (l)$ describes the interaction free 
energy of two planar surfaces at uniform  separation $l = const$. In general, 
the functional dependence of $V$ on $l$ will reflect various intermolecular 
forces such as van der Waals, electrostatic, or hydrophobic interactions. 
\cite{lipo115} In addition, 
the effective membrane potential also depends on external forces or constraints such 
as an applied osmotic pressure or  the confinement via another wall that provides an 
upper bound for the separation field $l$. 

It is useful to distinguish two different classes of membrane potentials 
corresponding to  membrane confinement and membrane adhesion, respectively.
 
{\em Membrane confinement} is described by  potentials $V_\me(l) $ that 
do not attain a finite value in the limit of large $l$. Simple examples are provided by 
(i) a confining wall at $l\equiv L$ which implies $V_\me = \infty$ for 
$l > L$;  (ii) a harmonic potential as given by $ \half G (l - l_*)^2$; and 
(iii) an osmotic pressure $P$ which implies that
$V(l)$ contains  the linear term  $P l$ for $l > 0$.  

{\em Membrane adhesion}, on the other hand, corresponds to  potentials
$V_\me(l) $ that (i) have at least one attractive potential well  and (ii)  
attain a finite value in the limit of large
$l$. Since we may always replace $V_\me$ by $V_\me (l) - V_\me(\infty)$, it is 
sufficient to consider potentials that decay to zero for large $l$. 
All effective potentials that arise from intermolecular forces have this 
latter property. 

A confined membrane exhibits critical behavior as one reduces the strength of the 
confining potential. If this potential is symmetric, i.e., if we can define a shifted
separation field $l^\prime \equiv l - l_o$ 
in such a way that $V_\me(- l^\prime) = V_\me (l^\prime)$, the average value
$\langle l^\prime \rangle $ is always zero  but the variance 
$\langle ( l^\prime - \langle l^\prime \rangle)^2 \rangle $ diverges as one reduces
the confining potential. This  {\em delocalization} behavior  is  obtained
(i) for a confining wall as one  moves this wall to larger values of $l = L$  and (ii) for a harmonic
potential  as one  decreases the strength $G$ of this potential. For an
asymmetric  potential, on the other hand, the confined membrane
also unbinds from the second surface at $l=0$ as one decreases the strength of the 
confining
potential. One example is provided by an external osmotic pressure $P$: in the limit of small pressure
$P$,  one obtains the  {\em complete unbinding } behavior
$\langle l^\prime \rangle 
\sim 1/ P^{1/3}$. \cite{lipo33} 

An adhering membrane, on the other hand, exhibits critical behavior as one 
effectively reduces the attractive part of the effective membrane potential
$V_\me(l)$. 
Since the membrane configurations are 
governed by $V_\me(l) / k_B T$, this reduction can be most easily obtained by raising 
the temperature $T$. In the absence of any confining potential, the membranes then 
undergo an {\em unbinding transition} at a characteristic  temperature $T = T_u$. 
The nature of this transition depends on the functional dependence of the 
effective membrane 
potential $V_\me$ on the separation field $l$. 
The unbinding transition is {\em continuous} if  $V_\me(l)$ has 
a single potential well but no potential barrier.  \cite{lipo33,lipo57}
Such an effective membrane potential
arises, e.g.,   from the interplay of van der Waals and hydration forces. 
If $V_\me(l)$ contains both a potential well and a potential barrier, this transition is
continuous for sufficiently low potential barriers but discontinous for 
sufficiently high barriers. \cite{lipo118,lipo124}

Unbinding transitions  have been observed 
experimentally by Mutz and Helfrich \cite{mutz89} for glycolipid bilayers
 and by Pozo-Navas {\em et al.} \cite{pozo03} for   bilayers composed of two phospholipids. 
In both cases, the  composition of the bilayer membranes was presumably uniform.

\subsection{Discretized models for uniform membranes}
\label{S.DiscreteModels}

In order to include the anchored molecules in the theoretical description, 
we will first discretize the uniform  membrane. A convenient discretization 
is provided by  a square lattice within the planar reference plane, see 
Fig.~\ref{figureDiscreteModel}. The corresponding lattice parameter is denoted
by $a$. In this way, the 2--dimensional coordinate $x$ is replaced by 
a discrete set of lattice sites labeled by the index $i$.  
The membrane configurations are now described in terms of separation 
fields $l_i$ associated with the lattice sites $i$, and the membrane is divided
up into discrete membrane patches, each of which has projected area $a^2$, 
see  Fig.~\ref{figureDiscreteModel}. 

Since the elastic part of the effective Hamiltonian  depends on 
the derivatives of $l$ with respect to the coordinates $x_1$ and $x_2$, we have to 
discretize these derivatives as well. For the excess area term $\sim
(\nabla l)^2 $, we will use the  discretization 
\be 
(\nabla_d l_i)^2 \equiv [l(x_1+a,x_2) - l(x_1,x_2)]^2 + 
[l(x_1,x_2+a) - l(x_1,x_2)]^2
\label{DiscNabla^2}
\ee
where $x_1$ and $x_2$ denote the Cartesian coordinates of the lattice site $i$. 
Likewise, the discrete Laplacian  is taken to be \cite{lipo57}
\be
\nabla_{d}^2 l_{i} \equiv  l (x_1+a,x_2) + l (x_1-a,x_2) + l (x_1,x_2+a)+ 
l(x_1,x_2-a) - 4l(x_1,x_2)
\label{DiscLaplacian}
\ee

The elastic Hamiltonian now has the form 
\be
\cH_\el \{l \} = \sum_i  [ \half \sigma  \left(\nabla_d l_i\right)^2 +
  \frac{1}{2 a^2} \, \kappa \,  (\Delta_d l_i)^2  ]
\label{HElasticDis}
\ee
and the  interaction Hamiltonian becomes
\be
\cH_{\rm in} \{l\}=    \sum_i   V(l_i)
\label{HInDisc} 
\ee
which represents a summation over all membrane patches $i$ with 
potential energies
\be
V(l_i) \equiv a^2 V_\me (l(x_1, x_2))
\ee
where $x_1$ and $x_2$  denote again the Cartesian coordinates of the lattice site i. 
Note that $V(l_i)$ and  $V_\me (l)$ have the dimensions of energy and energy per area, 
respectively.  

Realistic estimates of the entropy and free energy of  the membranes require that the lattice
constant $a$ is equivalent to the smallest possible wavelength for bending fluctuations of the
membranes.  Computer simulations with molecular membrane models indicate that this size is
somewhat larger than  the thickness of the lipid bilayer and of the order of 6 nm
\cite{lipo137,lipo154}. 

\subsection{Lattice gas models: General form}

Next, we include the membrane-anchored molecules, that may act as stickers
or repellers, into the theoretical modelling. In general, the membrane may contain 
$K$ different types of such molecules which will be distinguished by the index
$k=1, ..., K$. Since all membrane-anchored molecules undergo lateral diffusion 
along the membrane, these molecules can  form  many
different  {\em spatial patterns}. 
In order to describe these   patterns, we now introduce composition variables $n_i$
for all  lattice sites $i$. Each composition variable can  attain the values 
$n_i = 0, 1, ..., K$. If the membrane patch  $i$ contains the membrane-anchored
molecule of type $k$, this patch is characterized by $n_i = k$. If the patch does not 
contain any of the $K$ membrane-anchored components, the 
composition variable $n_i$
has the special value $n_i =0$.  

In the absence of the second membrane or surface, the concentrations of the
$K$ species of 
membrane-anchored molecules are governed by  $K$ chemical potentials  $\mu_k$ 
with $k=1, 2, \dots, K$. In addition, the cis-interaction between
one molecule of species $k$ located at lattice site $i$ 
and another molecule of species $k^\prime$ located at site $j$ is described by 
the pair-potential $W_{ij}^{k,k^\prime}$ which is negative for 
attractive cis-interactions. Thus, the configurations of the composition variables
$n_i$ are governed by the cis-interaction part of the effective 
Hamiltonian as given by
\be
\cH_\ci \{n\} = - \sum_i \sum_{k=1}^K   \mu_k \delta_{k,n_i}
 +  \sum_{\langle ij \rangle} \sum_{k=1}^K \sum_{k^\prime=1}^K  \delta_{k,n_i} 
  \delta_{k^\prime,n_j}W_{ij}^{k,k^\prime}
\label{HComp}
\ee
with the Kronecker symbol $\delta_{k,n}$ which is defined by $\delta_{k,n}=1$
for $k = n$ and 
$\delta_{k,n}=0$ otherwise.  The symbol $\langle ij \rangle$ indicates a summation 
over all pairs of lattice sites
$i$ and $j$.  Note that the chemical potential term alone 
already embodies the hardcore interactions between two neighboring membrane-anchored 
molecules  because of the underlying lattice.

The configuration of the adhering membrane is now described both by its separation field
$l_i$ and by its composition variables $n_i$. These two degrees of freedom are governed
by the effective Hamiltonian
\be
\cH \{l, n\}= \cH_\el \{l, n \} + \cH_{\rm in} \{l,n\}  
\label{HGeneral}
\ee
where the elastic part is now given by 
\be
\cH_\el \{l, n \} = \sum_i  [ \half \sigma  \left(\nabla_d l_i\right)^2 +
 \frac{1}{2 a^2} \, \kappa_i \, (\Delta_d l_i)^2  ]
\label{HElasticGeneral}
\ee
where the bending rigidities $\kappa_i$ depends, in general, on the composition
variable
$n_i$. In the following sections, we will mostly consider the simplified 
situation characterized by an $n$--independent bending rigidity 
$\kappa_i = \kappa$; an exception
is  section \ref{sectionStiffStickers} where the presence of sticker molecules leads to 
more rigid membrane patches. 

The interaction part of the effective Hamiltonian (\ref{HGeneral}) 
 has the  more general form
\be
\cH_{\rm in} \{l,n\} =  \cH_\ci \{n\} + \cH_\tr \{l, n\}    
\ee
which consists of the cis-interaction part $\cH_\ci $ as given by (\ref{HComp})
and the {\em trans}-interaction part $\cH_\tr$ which describes
the various trans-interactions between 
the two surfaces as mediated by 
the different membrane-anchored molecules. 

The membrane patch $i$ experiences the  trans-interaction $V_k(l_i)$
if the patch contains  a membrane-anchored molecule of type $k$,  and 
the trans-interaction  $V_0(l_i)$ if this patch 
does not contain any membrane-anchored molecule. The  total 
 trans-interaction part of the effective Hamiltonian is then given by 
\be
\cH_\tr \{l, n\} \equiv \sum_i   \sum_{k=0}^K \delta_{k,n_i} V_k (l_i) 
\label{HInGeneral}
\ee
with the Kronecker symbol $\delta_{k,n} $ as before. 

So far, we have not specified the projected area of the membrane-anchored molecules. 
If this projected area does not exceed the area $a^2$ of the membrane 
patches, the short-ranged repulsive cis-interactions between the 
membrane-anchored molecules are
incorporated by the underlying lattice and the  composition variables 
 $n_i$.  Thus, if the membrane-anchored molecules do not experience attractive or 
long-ranged 
repulsive cis-interactions,   the interaction Hamiltonian 
attains the  simple form 
\be
\cH_{\rm in} \{l,n\} =  \sum_i   \sum_{k=0}^K \delta_{k,n_i} [ V_k (l_i)  - \mu_k  ]
\quad {\rm with} \quad \mu_0 \equiv 0 . 
\label{HInNoCis}
\ee
This form is particularly useful since one may now perform the partial summation 
over the composition variables $n_i$ in the partition function as will be explained 
in more detail below. Furthermore, we will typically assume that the trans-interaction 
$V_0(l_i)$ for membrane patches without any stickers or repellers corresponds
to a short-ranged repulsion that can be incorporated into the hardwall repulsion
that ensures $l_i \ge 0$ for all $i$. 

\subsection{Membranes with  sticker molecules}
\label{S.Membranes+Stickers}

\begin{figure}
\begin{center}
\resizebox{0.6\columnwidth}{!}{\includegraphics[angle=90]{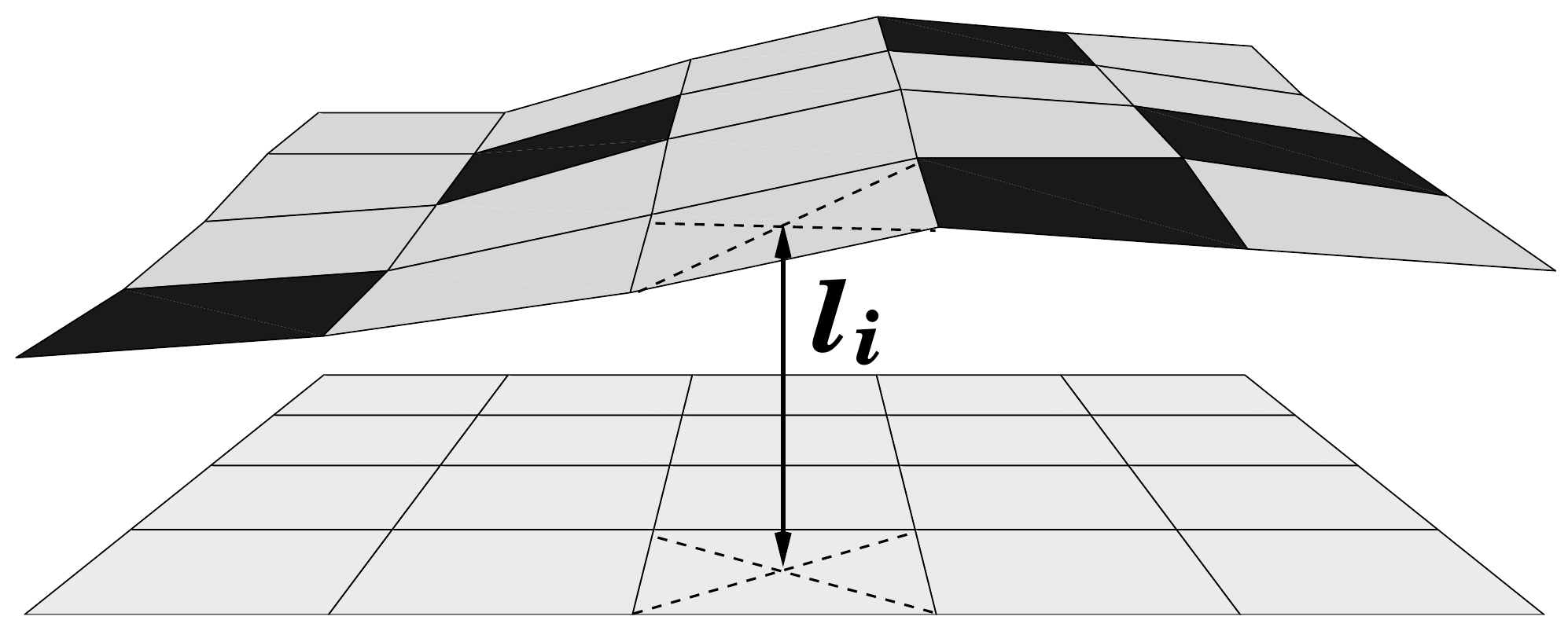}}
\caption{Segment of  deformable membrane in contact with a homogeneous and planar 
surface (such as another membrane supported on a planar substrate). 
The planar surface  provides a  `reference' plane which is discretized into a square lattice. 
The deformable membrane is composed of two types of  patches: Membrane 
patches without  a sticker molecule are grey wheras membrane patches with such a 
molecule are black. 
The configurations of the deformable membrane are described (i) by the separation field
$l_i$, which measures the local separation of the two surfaces,  and (ii) by the composition
variable $n_i$ with  $n_i=0$ for  sticker-free patches and  $n_i=1$ for patches 
with a sticker molecule. }
\label{figureDiscreteModel}
\end{center}
\end{figure}

The simplest example of adhesion mediated by membrane-anchored molecules is 
provided by membranes that contain only one species of sticker molecules.
\cite{lipo130,weikl00,weikl01} In this case, the composition variables $n_i$ attain
 only two values: 
 $n_i=1$ corresponding to  a sticker in   membrane patch  $i$,  and
$n_i=0$ corresonding to no sticker  in this patch, see
Fig.~\ref{figureDiscreteModel}. Therefore, the
composition variables $n_i$  are  now equivalent to occupation numbers for
the sticker molecules. 

Inspection of Fig.~\ref{figureDiscreteModel} shows that the 
 membrane models considered here represent   effectively two-component
systems. It is important to note, however, that  the two components do not 
correspond to 
individual lipid or sticker molecules, but  to lipid bilayer patches of size
$a^2$ with or without a sticker molecule. In the absence of the 
stickers,  the lipid bilayer patches need not consist of a single lipid species 
but are  assumed to have
a homogeneous or  uniform composition. In this sense, the models described here
correspond to 
multi-component membranes that can have many nonadhesive components but  
only one adhesive component. 

In section \ref{S.EntropicMechanisms}, we will consider several models 
of multi-component membranes with one species of membrane-anchored stickers
that  interact with an effectively homogeneous or uniform membrane. 
Such a situation arises if 
the stickers of the multi-component membrane bind directly to a lipid bilayer with 
uniform composition, or if the stickers bind to ligands 
 in the second membrane that are present at high concentrations. These models
can be extended to more complex situations in which both membranes contain 
 sticker  molecules  \cite{lipo130,weikl01}.

The effective Hamiltonian for a
multi-component membrane that interacts with a second, homogeneous membrane
via one species of stickers
can be obtained from the general expression (\ref{HGeneral}) with $K = 1$. 
One then obtains the effective Hamiltonian 
$\cH \{l, n\}= \cH_\el \{l, n \} + \cH_\ci \{n\} + \cH_\tr \{l, n\} $
with the elastic part $\cH_\el \{l, n \}$ as given by  (\ref{HElasticGeneral}).
   
The cis-interaction part of the effective Hamiltonian now has the relatively simple form
\cite{lipo130,weikl00,weikl01}
\be
\cH_\ci \{n\} =  - \sum_i \mu n_i +  \sum_{\langle ij \rangle} W_{ij} n_i n_j   
\label{HCompStickers} 
\ee
where  $\langle ij \rangle$ indicates a summation over all pairs of lattice 
sites $i$ and $j$, and the parameters $\mu \equiv \mu_1$ and
$W_{ij}
\equiv  W_{ij}^{1,1}$ represent the  sticker chemical potential and the sticker-sticker 
pair potential, respectively. Attractive and repulsive pair-potentials are 
described by $W_{ij} < 0$ and $W_{ij} > 0$, respectively. 
  
The trans-interaction part of the effective Hamiltonian is now given by 
\cite{lipo130}
\be
\cH_\tr \{l, n\} =  \sum_i \left[ (1- n_i) V_0 (l_i)+ n_i V_1 (l_i)  \right] , 
\label{HInStickers}
\ee
where $V_0$ is the trans-interaction of a  membrane patch without  sticker
whereas $V_1$ is the trans-interaction of a patch with  sticker. 

The interaction part of the effective Hamiltonian again simplifies if the 
projected area of the sticker molecules does not exceed the patch area
$a^2$ and if two sticker molecules 
 do not experience attractive cis-interactions or long-ranged repulsive
ones. If one can also neglect the  trans-interaction $V_0$ for
a sticker-free membrane patch, the interaction Hamiltonian is simply 
given by \cite{weikl00,weikl01}
\be
\cH_{\rm in} \{l, n \} = \sum_i  n_i  [ V_1 (l_i)  - \mu  ] . 
\ee

 In section \ref{sectionLargeStickers}, we will also
consider sticker molecules which cover several membrane patches. 
This implies that the projected area of the sticker molecule exceeds the
  patch area $a^2$ and that 
the smallest separation  of two sticker molecules is larger than the 
lattice parameter $a$.  This situation will be described by repulsive
pair-potentials $W_{ij} > 0$ which extend to next-nearest neighbor 
lattice sites in order to 
prevent the overlap of adjacent sticker molecules.

\subsection{Two types of membrane-anchored molecules}

Another relatively simple case is provided by multi-component membranes 
with two species of membrane-anchored molecules
\cite{weikl01,weikl02a,weikl04}. In section 
\ref{S.BarrierMechanisms}, we will discuss
membranes with one type of sticker and one type of repeller molecule. Likewise, 
one might consider membranes with two species of sticker molecules, say short and long
ones. In these cases, the composition variables $n_i$ can attain  three  different
values $n_i = 0, 1$, and 2 where the value $n_i = 0$ corresponds to a membrane
patch without any membrane-anchored molecule as before. 
For stickers and repellers, the values $n_i =1$ and $n_i =2$ are taken to indicate
a sticker  and a repeller molecule, respectively. 

It is again useful to consider the simplified situation 
in which (i)  the hardcore interaction between the stickers and repellers
represents their dominant cis-interaction, and (ii) the trans-interaction $V_0(l)$
for membrane patches without a sticker or repeller molecule is well described
by the hardwall potential which ensures  $l> 0$. The interaction part 
of the effective Hamiltonian then has the  simple form \cite{weikl01,weikl02a}
\be
\cH_{\rm in} \{l, n\} = \sum_i \left[ \delta_{1,n_i} \left(V_s(l_i)-\mu_s\right) + 
\delta_{2,n_i} \left(V_r(l_i)-\mu_r\right)\right]
\label{HInStickerRepeller}
\ee
as follows from   (\ref{HInNoCis}) with the sticker chemical potential $\mu_s \equiv \mu_1$, 
the repeller chemical potential  $\mu_r \equiv \mu_2$, the  
trans-interaction  $V_s \equiv V_1$ mediated by the stickers, and 
the trans-interaction $V_r \equiv V_2$ mediated by the repellers. 

The equilibrium phase behavior for a membrane with anchored stickers 
and repellers will be reviewed in section
\ref{sectionMobileRepellers}. The adhesion dynamics of a multi-component vesicle with
stickers and repellers is the topic of section \ref{sectionSR}. In the latter case,  
the unbound membrane segment which acts as a reservoir for the sticker and repeller
molecules will be taken into account explicitly, and the total number of membrane-anchored
molecules within the bound and unbound membrane segment will be kept constant.

As mentioned, the linear size $a$ of the membrane patches
corresponds to the smallest possible wavelength, or `cut-off' length, for bending
fluctuations of the membrane. This length is somewhat larger than the bilayer thickness
\cite{lipo137,lipo154}  
 and affects the entropic, fluctuation-induced mechanisms for phase separation
discussed in section  \ref{S.EntropicMechanisms}. 
However, the barrier  mechanisms for the lateral phase separation
of membranes with stickers and repellers as discussed in section 
\ref{S.BarrierMechanisms} are rather 
insensitive to the precise choice of
the cut-off length $a$, as follows from scaling estimates \cite{weikl02a}. 
We will use this property in order to choose  a larger patch size for the  T
cell adhesion model   \cite{weikl04} as reviewed in subsection \ref{sectionTcells}. 
This choice is convenient since one has to consider relatively
large cell-cell contact areas, and  two different
species of stickers which have a rather different linear size.  This larger choice for 
the size of the membrane patches also implies that these patches can 
contain more than one sticker or repeller.

\section{Theoretical methods}
\label{S.TheoreticalMethods}

\subsection{Monte Carlo simulations}

\subsubsection{Simple sampling and importance sampling}

In classical statistical mechanics, systems are characterized by their configurations $\{y\}$ and their configurational energy, or Hamiltonian, ${\cal H}\{y\}$. The thermodynamic properties of the system can be expressed as averages
\begin{equation}
\langle A(y) \rangle = \frac{1}{\cal Z} \sum_y A(y) e^{-{\cal H} \{y\}/k_B T}  
\label{average}
\end{equation}
where ${\cal Z}$ is the partition function
\begin{equation}
{\cal Z} = \sum_y  e^{-{\cal H} \{y\}/k_BT}
\end{equation}
For many systems of interest, partition functions and thermodynamic averages cannot be calculated exactly. These systems have to be studied  with approximate analytical methods or with numerical algorithms. Numerical methods face the problem that the systems of interest have usually many degrees of freedom and, hence, a large configurational space. These methods therefore can only 
probe a subset or `sample' of the full configurational space. 

In the {\em simple or random sampling} Monte Carlo method, a subset $\{y_1, \ldots, y_M\}$ of the configurations space is randomly selected. The average of a certain physical quantity, say $A$, 
is  then  estimated by
\begin{equation}
\overline{A(y)} = \frac{\sum_{i=1}^M   A(y_i) e^{-{\cal H}\{y_i\}/k_BT}}{\sum_{i=1}^M   e^{-{\cal H}\{y_i\}/k_BT}}
\end{equation}
In practice, the random  sampling method works only for systems with a flat energy landscape
for which all configurations have the same energy; one example is provided
by random walks on a lattice \cite{binder92}. Simple sampling is ineffective for other systems where only a relatively small fraction of configurations $y_i$ has a low energy  ${\cal H}\{y_i\}$ and, hence,  a large Boltzmann weight  $\exp[-{\cal H}\{y_i\}/k_BT]$. These configurations then dominate averaged
quantities  $\langle A(y) \rangle$
as  in (\ref{average}), but are hard to find from random sampling. 
 
 Another sampling method, which is more
efficient  than random sampling,  is the {\em importance sampling} Monte Carlo method. With this method, configurations are not randomly selected but `generated' successively via a Markov process. 
One important constraint on this Markov process is that the system relaxes, for long times,   towards the equilibrium distribution 
\begin{equation}
P_{eq} (y_i)= \frac{1}{\cal Z} e^{-{\cal H}\{y_i\}/k_BT}  
\label{eqdistribution}
\end{equation}
In this Markov process, a configuration  $y_i$ is generated from the preceding configuration $y_{i-1}$ with a certain transition probability \mbox{$W(y_{i-1}\to y_i)$}. The transition probabilities have to fulfill the  {\em detailed balance} conditions
\begin{equation}
P_{eq}(y_i) \,  W(y_i\to y_{j}) = P_{eq}(y_{j})  \, W(y_{j}\to y_i) 
\label{detailedbalance}
\end{equation}
which  ensure that the distribution of generated states relaxes towards the equilibrium distribution \cite{binder92,vanKampen92}.

Using the expression (\ref{eqdistribution}) for the equilibrium distribution, the detailed balance condition can   be rewritten in the form  $W(y_i\to y_{j})/W(y_{j}\to y_i)=\exp[-({\cal H}\{y_j\}
-{\cal H}\{y_i\})/k_B T]$ . This means that the ratio of the forward and backward transition rates between the states $i$ and $j$ only depends on the energy difference ${\cal H}_j-{\cal H}_i$. One  choice for the transition rates 
that obeys  detailed balance is provided by the Metropolis algorithm which is defined by 
the transition probabilities \cite{binder92}
\be
\begin{array}{llll}
W(y_i\to  y_j) & = & e^{-({\cal H}\{y_j\} -{\cal H}\{y_i\})/k_B T}  &
 {\rm for \, \,} {\cal H}\{y_j\} -{\cal H}\{y_i\}>0 \\[1ex] 
          &	= & 1       & {\rm otherwise} \ .
\end{array}   
\label{metropolis}
\ee

\subsubsection{Membrane simulations}

\begin{figure}
\begin{center}
\resizebox{0.35\columnwidth}{!}{\includegraphics[angle=-90]{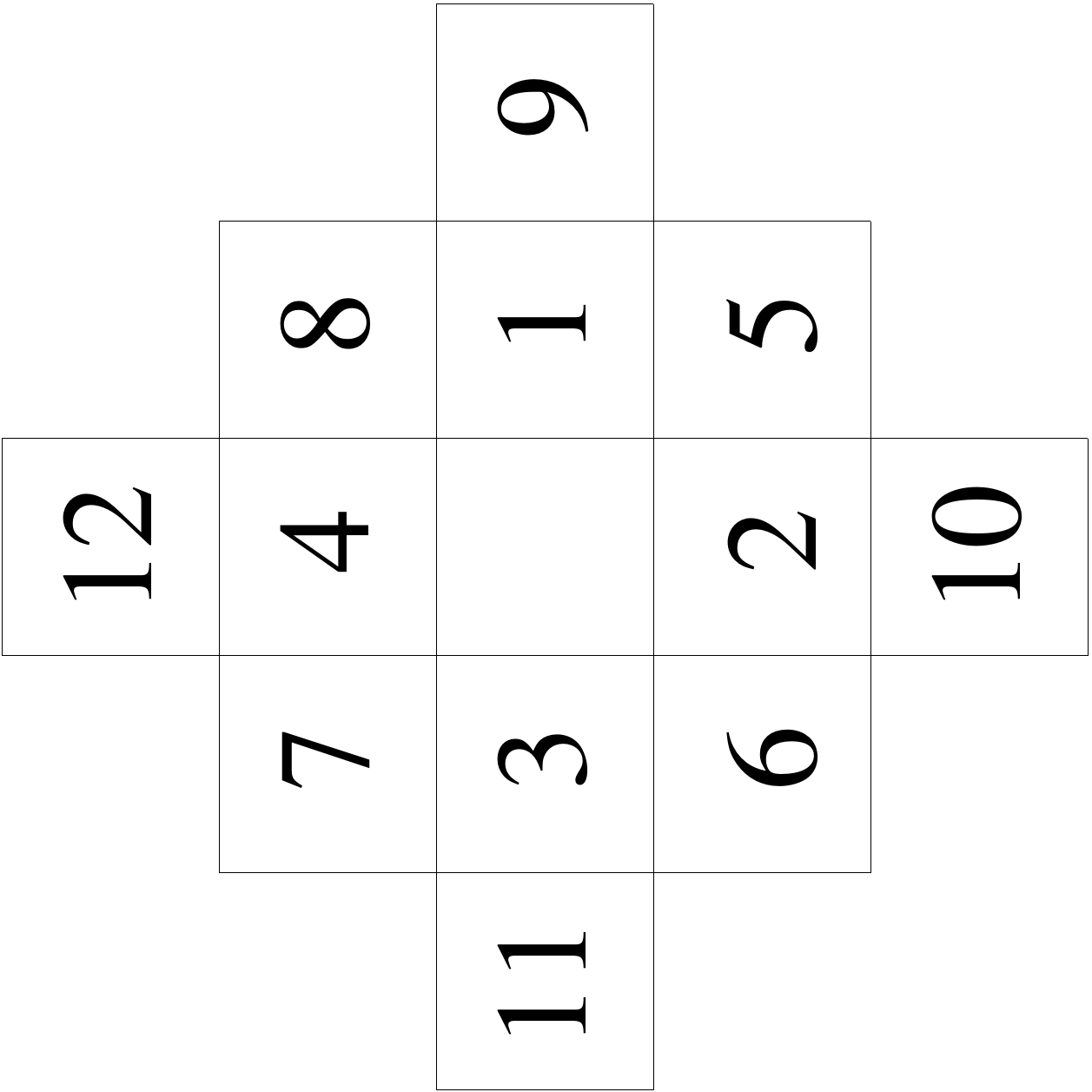}}
\caption{Neighbors of a site, or membrane patch, located in the center of the figure. Here, $1$ to $4$ are the nearest neighbors, $5$ to $8$ the next-nearest neighbors, and  $9$ to $12$ the next-next-nearest neighbors.
\label{figureNeighbors}}
\end{center}
\end{figure}

In order to  illustrate how Monte Carlo methods can be used to study the discrete membrane models introduced in subsection \ref{S.DiscreteModels} above, let us first consider two homogeneous and tensionless membranes with interaction potential  $V(l)$. It is convenient to introduce the rescaled separation field 
\be
z \equiv (l/a)\sqrt{\kappa/k_BT}
\ee
 where $a$ is the lattice parameter (or linear patch size)  of the discretized membrane  and $\kappa$ is the effective bending rigidity. Introducing the rescaled separation field $z$  simplifies the notation and  reduces the number of independent parameters. 
In terms of the rescaled separation field $z$, the Hamiltonian for the two membranes  has the simple form\footnote{Here and below, ${\cal H}\{z\}$ and $V(z_i)$ are dimensionless quantities given in units of the thermal energy $k_BT$.}
\begin{equation}
{\cal H}\{z\} = \sum_i \frac{1}{2} \left\{  (\Delta_d z_i)^2   + V(z_i)  \right\}
\end{equation}
where $\Delta_d$ is the discretized Laplacian. If $i1$, $i2$, $i3$ and $i4$ denote the four nearest neighbors of site $i$ on a square lattice, see Fig.~\ref{figureNeighbors}, the discretized Laplacian can be written as $\Delta_d z_i = z_{i1}+z_{i2}+z_{i3}+z_{i4}-4z_{i}$. Usually, a `new' membrane configurations in the Markov process is generated from an `old' configuration by attempting a local move $z_i = z_{\text{old}} \to z_i = z_{\text{new}}$ for a randomly selected lattice site $i$. Choosing a new value for the rescaled membrane separation $z_i$ at site $i$ affects not only 
the discretized Laplacian at this site, but also at the four nearest neighbor sites $i1$, $i2$, $i3$ and $i4$. Therefore, the energy difference between the `old' and `new' membrane configuration is
given by
 \be
\begin{array}{lll}
 \Delta {\cal H}& \equiv &  {\cal H}_{\text{new}} - {\cal H}_{\text{old}} 
\equiv {\cal H}|_{z_i =z_{\text{new}}} - {\cal H}|_{z_i =z_{\text{old}}} \\[1.5ex] 
          &	= &    \frac{1}{2} \left[(\Delta_d z_i)^2 + (\Delta_d z_{i1})^2+
  (\Delta_d z_{i2})^2 + (\Delta_d z_{i3})^2 + (\Delta_d z_{i4})^2\right]_{z_i =z_{\text{new}}} - \\[1ex]
          &    &   \,\frac{1}{2} \left[(\Delta_d z_i)^2 + (\Delta_d z_{i1})^2+
  (\Delta_d z_{i2})^2 + (\Delta_d z_{i3})^2 + (\Delta_d z_{i4})^2\right]_{z_i =z_{\text{old}}} + \\[1ex]
          &   &  V(z_{\text{new}}) - V(z_{\text{old}}) 
  \end{array}   
\ee

If $i5$ to $i8$ denotes the next-nearest neighbors of site $i$, and $i9$ to $i12$ the next-next-nearest neighbors (see Fig.~\ref{figureNeighbors}), the energy difference can be written as
 \begin{eqnarray}
 {\cal H}_{\text{new}} - {\cal H}_{\text{old}}  =  \hspace*{-.0cm}10 \left( z_{\text{new}} ^2 - z_{\text{old}}^2\right) + [z_{\text{new}} - z_{\text{old}}]
        [ -8 (z_{i1} +z_{i2} +z_{i3} +z_{i4})\nonumber\\
         + \; 2 (z_{i5} +z_{i6} +z_{i7} +z_{i8}) + z_{i9} +z_{i10} +z_{i11} +z_{i12}    ]
         + V(z_{\text{new}}) - V(z_{\text{old}}) \hspace*{0.2cm} \label{deltaH}
\end{eqnarray}
In the Metropolis dynamics (\ref{metropolis}), the move is always accepted if  ${\cal H}_{\text{new}} -{\cal H}_{\text{old}}$ is negative, but is only accepted with probability  $\exp[-({\cal H}_{\text{new}}-{\cal H}_{\text{old}})]$ if  ${\cal H}_{\text{new}} -{\cal H}_{\text{old}}$ is positive. Of course, these local moves have to be attempted at all sites or membrane patches, and typically a large number of Monte Carlo steps per site is required to reach the equilibrium distribution from a given initial configuration. 

It is important to choose new values $z_{\text{new}}$ for the rescaled separation in an unbiased way, 
e.g., using the rule
\begin{equation}
z_{\text{new}} = z_{\text{old}} + \delta z \,  \zeta 
\end{equation}
where $\delta z$ is the step size, and $\zeta$ is a random number between -1 and 1. Steps with negative values for $z_{\text{new}}$ are rejected since the two membranes cannot penetrate each other.

The relaxation time in units of Monte Carlo steps depends on the step size $\delta z$. If $\delta z$  is large, then only a small fraction of Monte Carlo steps will be accepted and the relaxation is slow. If $\delta z$  is small, most Monte Carlo steps will be accepted, but the relaxation is also slow because it takes a large number of steps per site to obtain a significantly different configuration. Usually, a suitable value for $\delta z$ can be found by trying to obtain an average acceptance rate between 0.4 and 0.5 for the Monte Carlo moves. 

The simulation of 
 multicomponent membranes requires  Monte Carlo moves of the concentration field $n$ in addition to the moves of the  separation field $z$. In the grandcanonical ensemble,  for example, a simple Hamiltonian for a membrane with stickers is given by%
\begin{equation}
{\cal H}\{z,n\} = {\cal H}_{el}\{z\} + \sum_i n_i\left[V(z_i)-\mu\right]  
\end{equation}
where $\mu$ is the chemical potential of the stickers. The local concentration $n_i$ adopts the values 0 or 1 indicating the absence or presence of  a sticker at site $i$. The Monte Carlo move  $n_i=n_{\text{old}}\; \to\; n _{\text{new}} = 1-n_i$ attempts to remove stickers from sites with $n_i=1$, and to add stickers at `empty' sites with $n_i=0$. The energy difference for these moves is 
\begin{eqnarray}
{\cal H}_{\text{new}} - {\cal H}_{\text{old}}  ={\cal H}|_{n_{\text{new}} = 1-n_i} - {\cal H}|_{n_{\text{old}} =n_i} = (1 - 2 n_i)  [V(z_i) - \mu] 
\end{eqnarray}
%

\subsection{Free energies of adhesion for homogeneous membranes}
\label{sectionFreeEnergies}

Thermally excited shape fluctuations lead to an entropic repulsion between membranes \cite{helfrich78}. If the membranes are bound together by attractive  trans-interactions, thermal fluctuations can cause an unbinding transition at a critical temperature $T_c$ 
\cite{lipo33}. 
 At temperatures $T$ below $T_c$, the attractive interaction between the membranes dominates, and the membranes are bound. At temperatures above $T_c$, the entropic repulsion dominates, and the membranes are unbound. Renormalization group calculations show that the unbinding transition is {\em continuous}, i.e., the average membrane separation $\langle l\rangle\sim |T-T_c|^{-1}$ diverges continuously when the critical temperature is approached from below . This continuous 
behavior has been confirmed by Monte Carlo simulations \cite{lipo57}. 

Monte Carlo simulations can also be used to determine the free energy of adhesion \cite{weikl00,weikl01}. The discretized Hamiltonian for two homogeneous membranes has the form 
as given by  (\ref{HElasticDis}) and (\ref{HInDisc}).  
In the following, the lateral tension $\sigma$ of the membranes is assumed to be zero and 
the attractive interaction potential of the two membranes is taken to  be a square well potential
with  potential depth $k_BT u<0$ and potential range $l_v$ as given by
\begin{eqnarray}
 V(l_i) = k_BT\, u \, \, \theta(l_v - l_i) &=k_BT u &\;\;\text{for $0\le l_i \le l_v$}\nonumber\\
                        &= 0 &\;\;\text{for $l_i > l_v$}  .
                        \label{hompot}
\end{eqnarray}
 After introducing the rescaled separation field $z\equiv (l/a)\sqrt{\kappa/k_BT}$, the system is described by  two dimensionless parameters, the  potential depth $u$ and the rescaled potential range $z_v=(l_v/a)\sqrt{\kappa/k_BT}$.
The hard wall potential which restricts the separation field $l_i$ to positive values is incorporated via 
a lower  bound for the $l_i$--integration. 

\begin{figure}
\begin{center}
\resizebox{0.6\columnwidth}{!}{\includegraphics[angle=0]{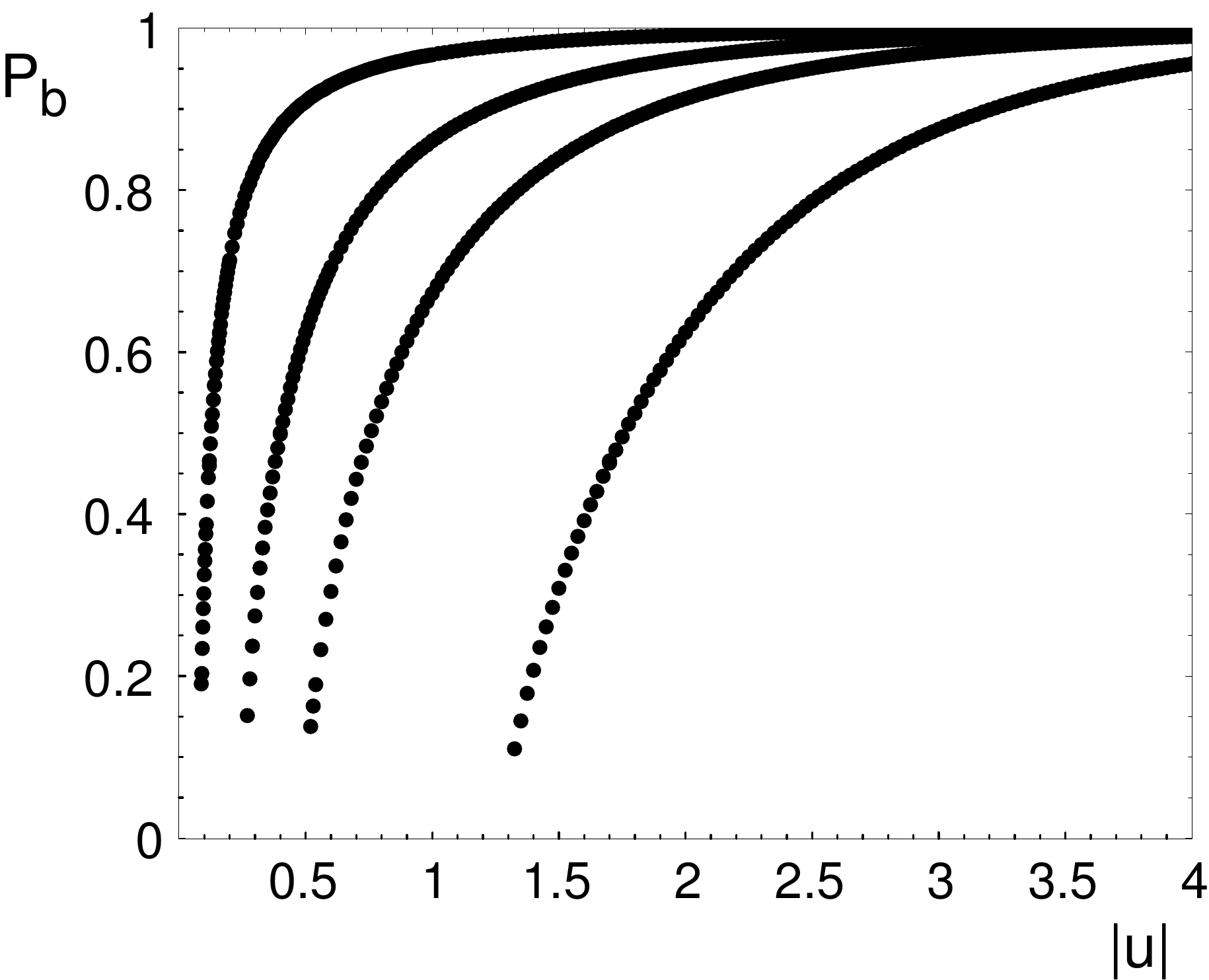}}
\caption{Contact probability $P_b$ of a homogeneous, tensionless membrane as a function of the depth $|u|$ of the square-well potential for the rescaled potential ranges $z_v=1$, 0.5, 0.3, and 0.1 (from left to right).
\label{figurePb}}
\end{center}
\end{figure}

In general, it is difficult to determine partition functions and, thus, free energies via  Monte Carlo simulations. However, the free energy of two adhering membranes can be determined {\it via} the standard method of thermodynamic integration. The quantity of interest here is the contact probability
\begin{equation}
P_b\equiv \langle\theta(l_v - l_i)\rangle
\end{equation}
which represents  the average value for the fraction of bound membrane segments, i.e.,  membrane segments with local separations $0< l_i<l_v$.  The contact probability is related to the free energy 
density 
\begin{equation}
F =- k_BT \ln \left[\prod_i\int_0^\infty dl_i\right]
               e^{\textstyle -{\cal H}\{l\}/k_BT} / A , 
\end{equation}
which represents the free energy per membrane area $A$, via
\begin{equation}
P_b = \frac{a^2}{k_BT} \frac{\partial F}{\partial u} . 
\end{equation}
The contact probability can be determined by  Monte Carlo simulations, see Fig.~\ref{figurePb}. For a given rescaled potential range $z_v$, the free energy of adhesion is then obtained from the Monte Carlo data for the contact probability  by the integration
\begin{equation}
F =  F_{ub} - \frac{k_BT}{a^2} \int_u^{u_*} du' P_b(u') \; .
\end{equation}
Here, $ F_{ub}$ denotes the free energy density of the unbound membrane, and $u_*$ is the critical potential depth. Close to the critical point with $P_b=0$, the correlation length and relaxation time of the membranes diverge. Therefore, small values of $P_b$ cannot be determined reliable in the simulations. But since $P_b$ is proportional to $(u-u_*)$ close to the critical potential depth $u_*$, the critical points and the full functions $P_b(u)$ can be obtained using a linear extrapolation of the data shown in Fig.~\ref{figurePb}. 

\subsection{Effective potential in the absence of cis-interactions}
\label{sectionEffectivePotential}

As explained in section \ref{S.Membranes+Stickers}, 
the simplest Hamiltonian for a discrete membrane with stickers has the form 
\be 
{\cal H}\{l,n\} = \cH_\el \{l\} + \sum_i n_i\left[V(l_i)-\mu\right] 
 \label{hamOne}
\ee
where the elastic part $\cH_\el $ is given by (\ref{HElasticGeneral}) with 
uniform bending rigidity $\kappa_i = \kappa$. 
The stickers  have the size of the lattice parameter $a$ and  cis-interactions between the stickers are absent. 
 Since the effective Hamiltonian in (\ref{hamOne}) is linear in $n$, the partial summation 
 over the sticker degrees of freedom can be performed in the partition function
\begin{equation}
{\cal Z}=\left[\prod_i \int_0^\infty\!\! dl_i\right] \left[\prod_{i}\sum_{n_i=0,1}\right] e^{-{\cal H}\{l,n\}/k_BT}
\end{equation}
which leads to
\begin{eqnarray}
{\cal Z}&=& \left[\prod_i \int_0^\infty\!\!  dl_i\right]
     e^{ - {\cal H}_{el}\{l\}/k_BT} \prod_i \left(1 + e^{[\mu -
V(l_i)]/k_BT}\right) \nonumber\\
&=&\left(1 + e^{ \mu/k_BT}\right)^N \left[\prod_i \int_0^\infty\!\!  dl_i\right]
     e^{ -\big({\cal H}_{el}\{l\} + \sum_i V_{ef}(l_i)\big)/k_BT}
\end{eqnarray}
Here, $N$ is the number of lattice sites. For a square-well sticker potential $V(l_i)=U \theta(l_v-l_i)$,
the effective potential has the form
\begin{equation}
V_{ef}(l_i) = - k_BT\ln\frac{1+e^{[\mu - U]/k_BT}}{ 1+e^{\mu/k_BT}} \theta(l_v-l_i)
        \equiv U_{ef}\theta(l_v-l_i)  \label{Vefgen}
\end{equation}
This effective potential is again a square--well potential. It has the same potential range $l_v$ as the sticker potential, and an effective potential depth $U_{ef}$ which depends on the chemical potential $\mu$ and the binding energy $U$ of the stickers. 

\begin{figure}
\begin{center}
\resizebox{0.6\columnwidth}{!}{\includegraphics[angle=0]{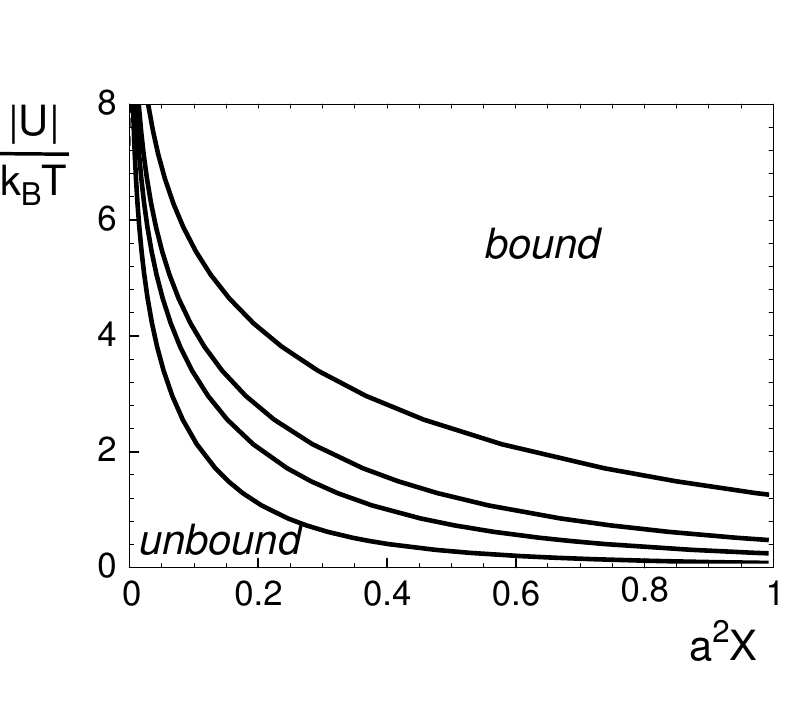}}
\caption{Unbinding lines for the rescaled potential ranges $z_v=1$, 0.5, 0.3, and 0.1 (bottom to top). The membranes are bound at large sticker concentrations $X$ and binding energies $|U|$, and unbound at small sticker concentrations and binding energies.
\label{figureUnbindingLines}}
\end{center}
\end{figure}

Summing out the stickers degrees of freedom in the partition function thus leads to the problem of {\em homogeneous} membranes with square-well interaction. The free energy per unit area ${\cal F} =-(k_BT/A)\ln {\cal Z}$ therefore can be obtained as in the previous section \ref{sectionFreeEnergies}: 
\begin{equation}
{\cal F}=-\frac{k_BT}{a^2}\ln\left[1+e^{\mu/k_BT}\right]+{\cal F}_{ub} -
\frac{k_BT}{a^2}
\int_u^{u_*} du' P_b(u') \label{fjoker}
\end{equation}
Here, $u=U_{ef}/k_BT$ is the dimensionless potential depth $u$, and $P_b(u)$ is the contact probability for homogeneous membranes. The effective potential depth $U_{ef}$ is defined in eq.~(\ref{Vefgen}). The sticker concentration $X\equiv-(\partial F/\partial \mu)=\langle n_i\rangle/a^2$ then follows as
\begin{equation}
X=\frac{1}{a^2}\left[ (1- P_b)\frac{e^{\mu/k_BT}}{1+e^{\mu/k_BT}} +
   P_b\frac{e^{[\mu-U]/k_BT}}{1+e^{[\mu-U]/k_BT}} \right]    \label{concentration}
\end{equation}
To obtain the critical sticker concentration $X_*$ at which the membranes unbind, we have to determine the critical chemical potential $\mu_*$ of the stickers. This critical chemical potential can be derived from the equation $u_*=U_{ef}(\mu_*)/k_BT$:
\begin{equation}
\frac{\mu_*}{k_BT} = \ln \frac{e^{|u_*|} - 1}{e^{|U|/k_BT} - e^{|u_*|}} \label{mustar}
\end{equation}
At the continuous unbinding point, the contact probability $P_b$ is zero. The critical sticker concentration therefore is 
\begin{equation}
X_* = \frac{e^{\mu_*/k_BT}}{a^2(1+e^{\mu_*/k_BT})}=\frac{e^{|u_*|} - 1}{a^2(e^{|U|/k_BT} - 1)} \label{Xstar}
\end{equation}
The resulting unbinding lines for various rescaled range potential $z_v$ are shown in Fig.~\ref{figureUnbindingLines}. The lines follow from eq.~(\ref{Xstar}) and the critical values $u_*=-1.25\pm 0.05$ for $z_v=0.1$,  $u_*=-0.47\pm 0.03$ for $z_v=0.3$, $u_*=-0.24\pm 0.02$ for $z_v=0.5$, and $u_*=-0.075\pm 0.010$ for $z_v=1$. These values are obtained from linear extrapolation of the MC data for the contact probablility $P_b$ shown in Fig.~\ref{figurePb}. 

The two-component membrane considered in this section does not exhibit lateral phase separation. This is a direct consequence of eq.~(\ref{concentration}) and the continuity of the contact probability $P_b$.

\subsection{Variational (mean-field) theory for cis-interactions}
\label{sectionMeanFieldTheory}

As mentioned before, a sticker molecule might also experience cis-interactions with other stickers in the same membrane. In general, these cis-interaction can be attractive or repulsive
and may be short- or long-ranged. If the trans-interaction $V_0$ for membrane patches
without stickers are again purely repulsive and short-ranged, 
the effective Hamiltonian has   the form
\be 
{\cal H}\{l,n\} = {\cal H}_{el}\{l\} + \sum_i n_i\left[V(l_i)-\mu\right]
      + \sum_{\langle ij\rangle} W_{ij} n_i n_j  \label{hamwith}
\ee
as follows from (\ref{HCompStickers}) and (\ref{HInStickers}). 
Here, $W_{ij}$ is the cis-interaction energy between two stickers in the membrane patches $i$ and $j$. The summation index $\langle ij\rangle$ indicates a summation over all pairs of membrane patches. In the following, we consider the short-range interaction energy
\begin{eqnarray}
 W_{ij}  & = W &\;\; \text{for nearest neighbors $i$, $j$}\nonumber\\
                        &= 0 &\;\;  \text{otherwise} \label{cisinteraction}
\end{eqnarray}
which only has a nonzero value $W$ for two stickers in adjacent membrane patches.  

The cis-interaction term in the Hamiltonian (\ref{hamwith}) contains an expression $n_in_j$ which is quadratic in the concentration field. Therefore, a direct summation of the sticker degrees of freedom as in the previous section is not possible. Approximate analytical methods addressing such situations are mean-field theories, which are sometimes also called self-consistent field theories. Mean--field theories can be derived in a systematic way from the variational principle \cite{callen85}
\begin{equation}
{\cal F}\le {\cal F}_o + \frac{1}{A}\langle {\cal H} - {\cal H}_o \rangle_o
    \label{variationalPrinciple}
\end{equation}
Here, ${\cal H}_o$ typically is a Hamiltonian which is linear in the considered field. For such a Hamiltonian, the degrees of freedom can be summed out exactly in the partition function. The variational principle states that the free energy ${\cal F}$ of a system with Hamiltonian $\cal H$ is smaller than or equal to the free energy of a system with Hamiltonian ${\cal H}_o$ {\em plus} the average value of the `perturbation'  ${\cal H} - {\cal H}_o$. This average value is calculated in the system with Hamiltonian ${\cal H}_o$, which is indicated by the subscript 0 of the 
brackets $\langle \cdots \rangle_o$. In the present situation, the Hamiltonian $H_o$ is 
\begin{equation}
{\cal H}_o\{l,n\}\equiv {\cal H}_{el}\{l\}+\sum_i n_i(V(l_i)-\mu +B) \label{Ho}
\end{equation}
where $B$ is a variational parameter and $A$ the membrane area. The effective Hamiltonian ${\cal H}_o$ is linear in the sticker concentration field $n$ and corresponds to stickers with $W=0$ and the shifted
chemical potential $\mu-B$. Therefore,to obtain the free energy $F_o$ we simply have to replace $\mu$ by $\mu-B$ in eq.~(\ref{fjoker}).  Evaluation of $\langle {\cal H} - {\cal H}_o\rangle_o$ now leads to
\begin{eqnarray}
{\cal F} &\le& {\cal F}_o - \frac{B}{A}\sum_i \langle n_i \rangle_o
       +\frac{1}{A} \sum_{\langle ij\rangle}W_{ij}  \langle n_i n_j\rangle_o\\
&\le& {\cal F}_o - \frac{B}{a^2} \langle n_i \rangle_o +  \frac{2W}{a^2}
    \langle n_i \rangle_o^2 \equiv {\cal F}_>    \label{Flarger}
\end{eqnarray}
In deriving the second inequality, we made use of the relation $\langle n_i n_j\rangle_o \ge \langle n_i\rangle_o^2$. The relation results from the fact the correlation function $\langle n_i n_j\rangle_o -\langle n_i\rangle_o^2$ is nonnegative because of fluctuation-induced attractive interactions between bound stickers. These fluctuation-induced interactions will be discussed in section \ref{sectionEntropicInteractions}. The factor 2 in the term $2W/a^2$ is the number of nearest-neighbor patches multiplied with $1/2$ to avoid a double-count of neighbor pairs.

The variational principle (\ref{variationalPrinciple}) tells us that the best approximation for the free energy ${\cal F}$ is the minimum of ${\cal F}_>$ with respect to the variational parameter $B$. Thus, we are looking for a value of $B$ satisfying
\begin{equation}
\frac{\partial {\cal F}_>}{\partial B} = \frac{\partial {\cal F}_o}{\partial B} - \langle n_i \rangle_o
     +   (4 W \langle n_i \rangle_o - B) \frac{\partial \langle n_i \rangle_o}{\partial B} =0 \label{Flargermin}
\end{equation}
Since  $\partial {\cal F}_o/\partial B =-\partial {\cal F}_o/\partial \mu= \langle n_i \rangle_o$, the minimization leads to the self-consistency equation
\begin{equation}
B=4 W\langle n_i \rangle_o \label{selfconsistency}
\end{equation}
Another name for this equation is `mean-field equation' since it is also obtained in the simple approximation
$\sum_{i,j}W_{ij} n_i n_j \approx \sum_{i,j}W_{ij} n_i \langle n_j \rangle = 4 W \langle n_j \rangle \sum_i n_i$
in which the effect of the four neighbors $n_j$ on $n_i$  is taken into account as the `mean field'  $4 W \langle n_j \rangle$.  However, to derive the phase behavior, we also need the free energy (\ref{Flarger}). 
A first-order phase transition corresponds to two solutions $B_1(\mu)$ and $B_2(\mu)$  of the self-consistency equation (\ref{selfconsistency}). At the transition point, the free energies  ${\cal F}_>(B_1(\mu))$ and ${\cal F}_>(B_2(\mu))$ of the two coexisting phases have to be equal. From the latter condition, one can derive the transition value $\mu_t$ of the chemical potential.

\section{Entropic mechanisms for domain formation}
\label{S.EntropicMechanisms}

\subsection{Entropic interactions between bound stickers}
\label{sectionEntropicInteractions}
Membranes that are confined by external forces or constraints loose 
configurational entropy. This entropy loss is proportional to the projected
area of the membrane and may be viewed as as effective repulsive force
as first proposed  by Helfrich   \cite{helfrich78,helfrich84}. For a fluctuating
membrane in contact with another surface,  this fluctuation-induced repulsion behaves as 
\begin{equation}
V_{fl} \approx  c_{fl}(k_BT)^2 / \kappa \, \bar{l}^2 
  \label{flucpot}
\end{equation}
for large average separations $\bar{l}$ between the membrane and the second surface 
where $\kappa$ is, in general, the effective bending rigidity for the separation 
field $l$ of the two surfaces. The dimensionless coefficient $c_{fl}$ has been 
determined by 
 Monte Carlo simulations and is found to be of the order of   0.1 \cite{lipo57,gompper89,janke89,lipo119}. The precise value  of $c_{fl}$ depends on the 
 confinement of the membranes
and varies  from $c_{fl}\simeq 0.08$ for a single membrane between parallel walls \cite{gompper89} to $c_{fl}=0.115\pm0.005$ for a membrane that is pushed by an  external pressure
against a wall \cite{lipo57,lipo119,lipo130}. In a more systematic treatment, this entropically induced interaction arises from the renormalization of the hardwall  potential $V_{\rm hw}(l)$
with  $V_{\rm hw}(l) = 0$ for $l > 0$ and $V_{\rm hw}(l) = \infty$ for $l < 0$ \cite{lipo33}. 

The fluctuation-induced repulsion as given by (\ref{flucpot})  has also been used 
to estimate the shape of a fluctuating membrane that is pinned to another membrane
or  surface by some external constraint 
\cite{barziv,menes97a,menes97b,weikl00b}. A detailed comparison between   
self-consistent calculations based on an effective Hamiltonian $\cH \{l\}$ with 
the membrane potential $V_\me(l) \equiv c_{fl}(k_BT)^2 / \kappa \,l^2$ 
and Monte-Carlo simulations of the fluctuating membrane subject to local  
pinning forces has shown that the dimensionless coefficient $c_{fl}$ depends both 
on the  boundary conditions imposed by the pinning forces and on the 
physical quantity under consideration \cite{weikl00b}. 

\begin{figure}
\begin{center}
\resizebox{0.45\columnwidth}{!}{\includegraphics{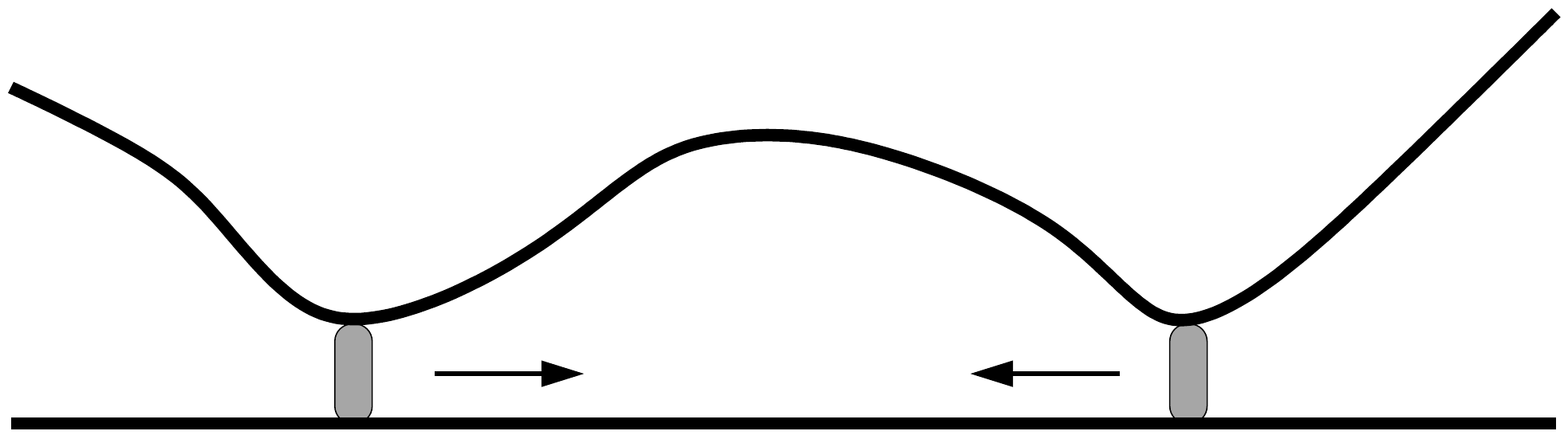}}
\hspace{1cm}
\resizebox{0.45\columnwidth}{!}{\includegraphics{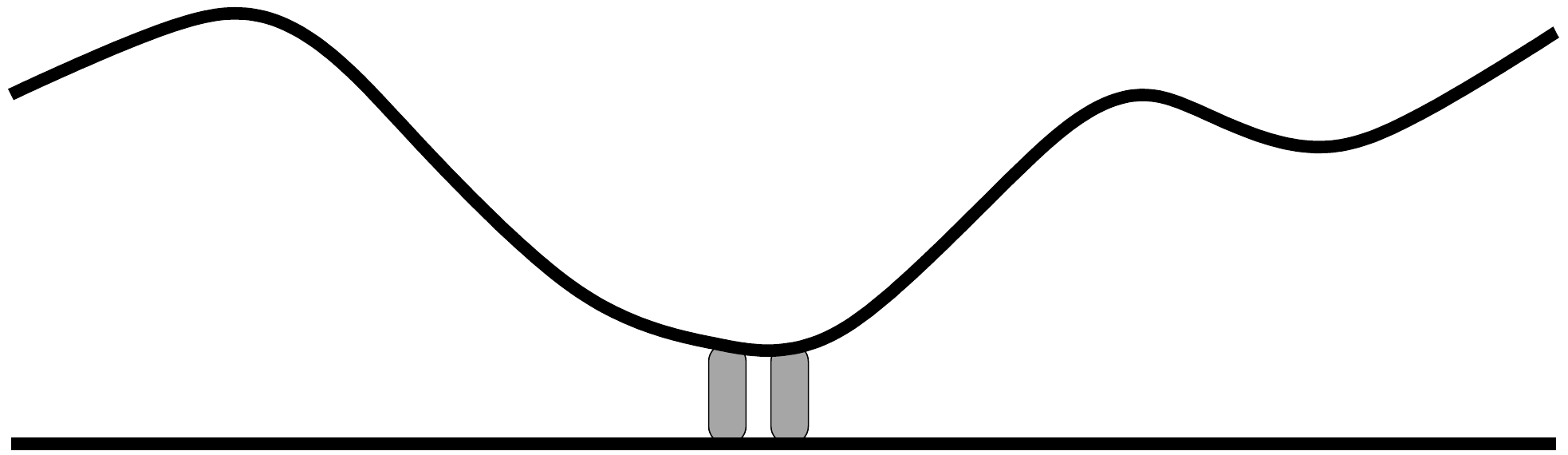}}
\caption{(Left conformation) A fluctuating membrane is pinned to a surface at two different spots by stickers that are far apart from each other. (Right conformation) If the two stickers are close together, the membrane is effectively pinned only at one spot, which costs less conformational entropy of the fluctuating membrane. The membrane fluctuations thus induce an attractive interaction between the stickers.
\label{figureEntropicStickerInt}}
\end{center}
\end{figure}

The membrane fluctuations also induce interactions between bound stickers. These interactions are attractive, as one can intuitively see by inspection of  Fig.~\ref{figureEntropicStickerInt}.  Suppose we have two fluctuating membranes with an average separation much larger than the sticker binding range. Locally clamping the two membranes together by a single bound sticker then costs a certain amount of work $w$ against the entropic repulsion of membranes. If we clamp the membranes together by two stickers which are far away from each other, we have to compress the membranes at two spots, which costs $2 w$. But if the two stickers are very close together,  we only have to compress the membranes at a single spot, with an entropic cost around $w$. Therefore, the right configuration in Fig.~\ref{figureEntropicStickerInt} is entropically preferred, since the left configuration constrains, or suppresses, the membrane fluctuations more strongly. The fluctuations thus induce an attractive interaction between the bound stickers.

In principle, the strength of these attractive interactions between bound stickers can be determined by integrating out the membrane fluctuations. More precisely, one has to integrate out the degrees of freedom for the separation field $l$ in the partition function, for fixed locations of the bound stickers. In practice, integrating out the membrane fluctuations is not possible in a rigorous way, at least for an arbitrary sticker configuration. Approximate scaling arguments have been used to estimate the entropic interactions for a regular array \cite{bruinsma94} and an isolated pair of bound stickers \cite{netz97}. Since many-sticker interactions such as screening effects clearly are important, the fluctuation-induced interactions can again {\em not}  be obtained as a simple sum of pairwise interactions.

In the following, we will focus on approaches which directly address the phase behavior of membranes with stickers. The central question in this section will be if the fluctuation-induced interactions between the stickers can be strong enough to induce lateral phase separation. We will see that the answer to this question depends on the type of stickers. 

\subsection{Small flexible stickers without cis-interactions}

\begin{figure}
\begin{center}
\resizebox{0.7\columnwidth}{!}{\includegraphics{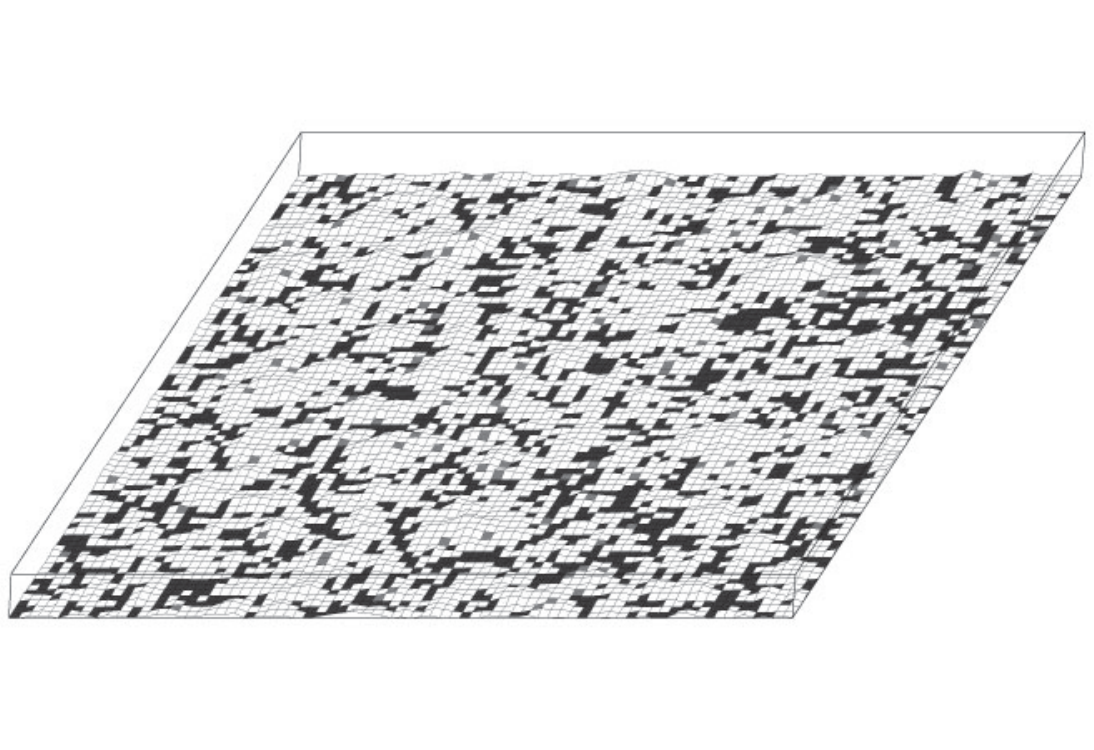}}

\vspace*{-1.3cm}
\resizebox{0.7\columnwidth}{!}{\includegraphics{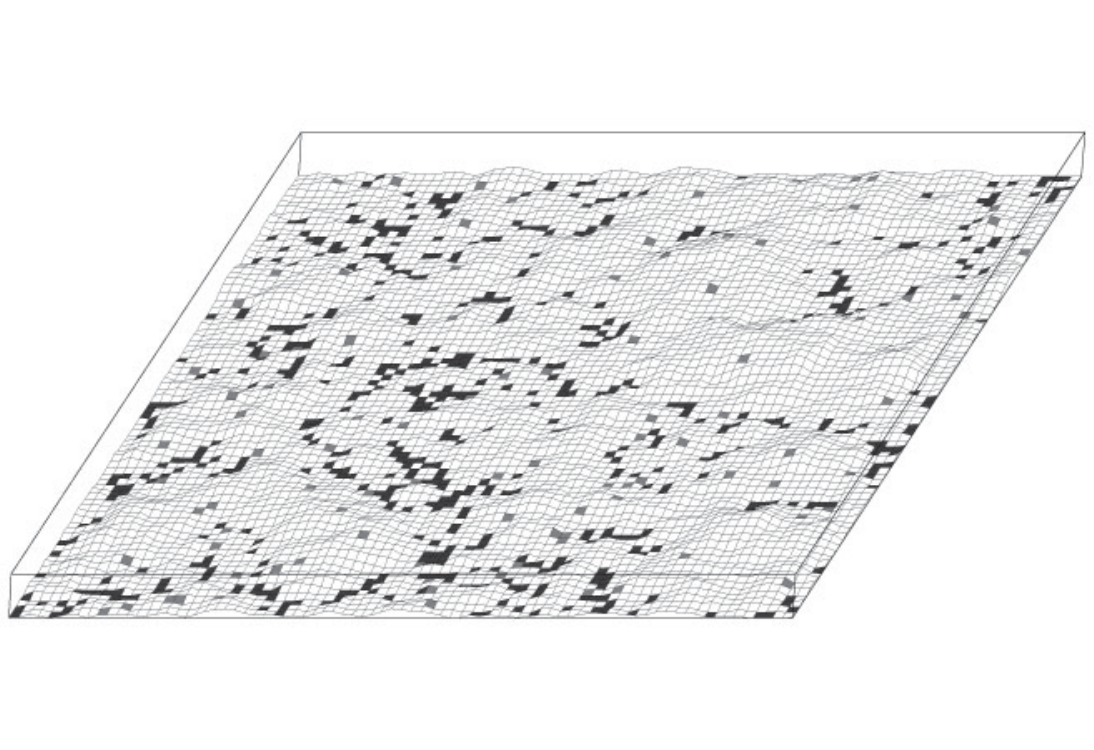}}

\vspace*{-1.3cm}
\resizebox{0.7\columnwidth}{!}{\includegraphics{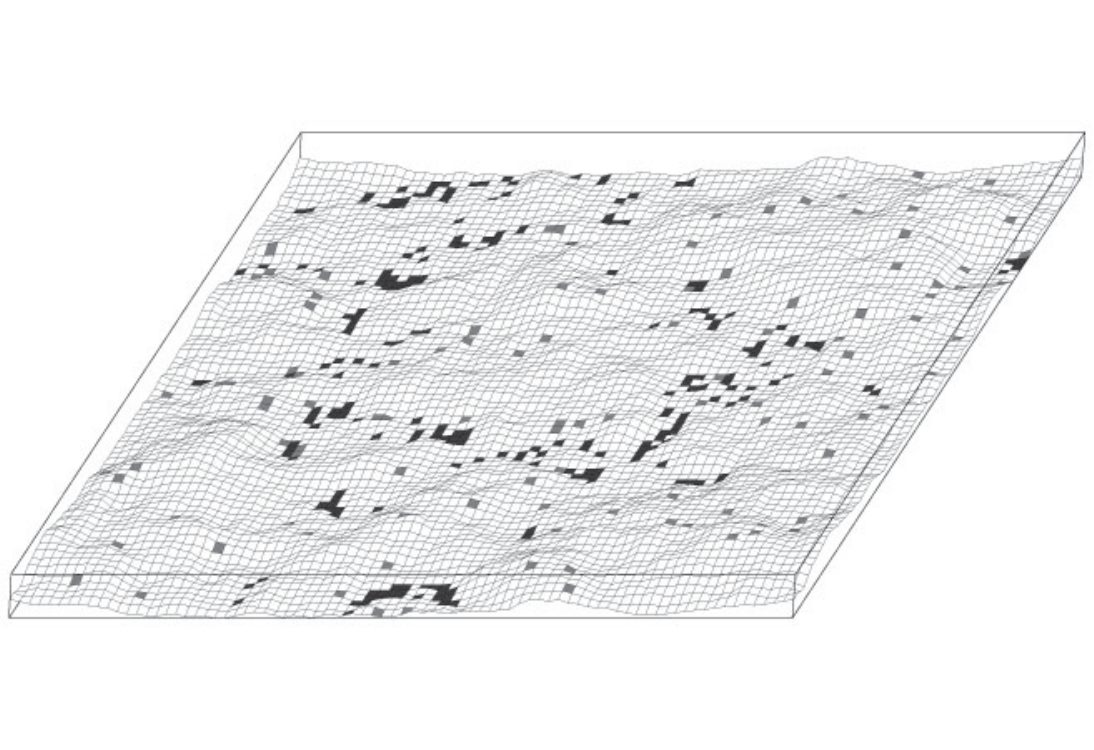}}
\caption{Typical Monte Carlo configurations of a two-component membrane with small, flexible stickers in the absence of cis-interactions. Bound stickers are black, unbound stickers gray, and membrane patches without stickers are white. The average separation of the membranes increases continuously as the sticker concentration is reduced. The bound stickers tend to form small clusters because of fluctuation-induced interactions. However, the membrane does {\em not} phase separate. 
\label{figureSnapshotsWithoutCis}}
\end{center}
\end{figure}

In the simplest case, the stickers are small and flexible and have no cis-interactions. In the grand-canonical ensemble, the Hamiltonian then has the form \cite{lipo130,weikl01}
\begin{equation}
{\cal H}\{l,n\} = {\cal H}_{el}\{l\} + \sum_i n_i\left[V(l_i)-\mu\right] \label{hamA}
\end{equation}
where ${\cal H}_{el}\{l\}$ is the elastic energy of the membranes, and $V(l_i)$ and $\mu$ are the trans-interaction potential and the chemical potential of the stickers. In section \ref{sectionEffectivePotential}, we saw that the sticker degrees of freedom in the partition function can be summed out exactly, because the Hamiltonian is linear in the concentration field $n$. Summing out the sticker degrees of freedom leads to a partition function for {\em homogeneous} membranes interacting with an effective potential $V_{ef}(l_i)$. If the sticker potential $V(l_i)$ is  a square-well potential, then $V_{ef}(l_i)$ is a square-well potential, too. It has the same potential range and an effective potential depth $U_{ef}$ which depends on the chemical potential $\mu$ and the binding energy $U$ of the stickers, see eq.~(\ref{Vef}). 

The unbinding transition of homogeneous, tensionless membranes with attractive and short-ranged interactions is continuous \cite{lipo33}, see also section \ref{sectionFreeEnergies}.  This implies that the contact probability $P_b$ , the fraction of bound membrane segments, continuously goes to the zero as the unbinding point is approached. The sticker concentration $X$ given in eq.~(\ref{concentration}) then is a {\em continuous} function of the chemical potential $\mu$ of the stickers. Therefore, lateral phase separation in sticker-rich and sticker-poor domains does not occur \cite{weikl00,weikl01}. In the grand-canonical ensemble, lateral phase separation corresponds to a {\em discontinuity} or jump in $X(\mu)$ at a certain value of the chemical potential.

The Monte Carlo snapshots shown in Fig.~\ref{figureSnapshotsWithoutCis} illustrate the continuous unbinding of the membranes. The average separation of the membranes increases continuously with decreasing sticker concentration, since membrane fluctuations become more and more pronounced. The membrane fluctuations also lead to small clusters of bound stickers. However, the fluctuation-induced interactions between the stickers do not cause lateral phase separation. 

\subsection{Stickers with cis-interactions}
\label{sectionAttractiveCisInteractions}

If the stickers now interact via cis-interactions, the Hamiltonian has the form \cite{weikl01}
\begin{equation}
{\cal H}\{l,n\} = {\cal H}_{el}\{l\} + \sum_i n_i\left[V(l_i)-\mu\right]
      + \sum_{\langle ij\rangle} W_{ij} n_i n_j  \label{hamwith}
\end{equation}
In the following, we consider the attractive and short-ranged cis-interactions 
\begin{eqnarray}
 W_{ij}  & = W &\;\; \text{for nearest neighbors $i$, $j$}\nonumber\\ &= 0 &\;\;  \text{otherwise} \label{cisinteraction}
\end{eqnarray}
with characteristic interaction strength $W<0$. 

\begin{figure}[t]
\begin{center}
\resizebox{\columnwidth}{!}{\includegraphics{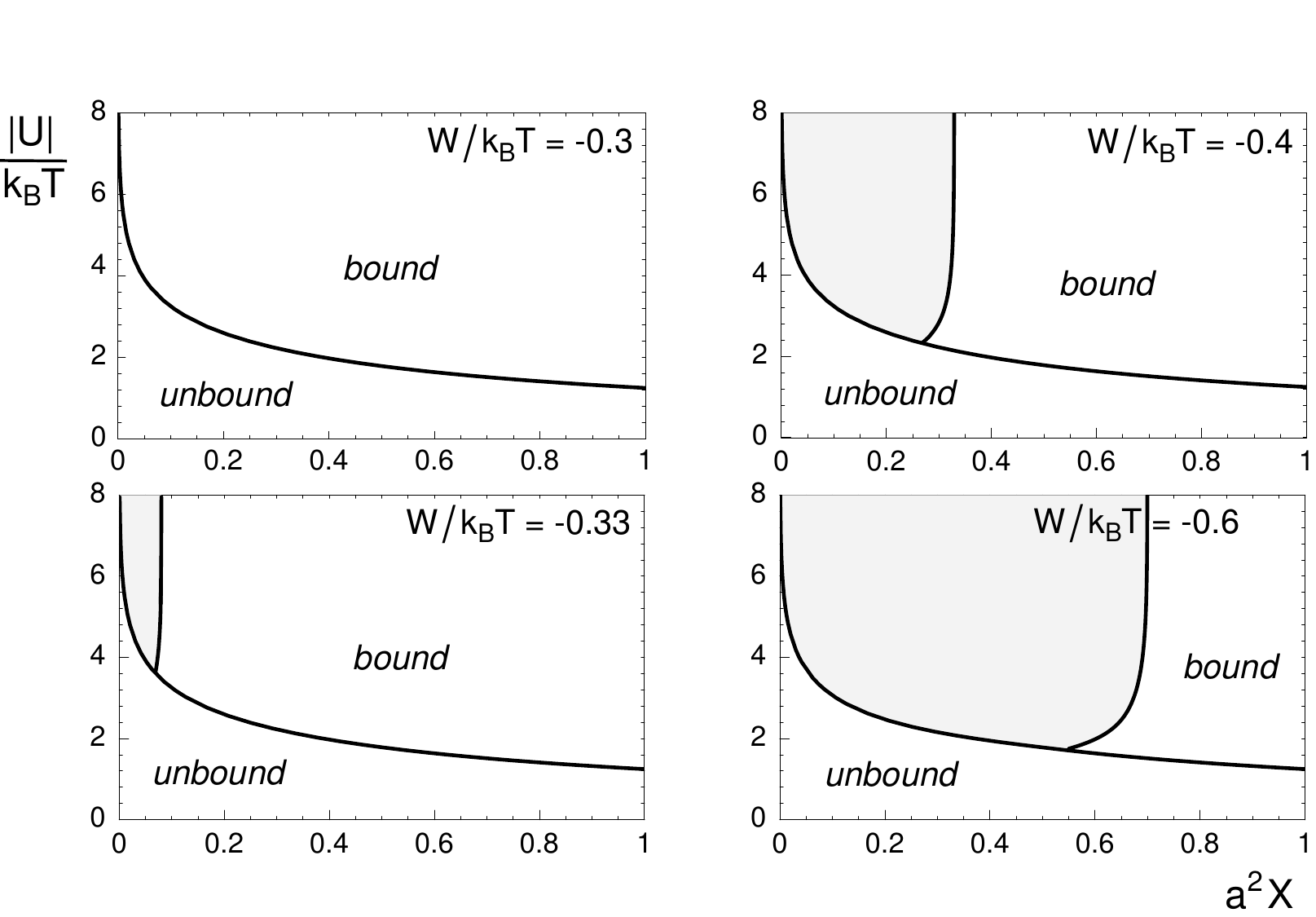}}
\caption{Mean-field phase diagrams for small, flexible stickers with attractive nearest-neighbor cis-interactions. Here, $W$ is the strength of the cis-interactions, $U$ is the binding energy, and $X$ is the concentration of the stickers. The phase diagrams at $W/k_B T = -0.33$, $-0.4$, and $-0.6$ contain shaded coexistence regions in which the membrane phase separates into a bound sticker-rich and an unbound sticker-poor phase.
\label{figureMfPhasediagrams}}
\end{center}
\end{figure}

Clearly, large absolute values $|W|$ of the interaction strength will lead to lateral phase separation into sticker-rich and sticker-poor domains. The central question here is how membrane fluctuations affect the critical interaction strength for lateral phase separation. An important reference value is the critical interaction strength $W_c$ for `non-fluctuating' membranes with constant separation $l_i$. In the absence of membrane fluctuations,  phase separation can only occur for interaction strengths with $|W|>|W_c|$. For constant membrane separation $l_i$, the Hamiltonian (\ref{hamwith})  reduces to the Hamiltonian of a 2-dimensional (2d) lattice gas. This lattice gas has the critical interaction strength  $W_c/ k_BT=-2 \ln(1+\sqrt{2})$. Mean-field theories, however, systematically underestimate the critical interaction strength and lead to $W_c/k_BT=-1$.

Mean-field phase diagrams for fluctuating membranes are shown in Fig.~\ref{figureMfPhasediagrams}. The rescaled potential range of the stickers here is $z_v=0.1$. The mean-field theory for the concentration field $n$ is described in detail in section \ref{sectionMeanFieldTheory}. The phase diagram at the interaction strength $W/k_BT=-0.3$ contains a bound and an unbound phase which are separated by a single line of continuous unbinding transitions. The membrane is bound for high concentrations or high binding energies $|U|$ of the stickers, and unbound for low concentrations $X$ or binding energies. For larger absolute values $|W|$ of the cis-interaction strength, the diagrams contain two-phase regions, shaded in gray. In the two-phase regions, an unbound phase with a low concentration of stickers and a bound phase with a higher sticker concentration coexist. The coexistence regions end in tricritical points. For binding energies $|U|$ below the tricritical value $|U_{tc}|$, the unbinding transition of the membrane is continuous. For $|U|>|U_{tc}|$, the unbinding transition is discontinuous and is then coupled to the phase separation within the membrane. At large absolute values $|U|$ of the sticker binding energy, the sticker concentrations of the two coexisting phases vary only slightly with $U$ since the majority of the stickers is already bound.

The important point is that the phase separation occurs at interaction strengths $|W|$ which are significantly below the critical interaction strength $|W_c|$ of the 2d lattice gas. At $z_v=0.1$, the phase separation can be observed already at a fraction $W/W_c=0.31$ of the critical mean-field value $W_c/k_BT=-1$ for the 2d lattice gas. This means that the phase separation is predominantly caused by fluctuation-induced interactions between the stickers. 

\begin{figure}[t]
\begin{center}
\resizebox{0.5\columnwidth}{!}{\includegraphics{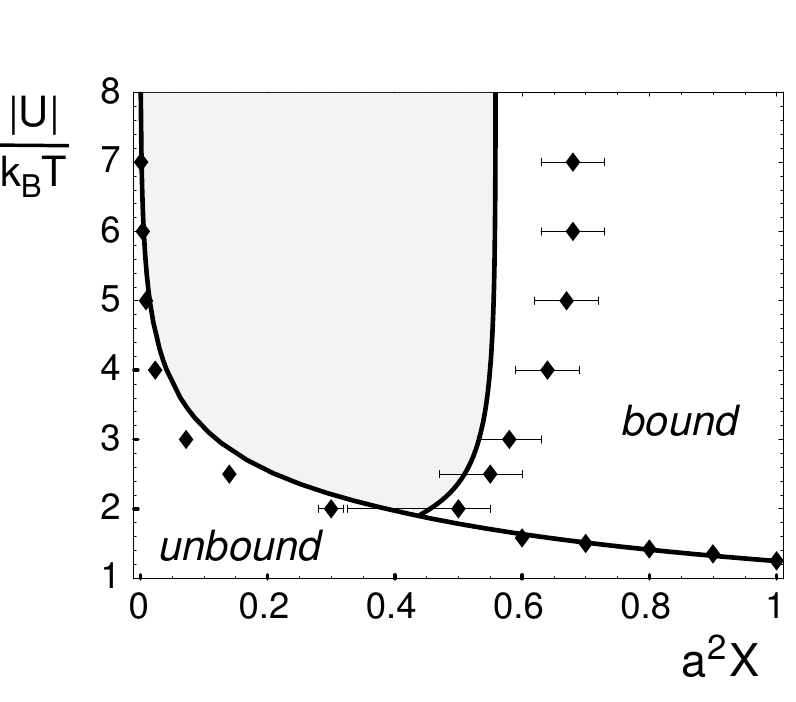}}
\caption{Comparison of Monte Carlo (data points) and mean-field phase diagrams (lines and shaded coexistence region). The cis-interaction strength is $W=0.5 W_c$, and the rescaled potential range is $z_v=0.1$. Here, $W_c$ is the critical interaction strength of the 2d lattice gas, which has the value $W_c/k_B T=-1$ in mean-field theory and $W_c/k_B T=-2\ln(1+\sqrt{2})$ for the Monte Carlo simulations. \label{figureMcMfPhasediagram}}
\end{center}
\end{figure}

Monte Carlo simulations confirm this result. Fig.~\ref{figureMcMfPhasediagram} shows a comparison of phase diagrams from mean-field theory (lines) and Monte Carlo simulations (data points). The cis-interaction strength of the stickers here is $W=0.5 W_c$. In the Monte Carlo simulations, the sticker concentration $X=\langle n_i\rangle/a^2$ is determined as a function of the chemical potential $\mu$, for various values of the binding energy $U$. A first-order transition is reflected in a discontinuity of $X(\mu)$ at a certain value of the chemical potential. The two limiting values of $X$ at the discontinuity are the concentrations of the coexisting phases.  

\begin{figure}
\begin{center}
\resizebox{0.7\columnwidth}{!}{\includegraphics{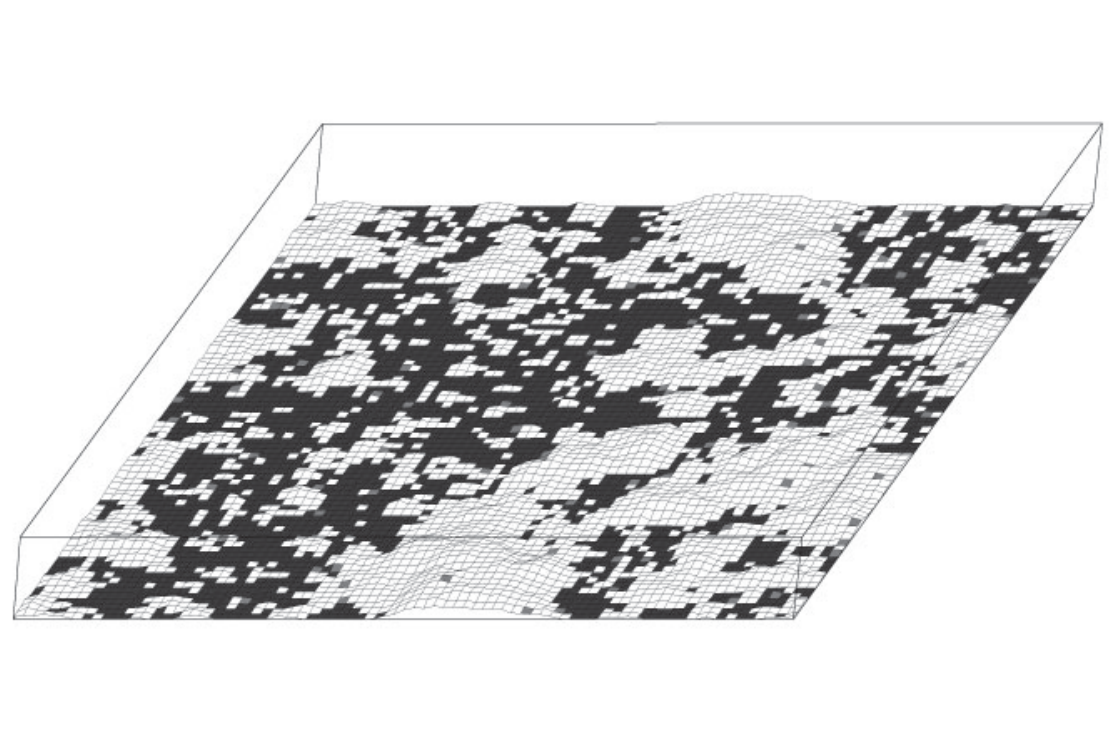}}

\vspace*{-1.3cm}
\resizebox{0.7\columnwidth}{!}{\includegraphics{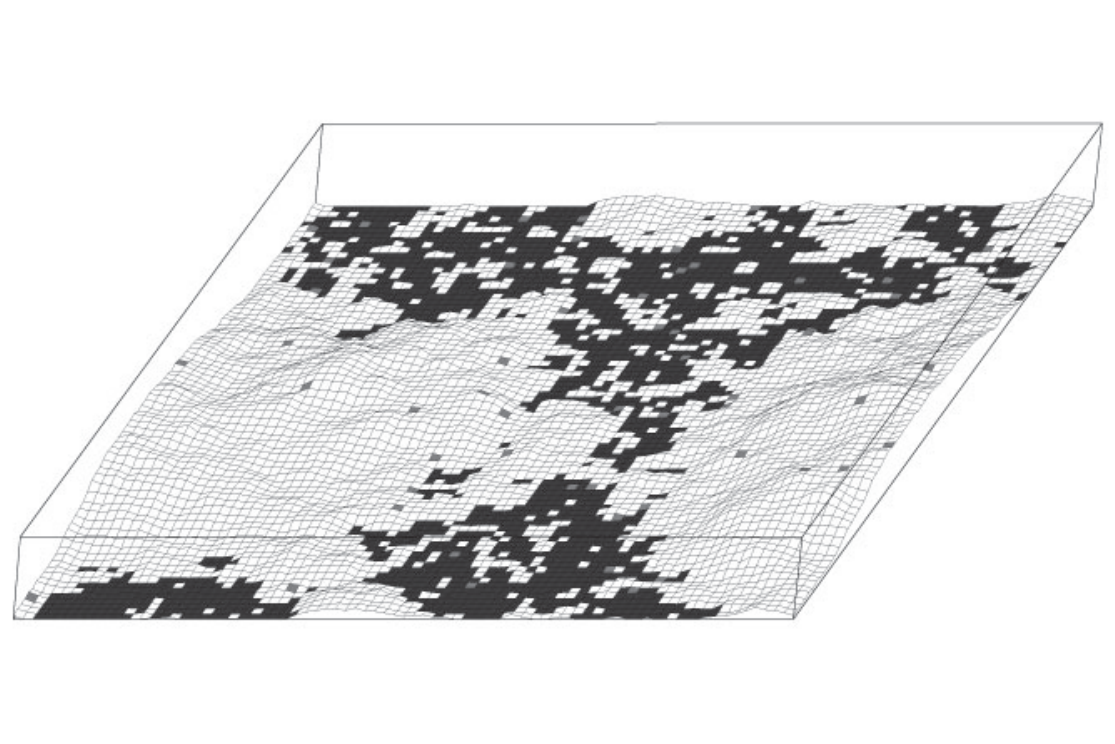}}

\vspace*{-1.3cm}
\resizebox{0.7\columnwidth}{!}{\includegraphics{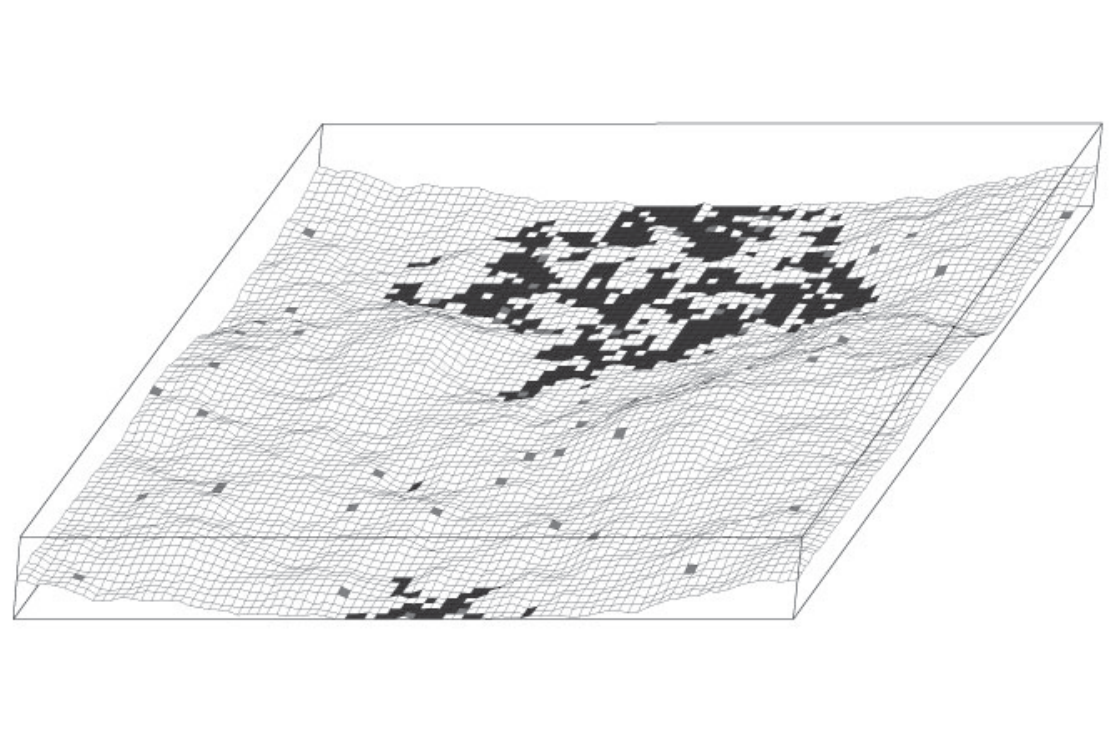}}
\caption{Typical Monte Carlo configurations of a two-component membrane containing small, flexible stickers with cis-interaction strength $W=0.5 W_c=-k_BT\ln (1+\sqrt{2})$, binding energy $U=-5 k_B T$ and rescaled potential range $z_v=0.1$. Bound sticker are black, unbound stickers gray, and membrane patches without stickers are white. The fixed sticker concentration $X$ is $0.45/a^2$, $0.3/a^2$, and $0.1/a^2$ (bottom to top). The membrane phase separates into a bound phase with high sticker concentration and an unbound phase with low sticker concentration.\label{figureSnapshotsWithCis}}
\end{center}
\end{figure}

The Monte Carlo snapshots of Fig.~\ref{figureSnapshotsWithCis} illustrate the lateral phase separation. The snapshots are from canonical Monte Carlo simulations with fixed sticker concentration $X=0.45/a^2$, $0.3/a^2$, and $0.1/a^2$. The rescaled potential range and cis-interaction strength of the stickers have the same values as in Fig.~\ref{figureMcMfPhasediagram}, and the sticker binding energy is $U=-5k_B T$. The three snapshots thus are taken at points within the coexistence region of  Fig.~\ref{figureMcMfPhasediagram}. Since the overall sticker concentration is kept constant in these simulations, the membrane phase-separates into bound domains with high sticker concentration and unbound domains with small sticker concentration. The extent of the bound phase shrinks with decreasing sticker concentration.

The phase behavior strongly depends on the rescaled potential range $z_v$. At the short potential range $z_v=0.1$, phase separation can be observed for $W/W_c>0.31$ in the mean-field theory, and for $W/W_c>0.35\pm 0.05$ in Monte Carlo simulations \cite{weikl01}. At the larger potential range $z_v=0.5$, phase separation occurs for $W/W_c>0.80$ in mean-field theory and for $W/W_c>0.65\pm 0.05$ in the simulations. The fluctuation-induced interactions between stickers thus decrease with increasing potential range. The reason for this decrease is that bound stickers with larger potential range are less restrictive for membrane fluctuations than bound stickers with shorter potential range.

\subsection{Large stickers} 
\label{sectionLargeStickers}

To capture all possible membrane shape fluctuations, the linear patch size of the discretized membrane model has to be equal to the cut-off length for the fluctuations, see section \ref{S.DiscreteModels}.  So far, we have only considered stickers with a lateral extension which is smaller than or equal to a single membrane patch. But the lateral extension of large stickers may be bigger than the cut-off length for the membrane fluctuations. These stickers then occupy several membrane patches.

\begin{figure}
\begin{center}
\resizebox{0.24\columnwidth}{!}{\includegraphics{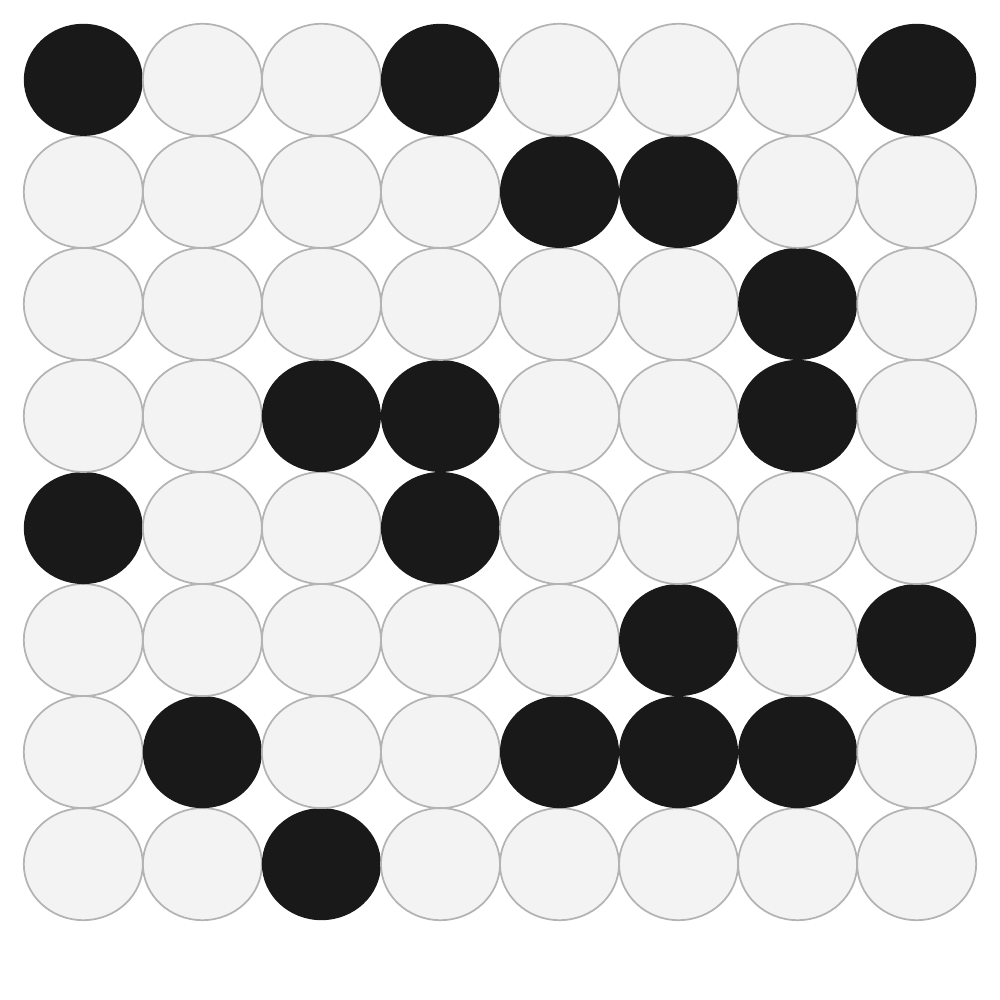}}
\hspace{1cm}
\resizebox{0.24\columnwidth}{!}{\includegraphics{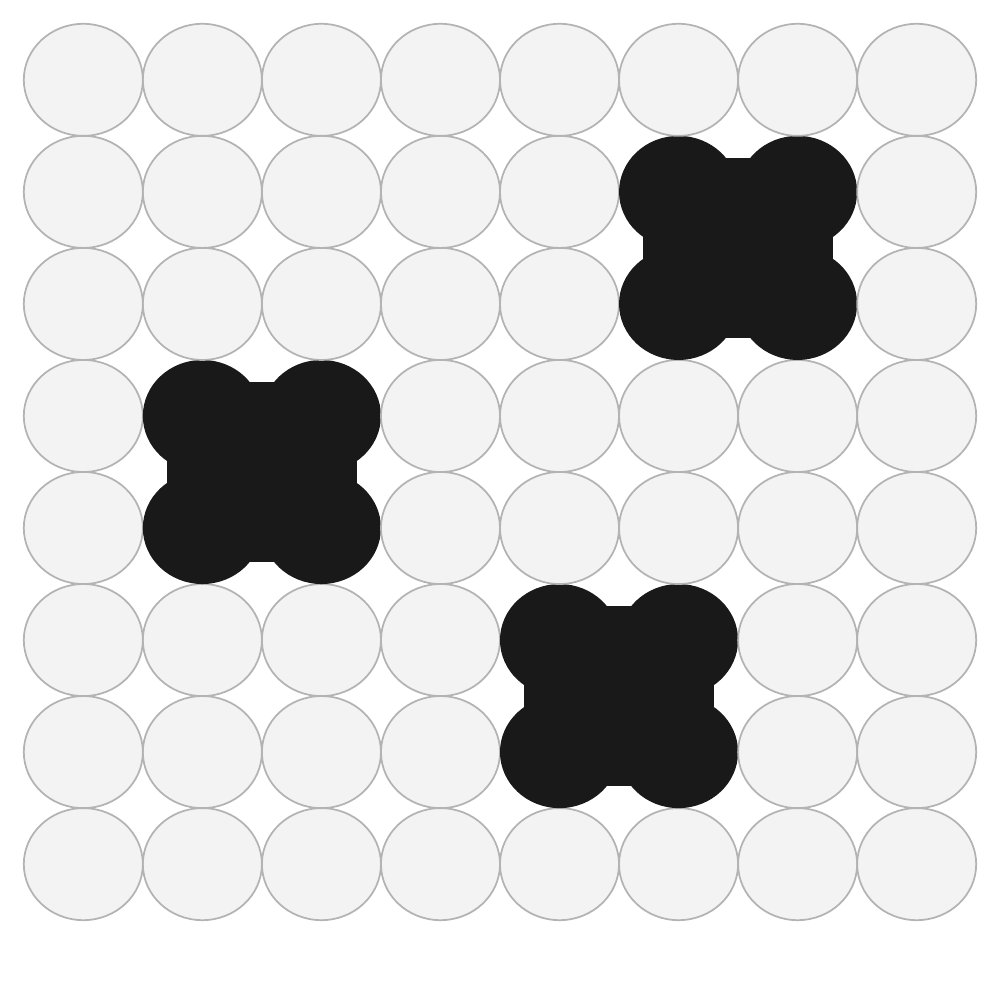}}
\hspace{1cm}
\resizebox{0.24\columnwidth}{!}{\includegraphics{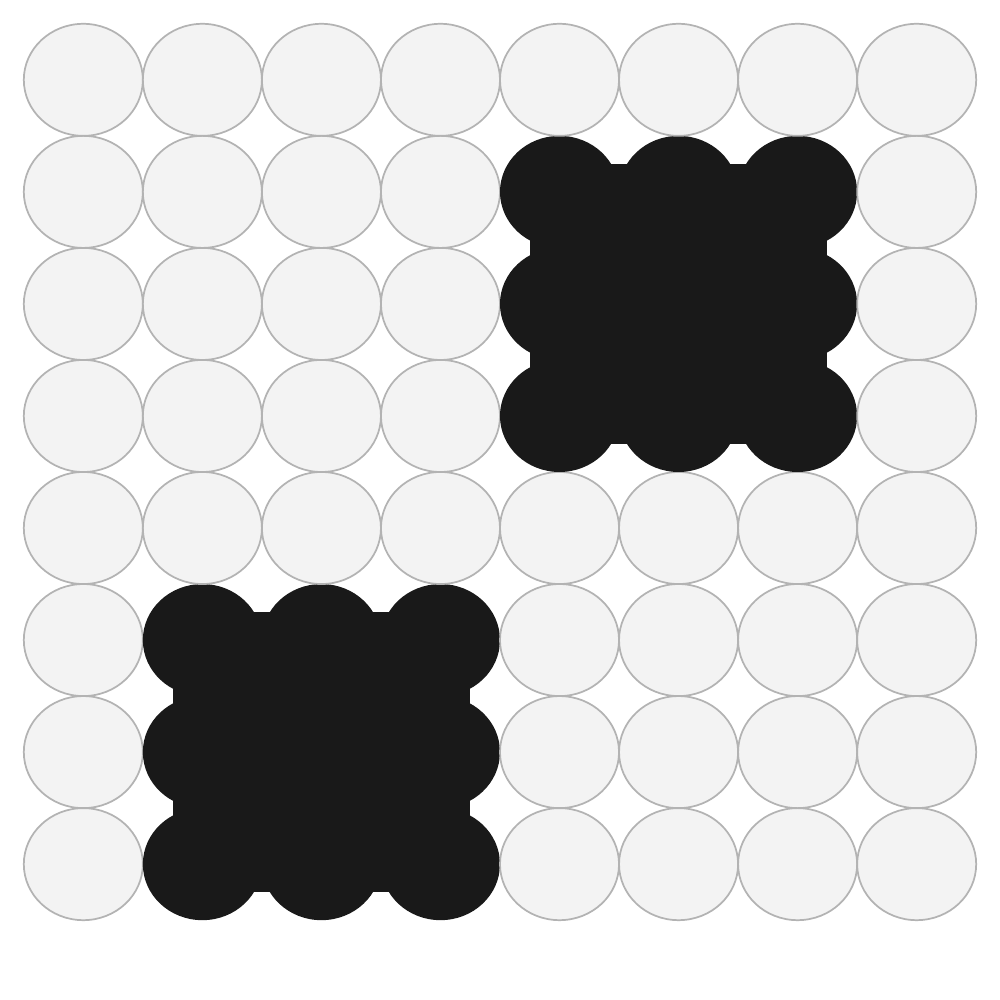}}
\caption{Membrane segments with: (left) small stickers of size $Q=1$ shown in black; (middle) quadratic stickers occupying $Q=2\times 2$ membrane patches; and (right) quadratic stickers occupying $Q=3\times 3$ membrane patches.
\label{figureCartoonLargeStickers}}
\end{center}
\end{figure}

A simple example are quadratic stickers which occupy $Q=2\times 2$ or $3\times 3$ lattice sites, see Fig.~10. If each of the $Q$ membrane patches occupied by a sticker molecule interacts with the second membrane via a square-well interaction with binding energy $U$ and potential range $l_v$, these large stickers can be seen as quadratic arrays, or clusters, of $Q$ stickers with the size of a single membrane patch. The Hamiltonian of a membrane with these stickers can be written in the form  \cite{weikl00}
\begin{equation}
{\cal H}\{l, n\}=  {\cal H}_{el}
  + \sum_i n_i\cdot(V_i(l) -\mu)+ \sum_{\langle ij\rangle} W_{ij} n_i n_j  \label{Hlarge}
\end{equation} 
with the sticker adhesion potential 
\begin{eqnarray}
V_i(l)= V(l_{i,1},\ldots,l_{i,Q}) = U \sum_{q=1}^Q \theta(l_v - l_{i,q})
\label{sumPotential}
\end{eqnarray}
Here, $\{(i,1),\ldots,(i,Q)\}$ denotes quadratic arrays of $Q=2\times 2$ or $3\times 3$ lattice sites. The site $i=(i,1)$, for example the central site of a sticker with size $Q=3\times 3$, indicates the sticker position. The cis-interactions in the Hamiltonian (\ref{Hlarge}) then are the repulsive hard-square interactions
\begin{equation}
W_{ij} =  \infty  \;\mbox{for}\; j \;\mbox{in}\; A^Q_i\; \mbox{and zero otherwise}
\label{hardsquare}
\end{equation}
where $A^Q_i$ denotes the `exclusion area' of an individual $Q$-sticker at lattice site $i$. The hard-square interactions prevent an overlap of the stickers.
%

\begin{figure}
\begin{center}
\resizebox{0.95\columnwidth}{!}{\includegraphics{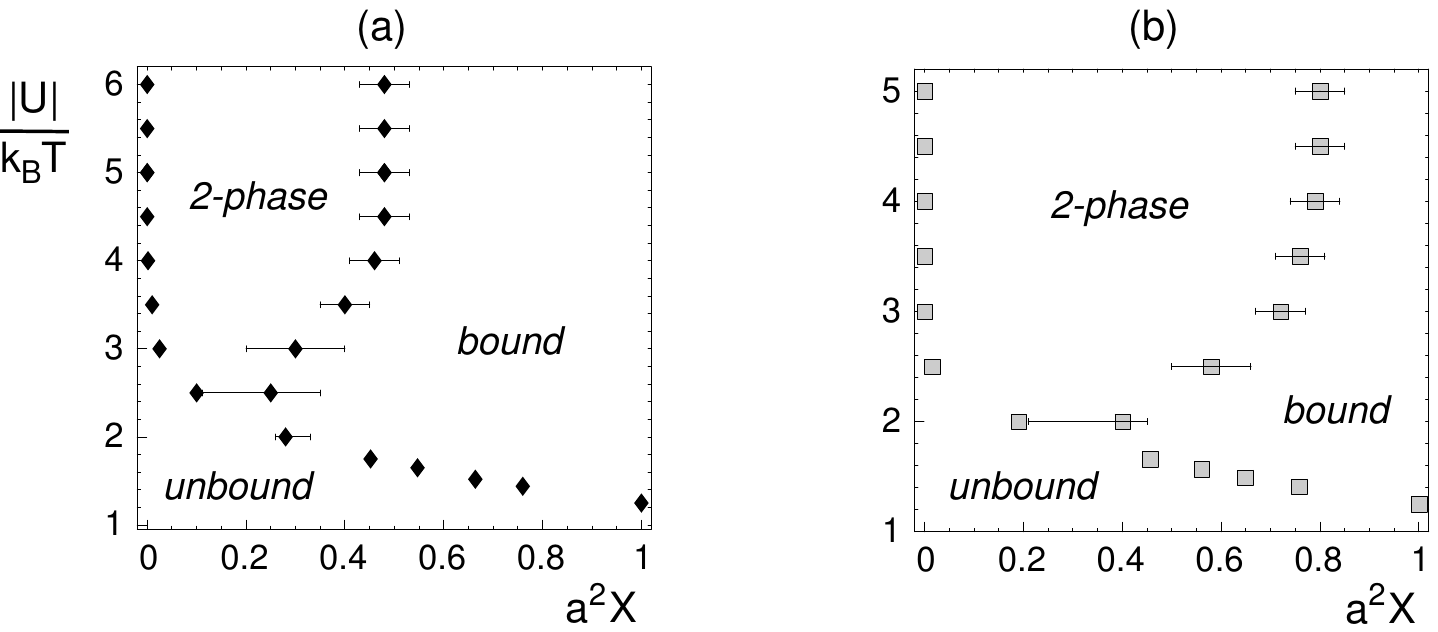}}
\caption{Phase diagrams for large quadratic stickers occupying (a) $Q=2\times 2$ and (b) $Q=3\times 3$ membrane patches. Each of the $Q$ membrane patches occupied by a single sticker interacts with the second membrane via a square-well potential with binding energy $U$ and rescaled potential range $z_v=0.1$.  The extent of the 2-phase region increases with the sticker size, which indicates an increase of the fluctuation-induced interactions between the stickers. 
\label{figurePhaseDiasLargeStickersOne}}
\end{center}
\end{figure}

Phase diagrams for large stickers are shown in Fig.~\ref{figurePhaseDiasLargeStickersOne}. The unbinding transition of the membranes is discontinuous for large values of $|U|$ and continuous for $|U|\lesssim 2 k_B T$. At large values of $|U|$, the membrane phase-separates into an unbound sticker-poor and a bound sticker-rich phase. The extent of the two-phase region increases with the sticker size $Q$. Since the cis-interactions (\ref{hardsquare}) of the stickers are purely repulsive, the phase separation is driven only by fluctuation-induced interactions between the stickers. These interactions increase with the sticker size (see Fig.~11), but decrease with the sticker potential range (see Fig.~12). The entropic interactions decrease with increasing potential range since bound stickers with larger potential range are less restrictive for the membrane shape fluctuations.

\begin{figure}[t]
\begin{center}
\resizebox{0.45\columnwidth}{!}{\includegraphics{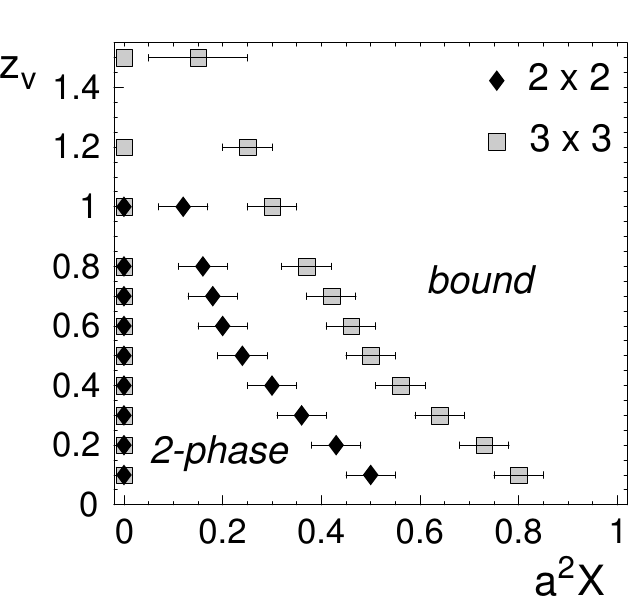}}
\caption{Phase diagrams for large quadratic stickers with size $Q=2\times 2$ and $Q=3\times 3$ as a function of the sticker concentration $X$ and the rescaled potential range $z_v$. The binding energy of a single sticker patch has the relatively large value $U=-5 k_B T$. The sticker concentration in the unbound phase (left data points) then is close to zero, see also Fig.~\ref{figurePhaseDiasLargeStickersOne}. The sticker concentration in the bound phase (right data points) decreases with increasing sticker potential range $z_v$ since the fluctuation-induced interaction becomes weaker.
\label{figurePhaseDiasLargeStickersTwo}}
\end{center}
\end{figure}

To understand the increase of the fluctuation-induced interactions with the sticker size, let us consider an arbritrary shape of the adhering membrane, and let us divide the membrane surface into two types of domains: (i) `Bound domains'  with a membrane separation smaller  than the sticker potential range $l_v$; and (ii) `unbound domains'  with a separation larger than $l_v$. Obviously, in order to  gain more adhesive energy one has to place more stickers into the bound domains. If these stickers have the size $Q=1$ of a single membrane patch, the maximal adhesive energy depends only on the total area of  the bound domains, but not on the  number of bound domains, or the shape of these domains. In contrast, if the stickers are larger and occupy several patches, the total adhesive energy depends on the detailed geometry of the bound domains.

As an example, consider stickers with size $Q=2\times 2$ and two bound domains which both have an area of $2 \times 3$ patches. As long as these two domains are disjoint, we need {\em four} $2 \times 2$ sticker  to obtain the maximal  adhesive energy from these two domains. In contrast, if we combine the  two bound domains into a single $2 \times 6$ bound domain, we need   only {\em three} such   stickers in order to gain the same adhesive energy. In general, if we cover the area of the bound domains with stickers occupying $Q > 1$ patches, many of these stickers will sit on the domain boundaries  and thus will not contribute to the  adhesive energy in the same way as those in the interior of the bound domains. This leads to an additional effective line tension of the boundaries between bound an unbound domains. This additional line tension favors the aggregation of bound domains and corresponds to an increase in the fluctuation-induced interactions.

\subsection{Rigid stickers} 
\label{sectionStiffStickers}

\subsubsection{Tensionless membranes}

Stickers with rather stiff `anchors' in the membrane may change the local membrane elasticity. If the stickers only affect the local bending rigidity $\kappa$ and do not change the modulus of Gaussian curvature, the discretized Hamiltonian has the form \cite{weikl01}
\begin{equation}
 H\{l, n\} =   \sum_i  \frac{\kappa}{2 a^2}\left(\Delta_d l_i\right)^2 
  + \sum_i  n_i\left(\frac{\kappa_s - \kappa}{2 a^2}\left(\Delta_d l_i\right)^2 +V(l_i)-\mu \right)
\label{stiffHam}
\end{equation}
in the absence of cis-interactions between the stickers. Here, $\kappa_s$ is the bending rigidity of membrane patches which contain stickers, and $\kappa$ is the rigidity of patches  without stickers, i.e.~the rigidity of the `bare' lipid bilayer.  

\begin{figure}[t]
\begin{center}
\resizebox{0.99\columnwidth}{!}{\includegraphics{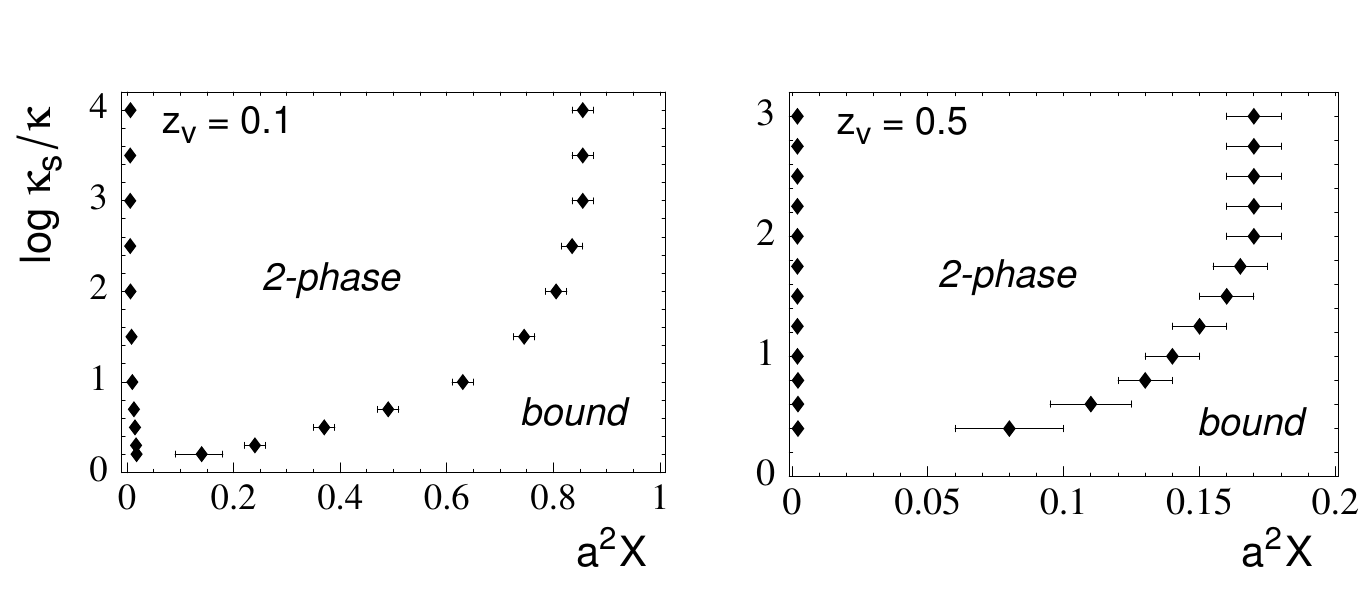}}
\vspace{-0.5cm}
\caption{Phase diagrams for rigid stickers as a function of the sticker concentration $X$ and the increased bending rigidity $\kappa_s$ of the membrane patches with stickers. The sticker rigidity $\kappa_s$ is given in units of the bending rigidity $\kappa$  for the lipid bilayer. The stickers have the binding energy $U=-5 k_B T$ and the rescaled potential ranges $z_v=0.1$ (left) and $z_v=0.5$ (right).  The data points represent the sticker concentrations in the two coexisting phases, a sticker-poor unbound and a sticker-rich bound phase.  At the relatively large sticker binding energy $U=-5 k_B T$ considered here, the sticker concentration in the unbound phase is very low. 
\label{figureStiffStickersOne}}
\end{center}
\end{figure}

\begin{figure}[t]
\begin{center}
\resizebox{0.5\columnwidth}{!}{\includegraphics{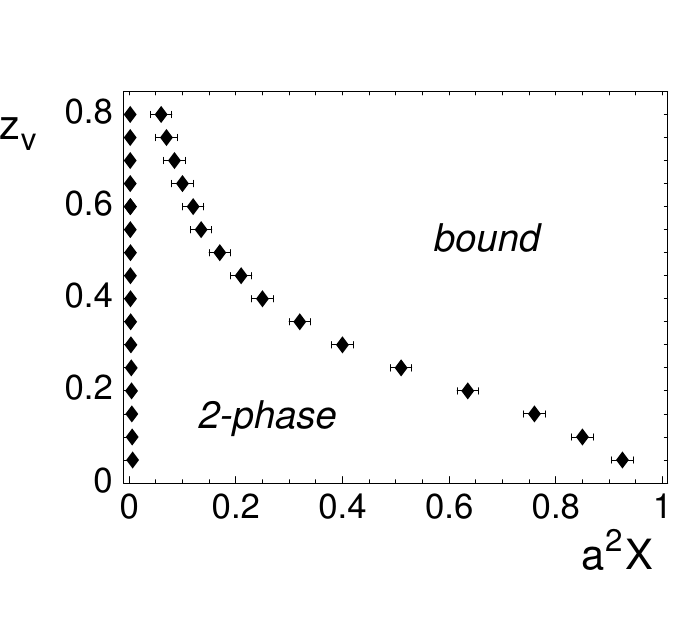}}
\caption{Phase diagrams for rigid stickers as a function of the sticker concentration $X$ and the rescaled potential range $z_v$. 
The sticker patches have the bending rigidity $\kappa_s=1000\kappa$, and the sticker binding energy is $U=-5k_B T$ as in Fig~\ref{figureStiffStickersOne}. The sticker concentration of the bound phase strongly decreases with increasing potential $z_v$, which indicates a decrease of the fluctuation-induced interactions between the stickers. 
\label{figureStiffStickersTwo}}
\end{center}
\end{figure}

Rigid  stickers aggregate also without any attractive cis-interactions. The fluctuation-induced interactions between rigid stickers thus are significantly enlarged. Similar to the case of large sticker considered in the previous section, the increased tendency for lateral phase separation can be explained by an additional effective line tension between bound membrane segments \cite{weikl01}.  Fig.~\ref{figureStiffStickersOne} shows how the phase behavior of the membranes depends on the rigidity $\kappa_s$ of the sticker patches. At the rescaled potential range $z_v=0.1$ of the sticker square-well potential, lateral phase separation occurs for $\kappa_s/\kappa >1.3\pm 0.3$ according to Monte Carlo simulations.  At $z_v=0.5$, the membranes phase-separate for $\kappa_s/\kappa >1.5\pm 0.5$. This means that the stickers aggregate already at rigidities $\kappa_s$ which are only slightly larger than bare lipid bilayer rigidity $\kappa$.  As in the previous sections, the fluctuation-induced interactions decrease with increasing sticker potential range $z_v$, see Fig.~\ref{figureStiffStickersTwo}.

\pagebreak

\subsubsection{Effect of  tension}
\label{sectionEffectOfTension}

Biological and biomimetic membranes are often under lateral tension. The tension suppresses membrane fluctuations and will therefore also affect fluctuation-induced interactions. A lateral tension $\sigma$ leads to the additional term 
\begin{equation}
{\cal H}_{\sigma}\{l\} = \sum_i \frac{\sigma}{2} \left(\nabla_d l_i\right)^2
\label{Hsigma}
\end{equation}
in the Hamiltonian. Here, 
\begin{equation}
(\nabla_d l_i)^2=(\nabla_d l_{x,y})^2=(l_{x+a,y} - l_{x,y})^2 + (l_{x,y+a} - l_{x,y})^2
\end{equation}
describes the local area increase of the curved membrane compared to a planar membrane configuration with constant separation $l$.

\begin{figure}
\begin{center}
\resizebox{0.55\columnwidth}{!}{\includegraphics{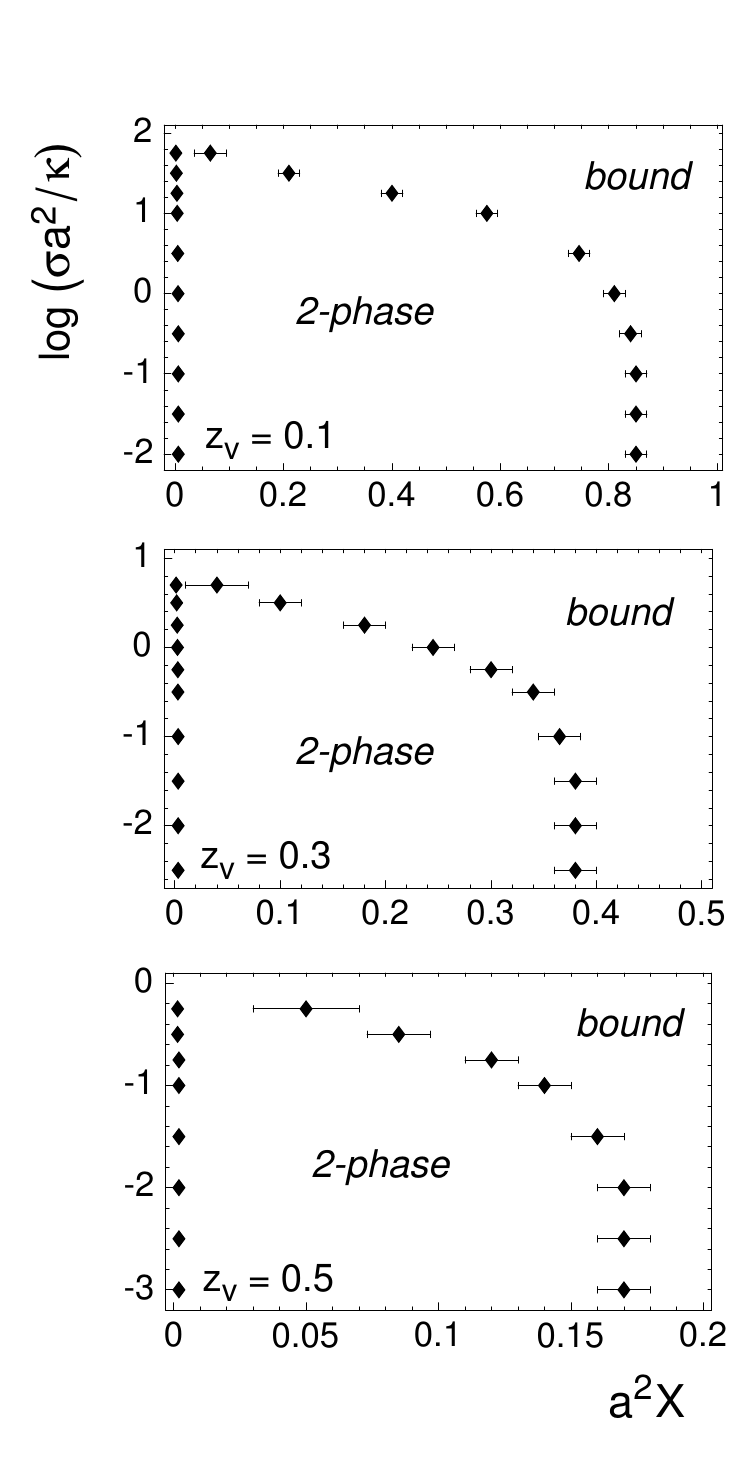}}
\caption{
Phase diagrams for  rigid stickers as a function of the sticker concentration $X$ and the dimensionless tension $\sigma a^2/\kappa$ at different rescaled potential ranges $z_v$. As in Fig.~\ref{figureStiffStickersTwo}, the binding energy of the stickers is $U=-5 k_B T$, and the increased bending rigidity of sticker patches is $\kappa_s=1000\kappa$. The extent of the coexistence regions decreases at large values of $\sigma$ where membrane fluctuations on the scale of the average distance between bound stickers are suppressed. These small-scale fluctuations cause the entropic interactions between the stickers. 
\label{figureStiffStickersThree}}
\end{center}
\end{figure}

Fig.~\ref{figureStiffStickersThree} shows how the phase behavior of membranes with rigid stickers depends on the rescaled lateral tension $\sigma a^2/\kappa$. The sticker ridigity $\kappa_s$ and binding energy $U$ are the same as in Fig.~\ref{figureStiffStickersTwo}.  At small lateral tensions $\sigma$, the concentrations of the coexisting phases agree with the tensionless case, see Fig.~\ref{figureStiffStickersThree}. At higher tensions, the width of the coexistence region decreases since shape fluctuations and, thus, fluctuation-induced interactions are suppressed.

However, the tensions at which the tendency for lateral phase separation clearly decreases are relatively high. Lipid membranes typically rupture at tensions $\sigma_r$ of a few millinewton per meter \cite{evans87}.  Taking $\sigma_r=4mN/m$, a membrane patch extension $a=5$~nm, and  the bending rigidity $\kappa=10^{-19}J$  leads to the estimate $\sigma_r a^2/\kappa=1$ or $\log_{10}(\sigma_r a^2/\kappa)=0$  for the maximum value of the reduced tension. The tensions that cause a significant decrease of the coexistence region thus are already close to or even above this estimate for the maximum value, see Fig.~\ref{figureStiffStickersThree}. 

To understand this behavior, one has to realize that membrane fluctuations are suppressed only on length scales larger than the crossover length $\sqrt{\kappa/\sigma}$. On smaller scales, thermal fluctuations are still governed by the bending energy. The decrease of the coexistence regions in the phase diagrams of Fig.~\ref{figureStiffStickersThree} sets in at values of the reduced tension $\sigma a^2/\kappa$ that correspond to crossover lengths of only a few lattice constants. The relevant fluctuations thus turn out to be fluctuations of the non-adhesive membrane segments between the small clusters of bound stickers, see Fig.~\ref{figureSnapshotsWithoutCis}. A decrease of fluctuation-induced interactions then occurs if the crossover length $\sqrt{\kappa/\sigma}$ is comparable to or smaller than the mean distance between the sticker clusters. This interpretation agrees with the observation that the influence of an increasing lateral tension is most pronounced for the rescaled potential range $z_v=0.5$. For this value of $z_v$, the sticker concentration of the bound phase at low tensions is smaller than in the other two cases $z_v=0.1$ and $z_v=0.3$, see Fig.~\ref{figureStiffStickersThree}. The average distance between the stickers in the bound phase therefore is larger than in the other cases.

\pagebreak

\section{Barrier mechanisms for domain formation}

\label{S.BarrierMechanisms}

\subsection{Membranes with stickers and mobile repellers}
\label{sectionMobileRepellers}

Biological membranes often contain repulsive glycoproteins. These `repellers' form a protective barrier, the glycocalyx. If a membrane contains both stickers and repellers, the positions of the molecules can be described by a discrete concentration field $n$ with three different values, e.g.~the values $n_i=1$ for membrane patches $i$ that contain stickers, $n_i=2$ for membrane patches with repellers, and $n_i=0$ for patches without stickers or repellers. The grand-canonical Hamiltonian of the membrane then can be written in the form \cite{weikl01,weikl02a}
\begin{equation}
{\cal H}\{l,n\} = {\cal H}_{el}\{l\} + \sum_i \left[ \delta_{1,n_i} \left(V_s(l_i)-\mu_s\right) + \delta_{2,n_i} \left(V_r(l_i)-\mu_r\right)\right]
\label{hamiltonianMobileRepellers}
\end{equation}
with the Kronecker symbol $\delta_{i,j}=1$ for $i=j$ and  $\delta_{i,j}=0$ otherwise. Here, $V_s(l_i)$ and $V_r(l_i)$ are the interaction potentials of stickers and repellers, and $\mu_s$ and $\mu_r$ the chemical potentials for the stickers and repellers. 

Since we have neglected cis-interactions between stickers or repellers, the Hamiltonian (\ref{hamiltonianMobileRepellers}) is linear in the concentration field $n$. Therefore, the degrees of freedom of the concentration field can be summed out exactly, see section \ref{sectionEffectivePotential}. This leads to the partition function 
\begin{equation}
{\cal{Z}} = \left( 1 + e^{\mu_s/k_BT} + e^{\mu_r/k_BT} \right)^N \nonumber \left[\prod_i \int_0^\infty\!\! dl_i\right]
                          e^{-\left[{\cal H}_{el}\{l\}+ \sum_i V_{ef}(l_i))\right]/k_BT}
\end{equation}
of two homogeneous membranes interacting via the effective membrane potential 
\begin{equation}
V_{ef}(l_i) = -k_BT \ln \frac{1 + e^{[\mu_s - V_s(l_i)]/k_BT} + e^{[\mu_r-V_r(l_i)]/k_BT}}{1 + e^{\mu_s/k_BT} + e^{\mu_r/k_BT}}
\end{equation}
Here, $N$ denotes the total number of membrane patches.

In the following, the stickers are characterized by a square-well potential with binding energy $U_s<0$ and range $l_v$, and the repellers by a square-barrier potential with barrier energy $U_r>0$ and range $l_r>l_v$. The range, or `size', of the repellers thus is larger than the sticker binding range. The effective potential then has the form
\begin{eqnarray}
V_{ef}(l_i)  =& U_{co} &  \hspace{0.2cm}  \text{for $0<l_i<l_s$}\nonumber\\
             =& U_{ba} &  \hspace{0.2cm}  \text{for $l_s<l_i<l_r$}\nonumber\\
            =&  0 & \hspace{0.2cm}     \text{for $l_r<l_i$}
      \label{potbarOne}
\end{eqnarray}
with the effective contact energy
\begin{equation}
U_{co}\equiv -k_BT \ln {\displaystyle\frac{1 + e^{(\mu_s - U_s)/k_BT} + e^{(\mu_r -U_r)/k_BT}}{1 + e^{\mu_s/k_BT} + e^{\mu_r/k_BT}}}
\end{equation}
and the effective barrier energy
\begin{equation}
U_{ba}\equiv -k_BT \ln {\displaystyle \frac{1 + e^{\mu_s/k_BT} + e^{(\mu_r -U_r)/k_BT}}{1 + e^{\mu_s/k_BT} + e^{\mu_r/k_BT}}}
\end{equation}
The effective barrier energy $U_{ba}$ is positive because of  $U_r>0$, and the effective contact energy $U_{co}$ is smaller than $U_{ba}$ because of $U_s<0$. Summing out the degrees of freedom of the concentration field thus leads to an effective potential with a potential minimum $U_{co}$ induced by the stickers  and a potential barrier $U_{ba}$ caused by the repellers.

Bound states of the membrane are only possible if the effective contact energy $U_{co}$ is negative and small enough to compensate the loss of configurational entropy, i.e.~the entropy difference between the bound and unbound state. This entropy difference arises since membrane fluctuations in the bound state are more restricted than fluctuations in the unbound state.
Suppose the barrier energy $U_{ba}$ is high and the membrane is confined to the potential well with width $l_s$ of the effective potential (\ref{potbarOne}). 
The entropy loss of the bound membrane may be estimated via 
$c_{fl} (k_BT)^2/\kappa l_s^2$  as follows from  (\ref{flucpot}) if one replaces
the average membrane separation ${\bar l}$ by the width $l_s$ of the potential well 
\cite{helfrich90}. The free energy difference per unit area between the bound and the unbound state of the membrane with adhesion potential (\ref{potbarOne}) can then be estimated as  $\Delta F=-|U_{co}| + c_{fl} (k_BT)^2/\kappa l_s^2$ \cite{lipo118}.  According to this  estimate, the membrane is bound for
\begin{equation}
|U_{co}| > c_{fl} (k_BT)^2/\kappa l_s^2 \label{criticalContactEnergy}
\end{equation}
since the free energy difference $\Delta F$ has to be negative to have a stable bound state.

\begin{figure}
\begin{center}
\resizebox{\columnwidth}{!}{\includegraphics{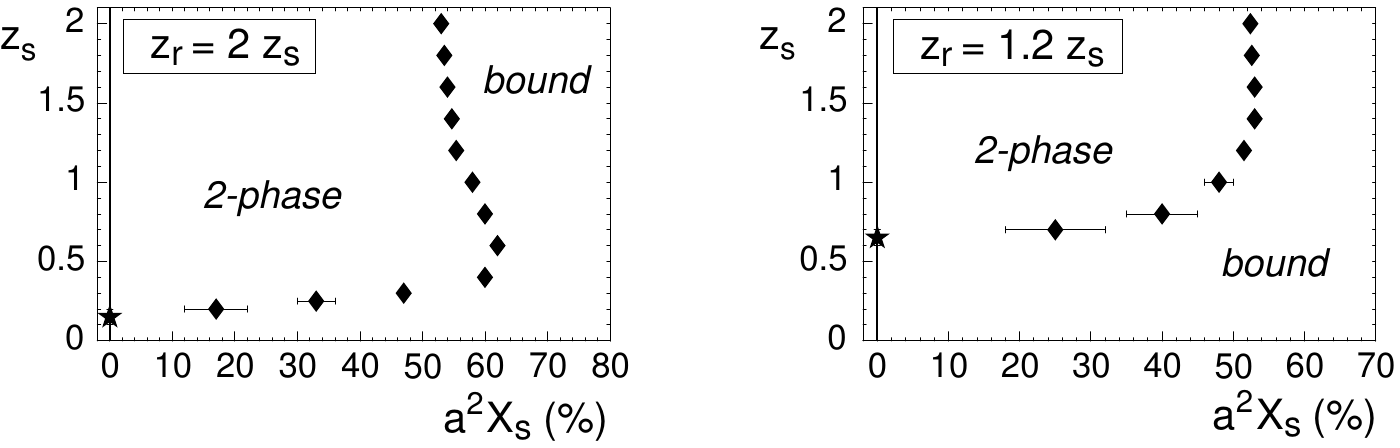}}
\caption{Phase diagrams for a membrane with stickers and repellers as a function of the sticker concentration $X_s$ and the rescaled potential ranges $z_s$ and $z_r$ of stickers and repellers. The chemical potential for the repellers is $\mu_r=0$,  which corresponds to a repeller concentration  $a^2 X_r=0.5$ in the unbound phase. At large values of $z_s$ and $z_r$, the repeller barrier with rescaled width $z_r-z_s$  is strong,  which leads to lateral phase separation and 2-phase coexistence. The data points correspond to the sticker concentrations in the bound phase. The sticker contration $a^2 X_s\simeq 10^{-4}$ in the unbound phase is given by the vertical lines. The critical points obtained from extrapolation are represented by stars.
\label{figurePhaseDiasStickersRepellers}}
\end{center}
\end{figure}

The character of the unbinding transition depends on the strength of the potential barrier. The barrier induces a line tension between bound and unbound membrane domains, simply because the membrane segments in the boundary region between these domains cross the potential barrier and thus have an unfavorable potential energy $U_{ba}$. According to scaling arguments \cite{lipo118,lipo124}, the unbinding transition is {\em discontinuous} for strong barriers with
\begin{equation}
 U_{ba}(l_r-l_s)^2 > c a^2 (k_B T)^2/\kappa  \label{criticalBarrier}
\end{equation}
and {\em continuous} for weak barriers with  $U_{ba}(l_r-l_s)^2 < c a^2 (k_B T)^2/\kappa$.  A discontinuous transition implies the coexistence of a bound phase with a high concentration of stickers and an unbound phase with a low sticker concentration. Sufficiently strong barriers therefore also lead to lateral phase separation and sticker aggregation. This barrier mechanism for lateral phase separation is weaker at higher temperatures $T$, in contrast to the entropic mechanisms discussed in chapter 4. Higher temperatures $T$ require larger barriers $U_{ba}$ for phase separation, see eq.~(\ref{criticalBarrier})

The coefficient $c$ in eq.~(\ref{criticalBarrier}) can be estimated from Monte Carlo simulations \cite{weikl02a}.  In the simulations, the sticker concentration $X_s\equiv\langle \delta_{n_i,1}\rangle/a^2$ is determined as a function of the chemical potential $\mu_s$  of the stickers. Lateral phase separation is reflected in a discontinuity of $X_s(\mu_s)$ at a transition value $\mu_s=\mu_s^*$. The two limiting values of $X_s$ at $\mu_s^*$ correspond to the sticker concentrations in the two coexisting phases.  Monte Carlo phase diagrams obtained from such simulations are shown in Fig.~\ref{figurePhaseDiasStickersRepellers}. The coefficient $c$ in eq.~(\ref{criticalBarrier})  can be estimated from the critical points for the 2-phase coexistence regions of the diagrams. These critical points are at $z_s^c=0.15\pm 0.05$ in the left and at $z_s^c=0.65\pm 0.05$ in the right diagram. In agreement with eq.~(\ref{criticalBarrier}), the two values differ by  a factor of 5 within the numerical accuracy, and thus reflect the same critical barrier  strength with $c=0.013\pm 0.005$.

\subsection{Membranes with stickers and generic repulsive interactions}

In the last section, we have considered the interplay of stickers and repellers during membrane adhesion. A related situation arises if stickers act against a generic repulsive interaction potential of the membranes. `Generic' means that the repulsive interaction is, or is taken to be, independent of the local composition of the membrane, in contrast to the `specific' sticker interactions.  One example for such a generic interaction is the electrostatic repulsion of equally charged membranes, provided the charge distributions can be approximated as uniform distributions. Another example is a  repulsive brush of immobilized repeller molecules. For cell membranes, immobilization of membrane molecules can arise from an anchoring to the cytoskeleton of the cell. Immobile repellers have approximately fixed positions in the membrane, in contrast to the mobile repellers considered in the last section which are free to diffuse within the plane of the membrane.

A discretized membrane with stickers and a generic potential $V_g(l_i)$ can be described by the Hamiltonian \cite{weikl02b}
\begin{equation}
{\cal H}\{l,n\} = {\cal H}_{el}\{l\} + \sum_i \left[V_g(l_i) + n_i \left(V_s(l_i)-\mu\right)\right]  \label{genham}
\end{equation}
if cis-interactions between the stickers are negligible. Summing out the sticker degrees of freedom in the partition function 
${\cal Z}$ as described in section \ref{sectionEffectivePotential} then leads to
\begin{equation}
{\cal Z}=   \bigg[\prod_i\int_{0}^{\infty} {\rm d}l_i\bigg]\exp\left[-\frac{{\cal H}_{el}\{l\}+\sum_i V_{ef}(l_i)}{k_BT}\right]
\end{equation}
with the effective potential
\begin{equation}
 V_{ef}(l) =  V_g(l) -k_BT\ln\left(1+\exp\left[ \frac{\mu - V_s(l)}{k_BT}\right] \right)
\label{Vef}
\end{equation}
%

\subsubsection{Stickers with square-well potential}

Let us first consider stickers which are again characterized by a square-well potential $V_s(l_i) = U\theta(l_s - l_i)$ with binding energy $U$ and range $l_s$, and a repulsive generic interaction which is characterized by a square-barrier potential $V_g(l_i) = U_{ba}\theta(l_{ba} - l_i)$ with barrier energy $U_{ba}>0$ and range $l_r$. If the range $l_r$ of the generic repulsion is larger than the sticker range $l_s$, the effective potential (\ref{Vef}) has the form
\begin{eqnarray}
V_{ef}(l_i) -V_o=& U_{co}&  \hspace{0.2cm}  \mbox{for \  $0<l_i<l_s$}\nonumber\\
             =& U_{ba}&   \hspace{0.2cm}  \mbox{for \  $l_s<l_i<l_r$}  \nonumber\\
            =&  0 & \hspace{0.2cm}     \mbox{for \ $l_r<l_i$} \label{potbarTwo}
\end{eqnarray}
with the contact energy
\begin{equation}
U_{co} = U_{ba} -k_B T\ln\frac{1+e^{(\mu - U)/k_BT}}{1+e^{\mu /k_BT}} \label{Uco}
\end{equation}
The constant term $V_o=-k_B T\ln\left(1+e^{\mu/k_B T}\right)$ depends only on the reduced chemical potential $\mu/k_BT$ of the stickers. 

This potential has the same form as the effective potential (\ref{potbarOne}) for a membrane with stickers and (mobile) repellers, which implies also the same phase behavior: The membrane is bound if the contact energy $|U_{co}|$ exceeds the threshold value estimated in eq.~(\ref{criticalContactEnergy}), and phase-separates if the barrier exceeds the critical strength estimated in eq.~(\ref{criticalBarrier}).

\subsubsection{Stickers with linear potential}

In the previous sections, we have seen that a barrier in the effective membrane potential causes lateral phase separation. This barrier mechanism for phase separation is rather general. So far, we have considered the interplay of attractive square-well and repulsive square-barrier potentials. For a deeper understanding of the mechanism, it is instructive to consider also the analytically convenient case in which the generic potential can be approximated by a harmonic potential 
\begin{equation}
V_g(l_i) = \frac{v_2}{2a^2} (l_i - l_o)^2  \label{harmpot}
\end{equation}
If a generic potential $V_g$ has a relatively deep minimum at a certain separation $l_o$ of the membranes, the harmonic approximation (\ref{harmpot}) can be justified by a Taylor expansion around the minimum. The prefactor in eq.~(\ref{harmpot}) is then $v_2 = a^2 (d^2 V_g/dl^2)|_{l_o}$. 
 
Let us further assume that the corresponding sticker potential $V_s(l)$ has an essentially constant gradient for those values of $l$ for which we can use the harmonic approximation (\ref{harmpot}) for the generic potential. In such a situation, we may truncate the expansion of the sticker potential in powers of  $l-l_o$ and use \cite{komura00,weikl02b}
\begin{equation}
V_s(l) = V_s(l_o) + \frac{\alpha (l - l_o)}{a} \label{linearpot}
\end{equation}
with  $\alpha \equiv a \partial V_s(l)/ \partial l |_{l_o} >  0$. This approximation may be valid for extensible, or `spring-like', sticker molecules that are irreversibly bound to the membranes and have an unstretched extension small compared to $l_o$. 

\begin{figure}
\begin{center}
\resizebox{0.5\columnwidth}{!}{\includegraphics{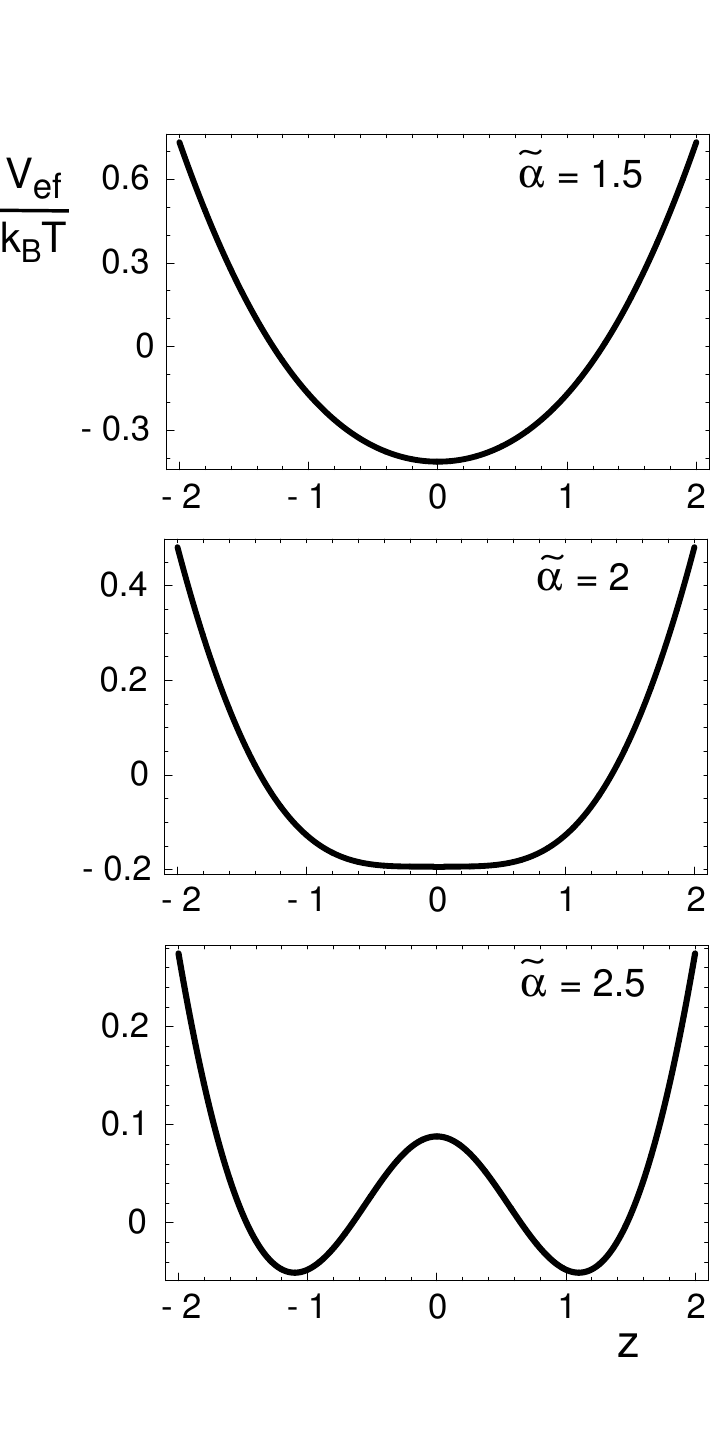}}
\caption{ The effective potential $V_{ef}$ as a function of the rescaled separation $z$ for three values of the coupling parameter $\tilde{\alpha}$. The analytical expression for $V_{ef}$ is given in (\ref{Vefstar}). The effective potential has a single minimum at 
$\tilde{\alpha}=1.5$ (top), and two degenerate minima at  $\tilde{\alpha}=2.5$ (bottom). At $\tilde{\alpha}=\tilde{\alpha}_c=2$ (middle), the potential undergoes a continuous bifurcation.
\label{figureDoubleWell}}
\end{center}
\end{figure}
 
 To simplify the notation,  we introduce here the dimensionless parameters
\begin{equation}
\tilde{\alpha} = \frac{\alpha}{\sqrt{v_2k_BT}} \hspace{0.5cm}
\mbox{and} \hspace{0.5cm} \tilde{\mu}=\frac{\mu - V_s(l_o)}{k_BT}
\label{AlphaReduced}
\end{equation}
and the rescaled separation field
\begin{equation}
z_i \equiv \tilde{\alpha}/2 + \sqrt{\frac{v_2}{k_BT}} \;\frac{l_i - l_o}{a}
\end{equation}
Let us first consider the special line in the $(\tilde{\mu}, \tilde{\alpha})$ parameter space given by 
\begin{equation}
 \tilde{\mu} = \tilde{\mu}_*\equiv -\tilde{\alpha}^2/2 \quad ,
\label{mustar}
\end{equation}
Along this line, the effective potential resulting from the eqs.~(\ref{Vef}),  (\ref{harmpot}) and (\ref{linearpot}) 
can be written in the form
\begin{equation}
\frac{V_{ef}(z)}{k_BT}\bigg|_{\tilde{\mu}= \tilde{\mu}_*} = \frac{z^2}{2}+\frac{\tilde{\alpha}^2}{8} -\ln\left[2\cosh(\tilde{\alpha}z/2)\right] \label{Vefstar}
\end{equation}
which is symmetric under the inversion $z\to -z$.  When one varies the parameter $\tilde{\alpha}$ while keeping $\tilde{\mu}=\tilde{\mu}_*(\tilde{\alpha})$, the effective potential exhibits a continuous bifurcation at the critical value $\tilde{\alpha} =\tilde{\alpha}_c = 2$, see Fig.~\ref{figureDoubleWell}. The potential has a single minimum for $\tilde{\alpha}<\tilde{\alpha}_c$, and two degenerate minima for $\tilde{\alpha}>\tilde{\alpha}_c$. The critical value $\tilde{\alpha}_c = 2$ of the bifurcation point can be directly  inferred from  the  second derivative of eq.~(\ref{Vefstar}): 
\begin{equation}
\frac{1}{k_BT}\frac{d^2 V_{ef}(z)}{d z^2} \bigg|_{\tilde{\mu}=\tilde{\mu_*}}= 1-\frac{\tilde{\alpha}^2}{4\cosh^2(\tilde{\alpha}z/2)}
\quad .
\end{equation}
For $z=0$, this expression is equal to $1-\tilde{\alpha}^2/4$, which  vanishes for $\tilde{\alpha} = \tilde{\alpha}_c=2$. 

{\em Limit of rigid membranes} --
 At large values of  the ratio $\kappa/v_2$, the membrane is quasi rigid. Thermally excited shape fluctuations of the membrane thus can be neglected. The free energy ${\cal F} = - (k_BT/A)\ln {\cal Z}$  per area $A$ is then simply given by $V_{ef}/a^2$, and the phase behavior can be determined by minimizing the effective potential. For $\tilde{\mu} = \tilde{\mu}_*(\tilde{\alpha})$ and $\tilde{\alpha}>2$, the effective potential (\ref{Vefstar}) is a symmetric double-well potential with two degenerate minima. As soon as the chemical potential  $\tilde{\mu}$ deviates from its  coexistence value  $\tilde{\mu} = \tilde{\mu}_*$, this symmetry is broken and the effective potential has a unique global minimum. The system thus exhibits a discontinuous transition when one changes the chemical potential from $\tilde{\mu} = \tilde{\mu}_* - \epsilon$ to $\tilde{\mu} = \tilde{\mu}_* + \epsilon$ for $\tilde{\alpha}>2$. Here, $\epsilon$ denotes a small chemical potential difference. In the limit of rigid membranes, the critical point for phase separation is identical with the bifurcation point of the effective potential at $\tilde{\alpha}_c = 2$ and $\tilde{\mu}_c=-\tilde{\alpha_c}^2/2=-2$.

The positions of the extrema  of the effective potential are determined by $dV_{ef}(z)/dz=0$. Along the  coexistence line with $\tilde{\mu}= \tilde{\mu}_*=-\tilde{\alpha}^2/2$, this leads to the transcendental equation
\begin{equation}
z = \frac{\tilde{\alpha}}{2} \tanh\left( \frac{\tilde{\alpha}z}{2} \right) \quad .
\label{trans}
\end{equation}
This equation  has the trivial solution $z=0$ for all values of $\tilde{\alpha}$. This solution corresponds to a minimum for $\tilde{\alpha} < \tilde{\alpha}_c=2$, and to a maximum for $\tilde{\alpha}>\tilde{\alpha}_c=2$. For $\tilde{\alpha}>\tilde{\alpha}_c=2$, eq.~(\ref{trans}) has two additional solutions corresponding to the two degenerate minima of the effective potential $V_{ef}$, see Fig.~\ref{figureDoubleWell}.

\begin{figure}[t]
\begin{center}
\resizebox{0.5\columnwidth}{!}{\includegraphics{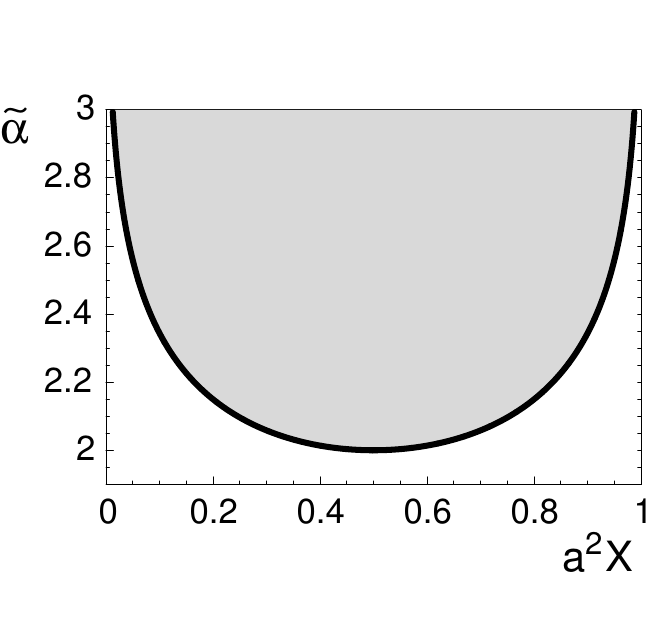}}
\caption{ Phase diagram for linear stickers in the absence of membrane fluctuations, depending on the sticker concentration $X$ and the reduced coupling constant $\tilde{\alpha}$. Within the grey two-phase region, a sticker-poor phase characterized by a relatively large membrane-surface separation coexists with a sticker-rich phase for which this separation is relatively small. The critical point is located at $a^2 X_c = 1/2$ and $\tilde{\alpha}_c = 2$.
\label{figurePhaseDiaDoubleWell}}
\end{center}
\end{figure}

Along the coexistence line, the sticker concentration 
\begin{equation}
X \equiv\langle n_i\rangle/a^2= -\frac{\partial F}{\partial \mu} = -\frac{1}{a^2}\frac{\partial V_{ef}}{\partial \mu}
\end{equation}
is given by
\begin{equation}
X \bigg|_{\tilde{\mu}= \tilde{\mu}_*}= \frac{1}{a^2}\frac{e^{-\tilde{\alpha}z_o}}{1+e^{-\tilde{\alpha}z_o}} \quad  .
\label{Xeins}
\end{equation}
The concentrations of the coexisting phases then are obtained by inserting the numerical solutions of the transcendental equation (\ref{trans}) into (\ref{Xeins}). The resulting phase diagram is shown in Fig.\ \ref{figurePhaseDiaDoubleWell}. Inside the shaded two-phase region, a sticker-poor phase with large membrane separation coexists with a sticker-rich phase with smaller separation.

{\em Flexible membranes} --
A flexible and, thus, fluctuating membrane can easily cross small barriers in the potential. First-order transitions then only occur if the barrier exceeds a critical height \cite{lipo118,lipo124}.  For a flexible membrane, the critical coupling constant $\tilde{\alpha}_c$ therefore will be larger than the bifurcation value $\tilde{\alpha}_c=2$ of the effective potential (\ref{Vefstar}).

\begin{figure}[t]
\begin{center}
\resizebox{0.5\columnwidth}{!}{\includegraphics{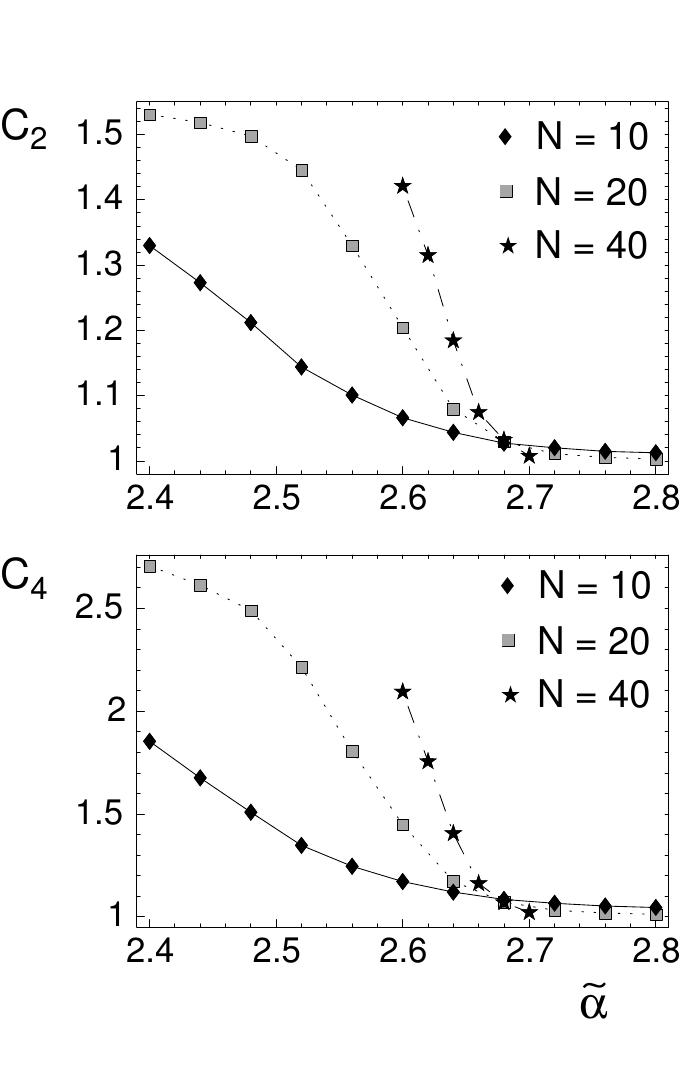}}
\caption{Monte Carlo data for the moments $C_2$ and $C_4$ defined in eq.~(\ref{C2C4Moments}) as a function of the reduced coupling constant $\tilde{\alpha}$. The ratio of the bending rigidity $\kappa$ and the strength $v_2$ of the generic harmonic potential (\ref{harmpot}) has the fixed value  $\kappa/v_2 = 1$. The membrane segments considered in the simulations consist of $N\times N$ membrane patches. 
\label{figureCumulants}}
\end{center}
\end{figure}

With Monte Carlo simulations, the critical point  can be determined {\it via} the moments
\begin{equation}
C_2 = \frac{\langle\bar{z}^2\rangle}{\langle |\bar{z}|\rangle^2}
\hspace{0.5cm} \mbox{and} \hspace{0.5cm}
C_4 = \frac{\langle\bar{z}^4\rangle}{\langle \bar{z}^2\rangle^2}
\label{C2C4Moments}
\end{equation}
Here,
\begin{equation}
\bar{z} = \frac{1}{N}\sum_{i=1}^N z_i
\end{equation}
is the spatially averaged order parameter, and $\langle\cdots\rangle$ denotes averages over all membrane configurations \cite{lipo124,binder92}. In principle, the values of these moments depend on the correlation length $\xi$ and the linear size $L$ of a membrane segment. But at the critical point, the correlation length $\xi$ diverges, and the values of the moments become independent of $L$ \cite{lipo124,binder92}. Therefore, the critical coupling constant $\tilde{\alpha}_c$  can be estimated from the common intersection points of the functions $C_2(\tilde{\alpha}$) and $C_4(\tilde{\alpha})$ at different values of $L$, see Fig.~\ref{figureCumulants}.

\begin{figure}
\begin{center}
\resizebox{0.5\columnwidth}{!}{\includegraphics{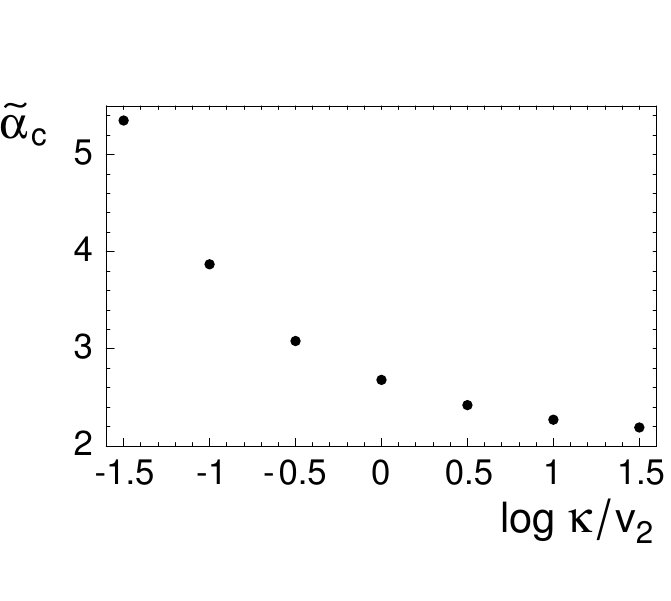}}
\caption{The critical values $\tilde{\alpha}_c$ of the coupling constant $\tilde{\alpha}$ as a function of $\kappa/v_2$. Here, $\kappa$ is the membrane rigidity, and $v_2$ is the strength of the generic harmonic potential. The coupling constant $\tilde{\alpha}$ is defined in eq.~(\ref{AlphaReduced}) and governs the strength of the linear sticker potential (\ref{linearpot}).  For large values of $\kappa/v_2$, the critical coupling constants tend towards the value $\tilde{\alpha_c} = 2$ of `rigid' membranes. The statistical errors here are smaller than the symbol sizes. 
\label{figureCriticalPoints}}
\end{center}
\end{figure}

Fig.~\ref{figureCriticalPoints} shows the critical rescaled coupling constant $\tilde{\alpha}_c$ as a function of the reduced rigidity $\kappa/v_2$. For large $\kappa/v_2$, $\tilde{\alpha}_c$ approaches the limiting value $\tilde{\alpha}_c=2$  of `rigid' membranes, see above. With decreasing $\kappa/v_2$, the membrane shape fluctuations become more pronounced and lead to an increase in the value of $\tilde{\alpha}_c$. As in section \ref{sectionMobileRepellers}, the membrane fluctuations thus reduce the tendency for lateral phase separation.

Lateral phase separation occurs for coupling constants $\tilde{\alpha}>\tilde{\alpha}_c$. In either of the two phases, the entire membrane then is located around one of the minima in the effective potential. In the sticker-poor phase, the membrane is found in the minimum with larger membrane separation. This minimum is dominated by the generic membrane potential and corresponds to a state of weak adhesion. In the sticker-rich phase, the membrane fluctuates around the minimum with smaller separation, which corresponds to a state of tight adhesion. In contrast, there is only a single phase for coupling constants $\tilde{\alpha}<\tilde{\alpha}_c$. For $2<\tilde{\alpha}<\tilde{\alpha}_c$ for example, the two minima of the effective potential are both populated by many different segments of the fluctuating membrane, which is then able to cross the potential barrier between the minima.

\section{Dynamics of domain formation during adhesion}
\label{S.Dynamics}

\begin{figure}
\begin{center}
\resizebox{0.45\columnwidth}{!}{\includegraphics{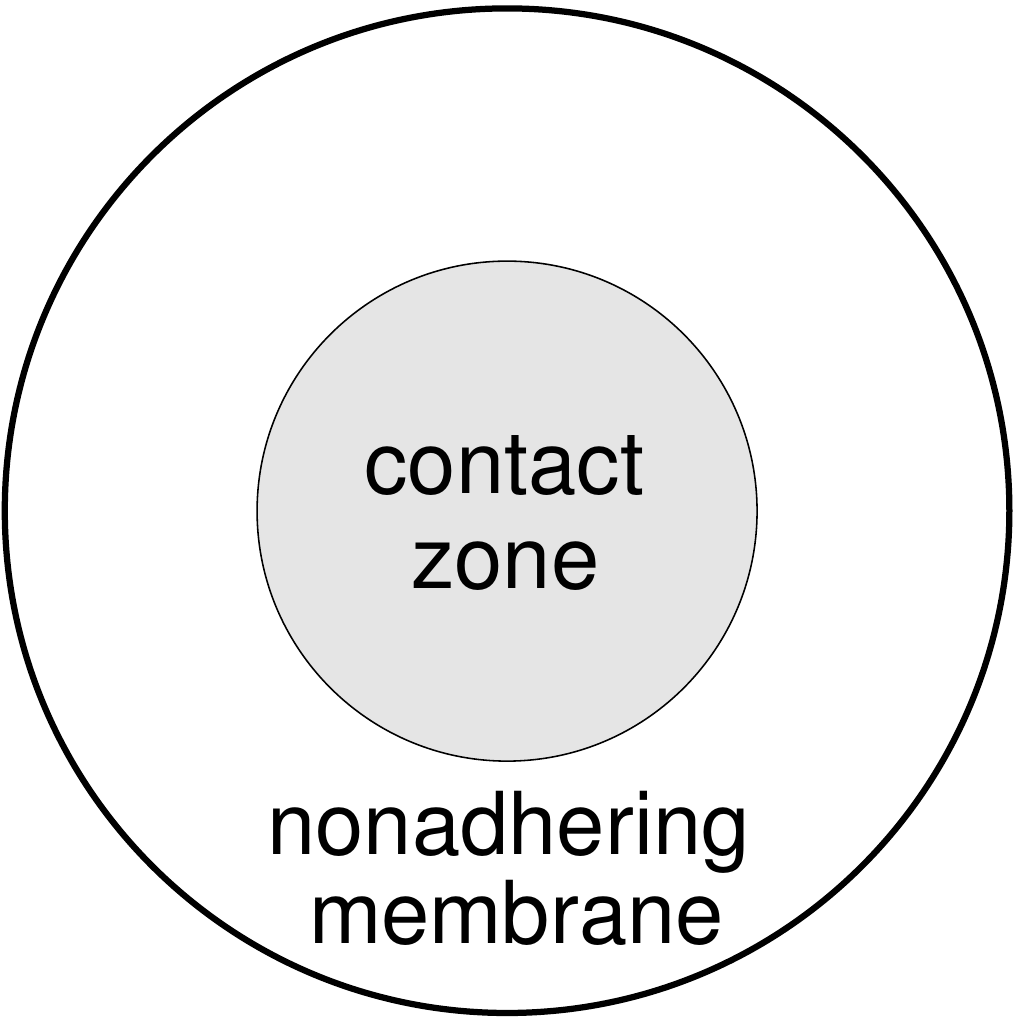}}
\caption{ 
`Cell' adhesion geometry: The circular contact zone is surrounded by a nonadhering membrane ring.
Stickers, or receptors, diffuse around in the whole membrane, but interact with the apposing membrane only within the contact zone.
\label{figureContactZone}}
\end{center}
\end{figure}

So far, we have focused on equilibrium aspects of the domain formation. In this section, we consider the adhesion dynamics, or in other words, the time-dependent evolution of the domain patterns. The models presented in this section mimick the adhesion geometry of cells or vesicles by dividing the membranes into two zones:  a contact zone, and a surrounding nonadhering membrane region. The stickers and receptors can diffuse in the whole membrane, but interact with the second membrane only within the contact zone. The problem of modeling the full shapes of cells or vesicles thus is avoided in these models. Instead, the contact zone is assumed to have an essentially circular shape and a constant area on the relevant time scales, see Fig.~\ref{figureContactZone}. For biomimetic vesicles with stickers and repellers, this adhesion geometry can result from fast initial gravity-induced adhesion of the vesicles on a supported membrane \cite{boulbitch01}. In the case of adhering T cells, initial adhesion is mediated by relatively long integrins. Experimental pictures show that the contact zone of T cells fully develops in less than 30 seconds \cite{grakoui99}.

\subsection{Adhesion of vesicles with stickers and repellers}
\label{sectionSR}

Within the contact zone, the membrane of an adhering vesicle with stickers and repellers can be described by the Hamiltonian \cite{weikl02a}
\begin{equation}
{\cal H}\{l,n\} = {\cal H}_{el}\{l\} + \sum_i \left[ \delta_{1,n_i} V_s(l_i) + \delta_{2,n_i} V_r(l_i) \right]
\label{SRhamiltonian}
\end{equation}
where $\delta_{i,j}$ is the Kronecker delta. In the nonadhering membrane region surrounding the contact zone (see Fig.~\ref{figureContactZone}), the stickers and repellers do not interact with the second membrane.  In this region, the configurational energy of the stickers and repellers thus is constant, i.e.~independent of the sticker and repeller positions, and independent of the membrane shape. The membrane shape therefore is only modeled explicitly  within the contact zone {\it via} the separation field $l$. In this model, we use `free' boundary conditions of the separation field $l$ at the contact zone rim, i.e.~an unconstrained boundary separation. The Hamiltonian (\ref{SRhamiltonian}) is the canonical equivalent of the grand canonical Hamiltonian (\ref{hamiltonianMobileRepellers}) of section \ref{sectionMobileRepellers}. We consider now the canonical ensemble since the total numbers of stickers and repellers in the whole vesicle membrane are constant. 

In the following, the sticker potential is a square-well potential with depth $U_s=-10 k_B T$ and rescaled range $z_s$, and the repeller potential is a square-barrier with height $U_r=10 k_B T$ and rescaled range $z_r> z_s$. For these relatively large energies of $10 k_B T$, the sticker binding is nearly irreversible, i.e.~the majority of stickers stays bound after first binding, and the repellers exclude membrane separations smaller than $z_r$ almost completely. The contact zone of the membrane here has the diameter 100 $a$, and the whole membrane has the diameter 200 $a$, see Fig.~\ref{figureContactZone}. 

\begin{figure}[t]
\begin{center}
\resizebox{0.95\columnwidth}{!}{\includegraphics{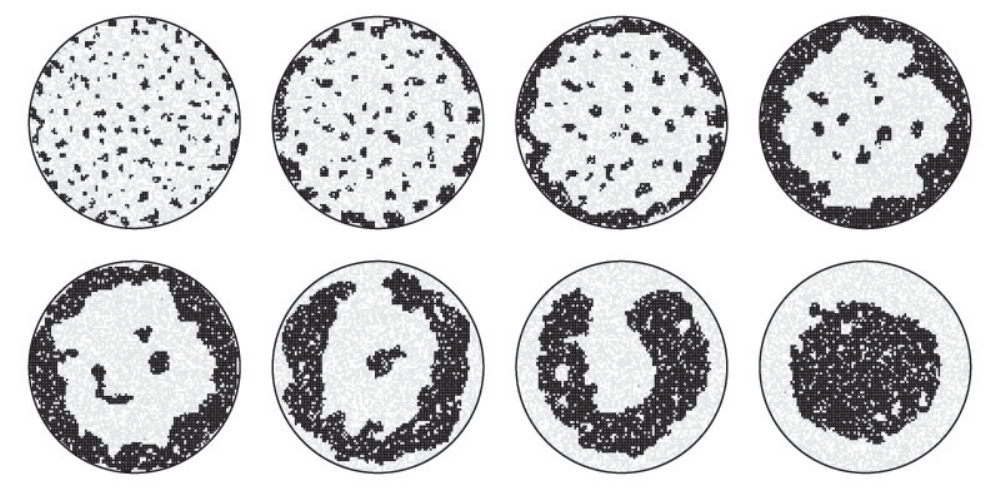}}
\caption{Typical pattern evolution in the dynamic regime (B). The potential ranges of stickers and repellers are $z_s=1.0$, $z_r=1.6$ and the overall concentrations of stickers and repellers are $a^2 X_s=0.1/a^2$,  $a^2 X_r=0.5/a^2$. Stickers are shown in black, repellers in grey. The circle represents the rim of the contact zone. Because of the diffusion of stickers into the contact zone, clusters at  the rim grow faster, forming rings at intermediate time scales.  The final configuration represents the equilibrium state. The snaphots are taken at  $10^3$, $4\cdot 10^3$, $10^4$, $4\cdot 10^4$,   $1.6\cdot 10^5$, $6.3\cdot 10^5$, $10^6$, and $4\cdot 10^6$ MC steps.
\label{figureSRone}}
\end{center}
\end{figure}

The formation and evolution of the domain patterns can be studied with Monte Carlo simulations. The Monte Carlo simulations presented below start from a random distribution of stickers and repellers, and a rescaled membrane separation $z_i = z_r$ in the  contact zone, with all stickers unbound.  A Monte Carlo step consists in attempts (i) to move each sticker and repeller to one of the 8 neighbor sites (lateral diffusion) and (ii)  to shift the rescaled membrane separation $z_i$ at every lattice site $i$ in the contact zone (shape fluctuations).

In this model, three different dynamic regimes of pattern formation can be observed. The dynamic regimes depend  on the characteristic lengths and the concentrations of stickers and  repellers: 

{\it Regime (A)}: Long repellers impose a strong barrier to sticker adhesion. The nucleation time for sticker binding therefore is large compared to typical diffusion times, and the membrane binds {\it via}  growth of a single
sticker nucleus. Such an adhesion behavior has been recently observed for biomimetic vesicles with PEG-lipopolymers as repellers  and integrins as stickers \cite{boulbitch01}.

\begin{figure}[t]
\begin{center}
\resizebox{\columnwidth}{!}{\includegraphics{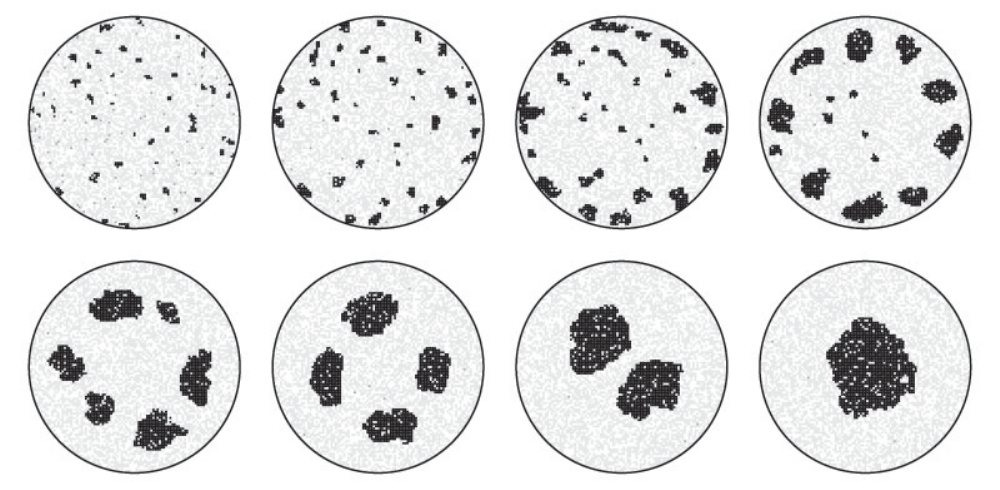}}
\caption{Typical pattern evolution in the dynamic regime (C). The potential ranges of stickers and repellers are 
 $z_s=1.0$, $z_r=1.8$, and the overall concentrations are  $X_s=0.04/a^2$, $X_r=0.5/a^2$. Stickers are shown in black, repellers in grey.  The sticker concentration is smaller than in Fig.~\ref{figureSRone}. Therefore, circular patterns of separate clusters emerge at intermediate times, instead of closed sticker rings as in Fig.~\ref{figureSRone}.  The snaphots are taken at $10^3$, $4\cdot 10^3$, $1.6\cdot 10^4$, $6.3\cdot 10^4$, $4\cdot 10^5$, $6.3\cdot 10^5$, $1.6\cdot 10^6$, and $4\cdot 10^6$ MC steps. 
\label{figureSRtwo}}
\end{center}
\end{figure}

{\it Regime (B)}: For short repellers, the nucleation time for sticker binding  is small,  and many nuclei of bound  stickers are formed initially. Since unbound stickers diffuse into the contact zone, nuclei at the rim of this zone grow faster.  At sufficiently high  sticker concentrations,  an intermediate ring of bound stickers is then formed, enclosing a central domain of repellers, see Fig.~\ref{figureSRone}. Later,  this pattern inverts, and a single central sticker cluster is surrounded by repellers. The central position of this sticker clusters here is caused by membrane shape fluctuation in the unbound repeller domain surrounding the cluster. The sequence of  patterns in this regime has a striking similarity to the pattern evolution observed during T cell adhesion (see Introduction and next section).

{\it Regime (C)}: In an intermediate regime, the sticker concentration is not large enough for the formation of a closed sticker ring from the initial nuclei. Instead, circular arrangements of separate  sticker clusters emerge at intermediate times, see Fig.~\ref{figureSRtwo}.

\begin{figure}[t]
\begin{center}
\resizebox{\columnwidth}{!}{\includegraphics{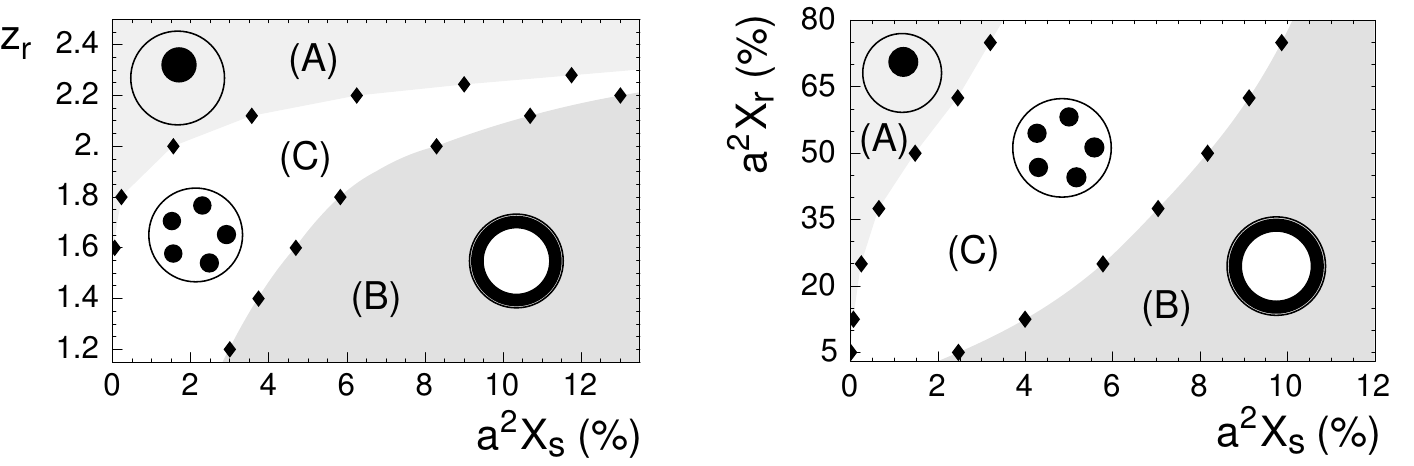}}
\caption{Dependence of the dynamic regimes (A), (B), and (C) on the sticker and repeller concentrations
$X_s$ and $X_r$ and the repeller range $z_r$. The sticker potential range here is $z_s=1.0$. The repeller concentration is $X_r=0.5/a^2$ in the left diagram. In the right diagram, the repeller range is $z_r=2.0$.  Characteristic intermediate
patterns are indicated for all three regimes.  In the upper left grey area (regime (A)), the maximal number of sticker clusters $N_{cl}^{max}$ during adhesion is smaller than 3. In the lower right grey area (regime (B)), the maximal
ring occupation $Y^{max}$ is larger than $0.8$. The nucleation time for sticker clusters increases with the length $z_r$ or concentration $X_r$ of repellers, and thus also the extent of the dynamic regime (A). 
\label{figureSRthree}}
\end{center}
\end{figure}

Two quantities are helpful to characterize the three different regimes (A), (B), and (C) systematically \cite{weikl02a}. The first quantity is the maximal number of bound sticker clusters $N_{cl}^{max}$ during the pattern evolution. The number of sticker clusters $N_{cl}$ first increases since new sticker clusters nucleate, and later decreases as a consequence of cluster coalescence. The second quantity is the maximal sticker occupation $Y^{max}$ in an outer ring of the contact zone. To define this quantity, the membrane  ring with distances  $40\;a<r<50\;a$ from the center of the contact zone is divided  into 100 equal segments,  each covering an angle $2\pi/100$. The ring occupations simply is the fraction of segments $Y$  which contain bound stickers.  The ring occupation $Y$ has a maximal value $Y^{max}$ at intermediate times when a ring of sticker clusters is formed. 

Appropriate values to describe the crossover between the three dynamic regimes are the maximal number of sticker clusters $N_{cl}^{max}=3$ and the maximal ring occupation $Y^{max}=0.8$ (see Fig.~24): Simulations with $Y^{max}>0.8$ show intermediate configurations  with a closed ring of bound stickers as in Fig.~\ref{figureSRone} (dynamic regime (B)).  For $N_{cl}^{max}<3$, on the other hand, adhesion proceeds by sticker condensation mostly around a single, dominant nucleus. For $N_{cl}^{max}>3$ and $Y^{max}<0.8$, configurations with a circular arrangement of separate clusters emerge as in Fig.~\ref{figureSRtwo} (dynamic regime (C)).

\pagebreak

\subsection{Adhesion of T cells}
\label{sectionTcells}

\subsubsection{Model}

\begin{figure}
\begin{center}
\resizebox{0.7\columnwidth}{!}{\includegraphics{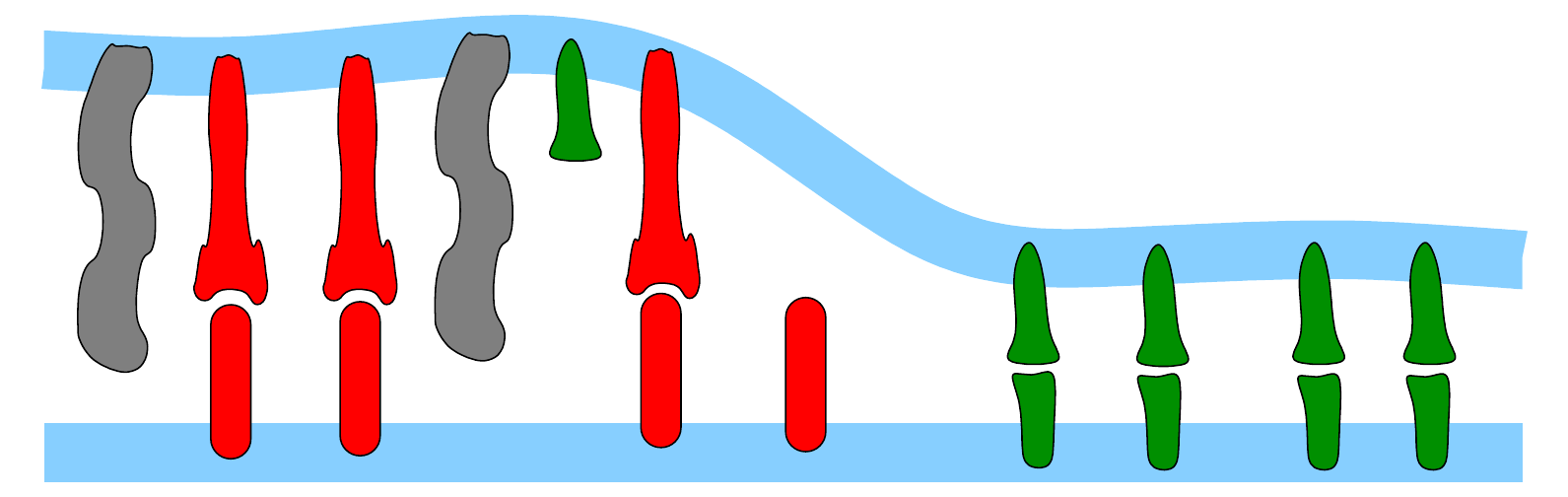}}
\end{center}
\caption{Cartoon of a T cell membrane (top) adhering to an APC membrane (bottom). The T cell membrane contains the T cell receptor TCR (green) and the receptor LFA-1 (red). The APC membrane contains the corresponding ligands MHCp (green) and  ICAM-1 (red). Both membranes contain repulsive glycoproteins (grey). Because of the different lengths of bound TCR/MHCp complexes, LFA-1/ICAM-1 complexes, and glycoproteins, the membrane phase separates into domains. \label{Tcartoon}} 
\end{figure}

Helper T cells mediate immune responses to antigen-presenting cells (APCs), see Introduction. The cell adhesion model presented here considers two apposing membranes with different concentration fields. The first membrane represents the T cell and contains the receptors TCR and LFA-1. The second membrane represents the APC and contains the ligands MHCp and ICAM-1 (see Fig.~\ref{Tcartoon}). 

To keep MC simulations of the relatively large cell membranes and contact zones tractable, the value $z=1$
of the rescaled membrane separation field  $z=(l/a)\sqrt{\kappa/T}$ corresponds in this model to a length of 20 nm. This results in the relation $a=20\sqrt{\kappa/(k_B T)}$~nm for the linear patch size. For the typical bending rigidities $\kappa_1=\kappa_2=25 k_B T$ of the two biomembranes \cite{seifert95}, the effective rigidity $\kappa$ has the value $12.5 k_BT$, and the linear patch size is $a\simeq 70$ nm. Monte Carlo simulations with smaller patch sizes should lead to comparable results, but require significantly longer computation times. 

Since the membrane patch size $a$ is relatively large, the model allows several molecules in a single patch. The local composition of the T cell membrane is then described by the numbers $n_i^T$ of TCRs, $n_i^L$ of LFA-1, and $n_i^{Gt}$ of glycoproteins in each membrane patch $i$. Correspondingly, the composition of the APC membrane is given by the numbers $n_i^M$ of MHCp, $n_i^{I}$ of ICAM-1, and $n_i^{Ga}$ of glycoproteins in all patches. 
The overall configurational energy of the membranes in the contact zone is the sum of the elastic energy (\ref{elasticEnergy}) and the interaction energies of receptors, ligands, and glycoproteins \cite{weikl04}:
\begin{eqnarray}
{\cal H}\{l,n\} = {\cal H_\text{el}}\{l\} + \sum_i\Big[ \min(n_i^T,n_i^M)V_\text{TM}(l_i)
  +  \min(n_i^L,n_i^I)V_\text{LI}(l_i) \nonumber\\ + \left(n_i^{Gt}+n_i^{Ga}\right) V_G(l_i)\Big] \hspace*{2cm}
\label{totalEnergy}
\end{eqnarray}
Here, $V_\text{TM}(l_i)$ and $V_\text{LI}(l_i)$ are the attractive interaction potentials of TCR/MHCp and LFA-1/ICAM-1 complexes, $V_G(l_i)$ is the repulsive interaction potential of the glycoproteins. The term $\min(n_i^T,n_i^M)$ denotes the minimum of the numbers of TCR and MHCp molecules at site $i$. This minimum is equivalent to the number of interacting TCR/MHCp pairs in the apposing patches at site $i$. The elastic energy
\begin{equation}
{\cal H_\text{el}}\{l\} = \sum_i\left[(\kappa/2a^2)(\Delta_d l_i)^2 + (\sigma/2)(\nabla_d l_i)^2\right]
\label{elasticEnergy}
\end{equation}
of the model has a contribution from a lateral tension $\sigma$. In the simulations presented here, the value for the lateral tension is  $\sigma=0.1 \kappa /a^2\simeq 2\cdot 10^{-6}N/m$, which is within the range of values measured for Dictyostelium discoideum cells  \cite{simson98}.

The receptor-complexes can only form if the membrane separation is in an appropriate range. The length of the TCR/MHCp complexes is about 15~nm, while the LFA-1/ICAM-1 complexes have a length of about 40~nm \cite{dustin00}. The membrane within a patch is `rough' because of the thermal fluctuations on length scales smaller than the linear extension $a\simeq 70$~nm of the patches. Therefore, receptor/ligand complexes can also form if the  separation of two apposing patches deviates slightly from the precise lengths $z_{TM}$ and $z_{LI}$ of the complexes. In the model, the interaction potential of TCR and MHCp is characterized by the square-well potential
\begin{eqnarray}
 V_{TM} &=& -U_{TM}  \hspace{0.3cm} \text{for} \hspace{0.2cm} 10\text{~nm} < l_i < 20\text{~nm},\nonumber\\
      &=& 0 \hspace{0.9cm} \text{otherwise} \label{potTM}
\end{eqnarray}
 and the interaction potential of ICAM-1 and LFA-1 is
\begin{eqnarray}
 V_{LI} &=& -U_{LI}  \hspace{0.3cm} \text{for} \hspace{0.2cm} 35\text{~nm} < l_i < 45\text{~nm},\nonumber\\
      &=& 0 \hspace{0.9cm} \text{otherwise} \label{potLI}      
\end{eqnarray}
Here, $U_{TM}>0$ is the binding energy of a TCR/MHCp complex, and $U_{LI}>0$ the binding energy of LFA-1/ICAM-1.

The repulsive glycoproteins protruding from both membranes vary in size. However, many of these proteins have a length comparable to the length of the LFA-1/ICAM-1 complexes. These glycoproteins do not inhibit the binding of ICAM-1 and LFA-1, but impose a steric barrier for the formation of TCR/MHCp complexes. They are characterized by the potential
 \begin{eqnarray}
 V_{G} &=& U_G (l-l_G)^2    \hspace{0.3cm} \text{for} \hspace{0.2cm} l < l_G,\nonumber\\
      &=& 0 \hspace{0.9cm} \text{otherwise}
 \end{eqnarray}
with $U_G=10\kappa/a^2$ and $l_G=40$~nm. This potential aims to capture that a membrane patch of size $a$ containing a glycoprotein has to bend around this protein to achieve an overall patch separation smaller than the length of the glycoprotein.

\subsubsection{Adhesion dynamics without cytoskeletal transport}

We first consider the pattern formation in the absence of active forces which transport molecules in or out of the contact zone.  In the absence of active transport, the diffusive motion of the macromolecules is modeled as an unbiased hopping process between neighboring membrane patches as in section \ref{sectionSR}. Each receptor, ligand, or glycoprotein in a certain membrane patch can hop to one of the four nearest neighbor patches during a single time step. During a time step, we also attempt to shift the separation $l_i$ between apposing membrane patches in the contact zone by $d\cdot\zeta[-1,1]$ where $d$ is the step width $10$~nm, and $\zeta[-1,1]$ is a random number between $-1$ and $1$. 

\begin{figure}[t]
\vspace{-0.2cm}
\resizebox{\columnwidth}{!}{\includegraphics{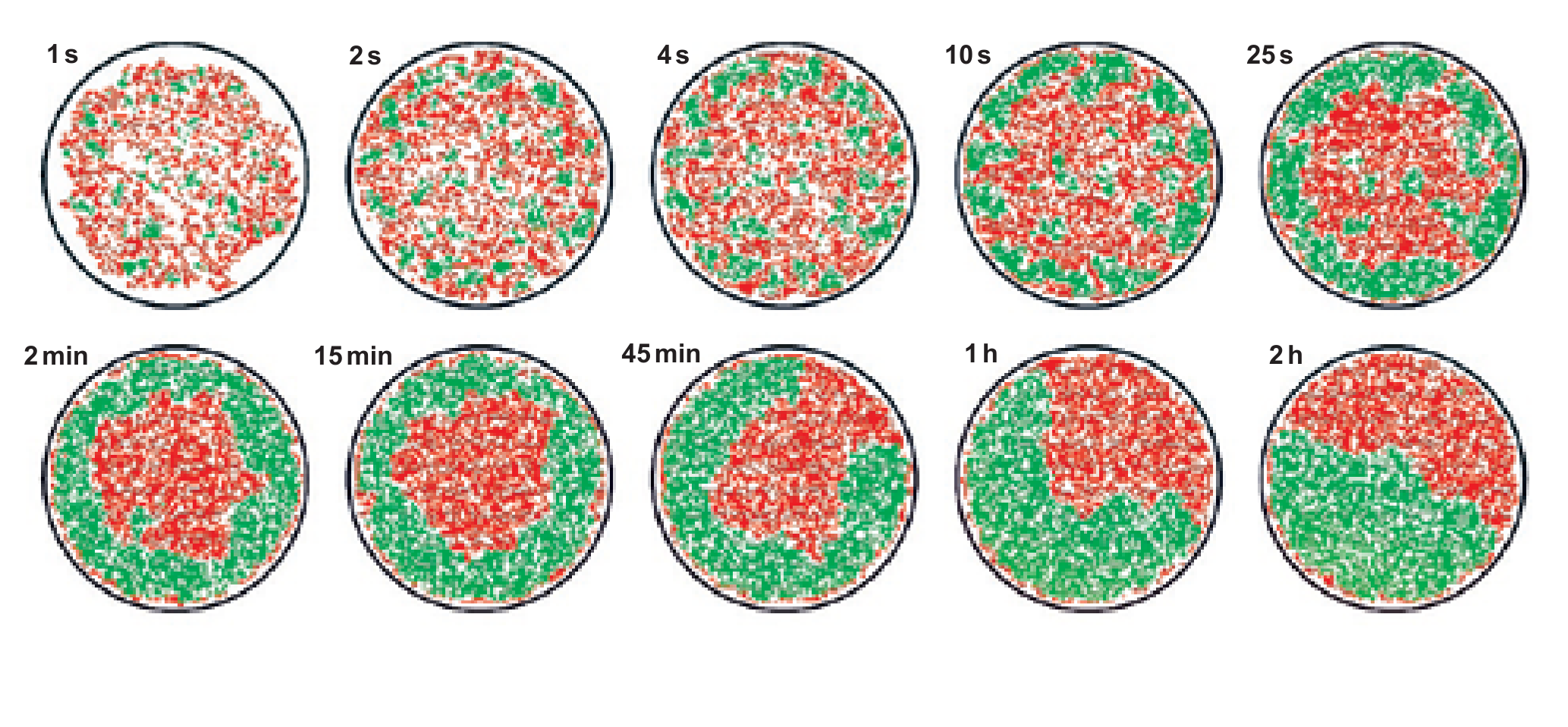}}
\vspace{-1.3cm}
\caption{Typical pattern evolution without active TCR transport in the dynamic regime 1. Membrane patches with bound TCR/MHCp complexes are shown in green, patches with bound LFA-1/ICAM-1 complexes in red. The black circle represents the contact zone rim. The effective binding energies of the TCR/MHCp and the LFA-1/ICAM-1 complexes are 
 $U_{TM}=6.5 k_B T$  and  $U_{LI}=3 k_B T$. The overall concentrations of TCR, ICAM-1, LFA-1, and glycoproteins in each of the membranes is $0.4 / a^2 \simeq 80$ molecules/$\mu$m$^2$ for a linear patch size $a\simeq 70$ nm, and the concentration of MHCp is $0.1 / a^2 \simeq 20$ molecules/$\mu$m$^2$.}
\label{figureTrings}
\end{figure}

A single Monte Carlo step roughly corresponds to 1 ms of real time. This time estimate can be derived from the 2-dimensional diffusion law $\langle x^2\rangle = 4 D t$ and the typical diffusion constant $D\simeq 1$~$\mu$m$^2$/s for membrane-anchored macromolecules. In a single Monte Carlo step, a free receptor, free ligand, or a glycoprotein moves a distance $a$ to a neighboring membrane patch, which corresponds to a diffusion time $t=a^2/(4 D)\simeq 1$ ms for $a=70$~nm. On the length scale of our patches, the diffusive motion of the macromolecules is slower than the relaxation of the membrane separation \cite{brochard75} and hence defines the time scale.

As initial conformation of the MC simulations, the separation profile is $l=l_o +c r^4$ where $r$ is the distance from the center of the contact zone, $l_o$ is 45~nm, and $c>0$ is chosen so that the separation at the rim of the contact zone with radius $r=45a$ is 100~nm (`clamped' boundary condition). This initial separation in the contact zone is larger than 45~nm, and thus beyond the interaction range of receptors, ligands, and glycoproteins. Initially, these molecules are taken to be randomly distributed within the whole membrane.

Since the length difference of the complexes leads to phase separation at the molecular concentrations considered below, the two types of receptor/ligand complexes have to `compete' for the contact zone. In general, the overall area of TCR/MHCp domains in the contact zone increases with the concentrations of TCR and MHCp molecules and with the effective binding energy $U_{TM}$.  However, if the molecular concentrations or the binding energy are too small, TCR/MHCp domains do not form, and the contact zone contains only bound LFA-1/ICAM-1 complexes. At molecular concentrations and binding energies where TCR/MHCp and LFA-1/ICAM-1 domains coexist, we observe two different regimes for the dynamics with clearly distinct patterns of TCR/MHCp domains at intermediate times. The pattern evolution roughly depends on the overall area of TCR/MHCp domains after initial relaxation.

\begin{figure}
\vspace{-0.2cm}
\resizebox{\columnwidth}{!}{\includegraphics{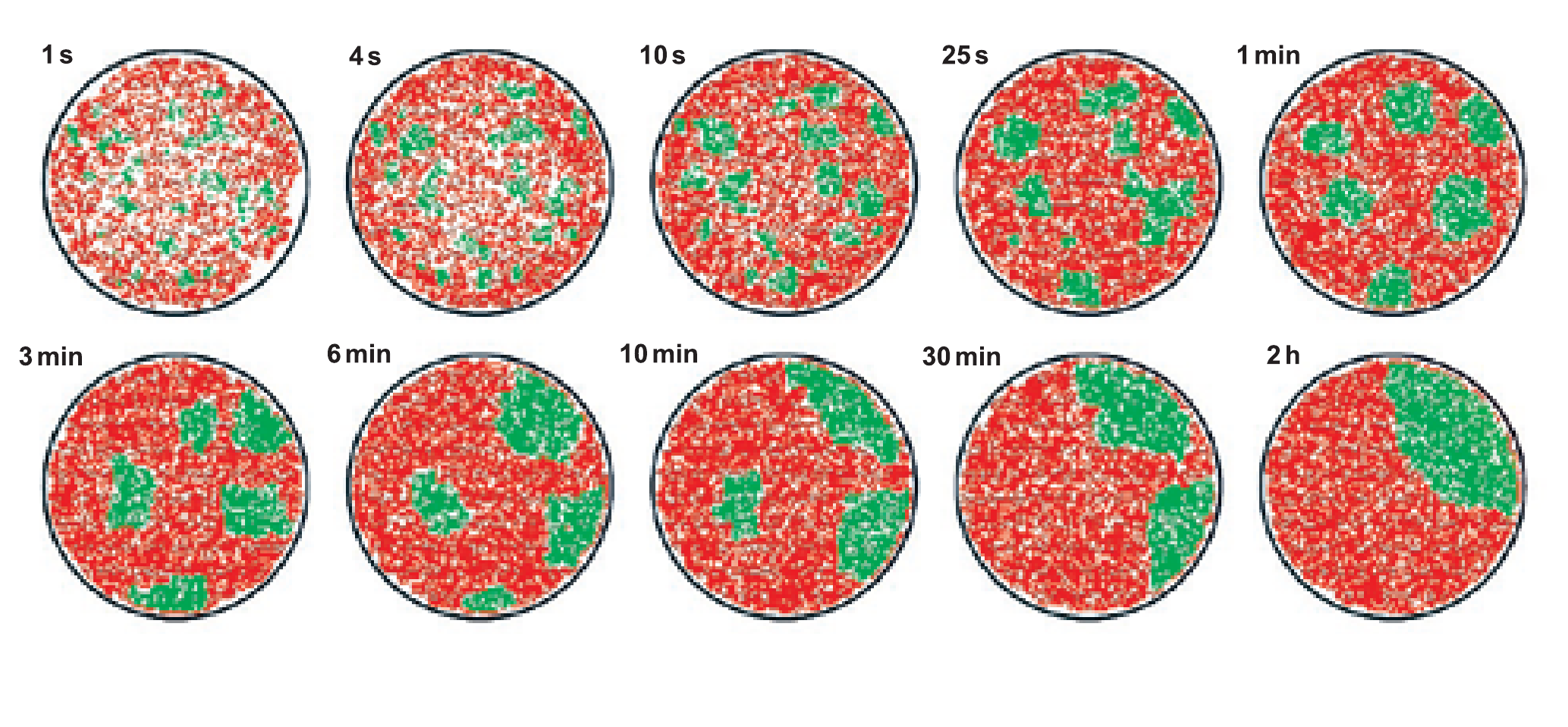}}
\vspace{-1.3cm}
\caption{Typical pattern evolution without active TCR transport in the dynamic regime 1. Membrane patches with bound TCR/MHCp complexes are shown in green, patches with bound LFA-1/ICAM-1 complexes in red. The effective binding energies are $U_{TM}=5.5 k_B T$ and $U_{LI}=4 k_B T$, and the molecular concentrations are the same as in Fig.~\ref{figureTrings}. 
}
\label{figureTmultifocal}
\end{figure}

{\it Regime 1}: If the overall area of TCR/MHCp domains is relatively large, we observe a characteristic ring-shaped TCR/MHCp domain at intermediate times, surrounding a central domain of LFA-1/ICAM-1 complexes, see Fig.~\ref{figureTrings} The ring finally breaks to form a single large TCR/MHCp domain. The pattern evolution in this regime is very similar to regime (B) of section \ref{sectionSR}. 

{\it Regime 2}: For smaller TCR or MHCp concentrations, or smaller effective binding energy, characteristic multifocal TCR/MHCp patterns emerge at intermediate times. A typical example is shown in Fig.\ \ref{figureTmultifocal}. The patterns in this regime are similar to those formed in regime (C) of section \ref{sectionSR}.

To distinguish the two dynamic regimes systematically, it is useful to consider again the maximal sticker occupation $Y^\text{max}$ in a peripheral ring of the contact zone, now with distances $35\;a<r <45\; a$ from the center. A pattern evolution with $Y^\text{max}<80$~\% typically has multifocal intermediates as in Fig.~\ref{figureTmultifocal} (Regime 2), while pattern evolutions with $Y^\text{max}>80$ exhibit the inverted synapse of T cells with a peripheral TCR/MHCp ring, see Fig.~\ref{figureTrings} (Regime 1). 

\begin{figure}[t]
\begin{center}
\resizebox{\columnwidth}{!}{\includegraphics{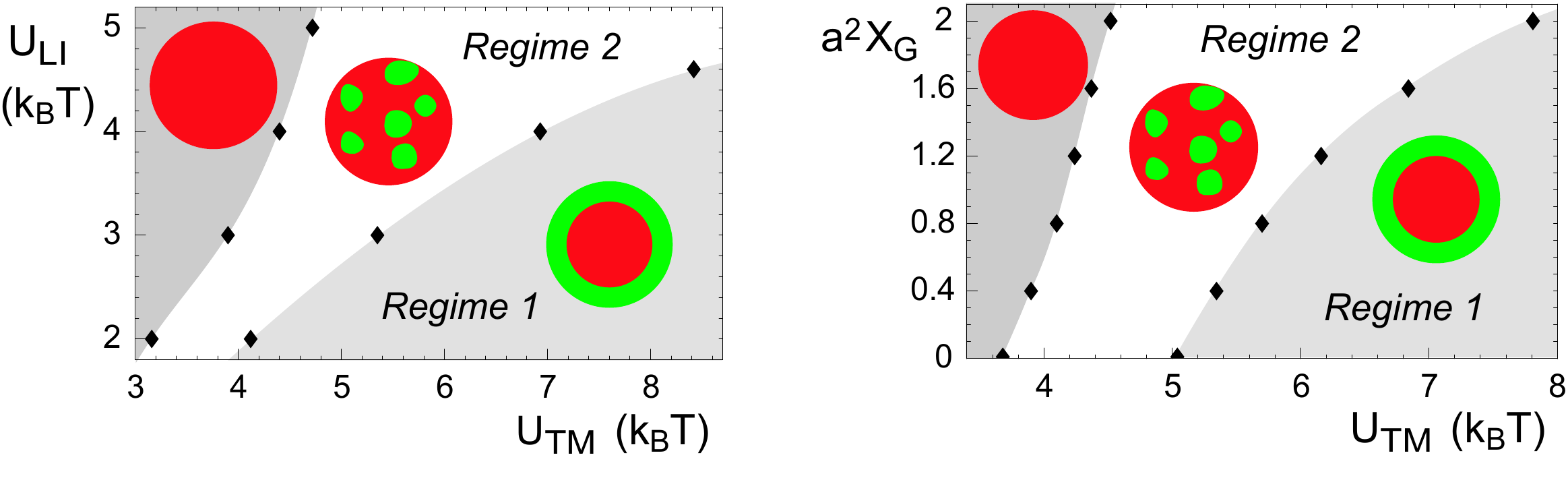}}

\caption{Dynamic regimes for T cell adhesion.  At large values of the binding energy $U_{TM}$ of the TCR/MHCp complexes, a peripheral TCR/MHCp ring emerges at intermediated times as in Fig.~\ref{figureTrings} (Regime 1). At medium values of $U_{TM}$, multifocal patterns as in Fig.~\ref{figureTmultifocal} arise at intermediate times (Regime 2). At small values of $U_{TM}$, TCR/MHCp domains in the contact zone do not form. The threshold for the formation of TCR/MHCp domains and the crossover between the two dynamic regimes depend on the binding energy $U_{LI}$ of LFA-1/ICAM-1 complexes and the glycoprotein concentration $X_G$ in both membranes. The  concentrations of TCR, LFA-1, and ICAM-1 are $0.4 / a^2 \simeq 80$ molecules/$\mu$m$^2$ and the concentration of MHCp is $0.1 / a^2 \simeq 20$ molecules/$\mu$m$^2$. In the left diagram, the glycoprotein concentration in each of the membranes is $X_G=0.4/a^2$. In the right diagram, the binding energy $U_{LI}$ of LFA-1/ICAM-1 complexes has the value $3k_B T$. The black diamonds in the figure represent data points obtained from Monte Carlo simulations. 
\label{figureTdiagrams}}
\end{center}
\end{figure}

The left diagram in Fig.~\ref{figureTdiagrams} shows how the dynamic regimes for pattern formation depend on the effective binding energies $U_{TM}$ and $U_{LI}$ of the TCR/MHCp and LFA-1/ICAM-1 complexes.  An increase in $U_{TM}$ in general leads to more TCR/MHCp complexes in the contact zone, while an increase in $U_{LI}$ leads to the binding of more LFA-1/ICAM-1 complexes. Instead of varying the effective binding energies $U_{TM}$ and $U_{LI}$, the numbers of bound receptor/ligand  complexes in the contact zone could also be changed by varying the overall concentrations of the receptors and ligands, with similar effects on the pattern formation.

The right diagram in Fig.~\ref{figureTdiagrams} shows the effect of the glycoprotein concentration $X_G$ on the adhesion dynamics. The length of the glycoproteins is compatible with the length of the LFA-1/ICAM-1 complexes. Hence, the glycoproteins can enter the red LFA-1/ICAM-1 domains in the contact zone, but are excluded from the green TCR/MHCp domains. The accessible membrane area for the glycoproteins increases with the fraction of LFA-1/ICAM-1 domains in the contact zone, and so does the entropy of the glycoprotein distribution. Therefore, an increase in the overall glycoprotein concentrations  leads to a larger fraction of red LFA-1/ICAM-1 domains in the contact zone, and thus has a similar effect as increasing the binding energy $U_{LI}$ of the LFA-1/ICAM-1 complexes. 

In both dynamic regimes of pattern formation, the coalescence of clusters finally leads to a single TCR/MHCp domain in our model. In the absence of active transport processes, we always observe that the final TCR/MHCp domain is in contact with the rim of the contact zone,  see Figs.~\ref{figureTrings} and \ref{figureTmultifocal}. This behavior can be understood from the line tensions at the domain boundaries and at the rim of the contact zone \cite{weikl04}. In these equilibrium conformations, the length of the boundary between the two domains is significantly shorter than in the case of a central circular TCR/MHCp cluster.

\subsubsection{Adhesion dynamics with active transport of TCRs}

In T cells, active processes transport receptors into the contact zone  \cite{wulfing98} and glycoproteins out of this region \cite{allenspach01,delon01}. The framework enabling these transport processes is the actin cytoskeleton, which polarizes during adhesion, with a focal point in the center of the contact zone \cite{alberts94,dustin98}. For TCRs, the transport is mediated by myosin, a molecular motor protein binding to the actin filaments. The transport of TCRs can be modeled as a directed diffusion. In the model, each TCR molecule is simply assumed to experience a constant force which is directed towards the center of the contact zone midpoint. This force corresponds to an additional term $F\cdot r$ in the configurational energy of each TCR where $F$ is the magnitude of the force and $r$ the distance of the receptor from the center of the contact zone. Under the influence of this force, diffusive steps bringing TCRs closer to the focal point of the cytoskeleton in the center of the contact are, in general, more likely than diffusive steps in the opposite direction.

\begin{figure*}
\resizebox{\columnwidth}{!}{\includegraphics{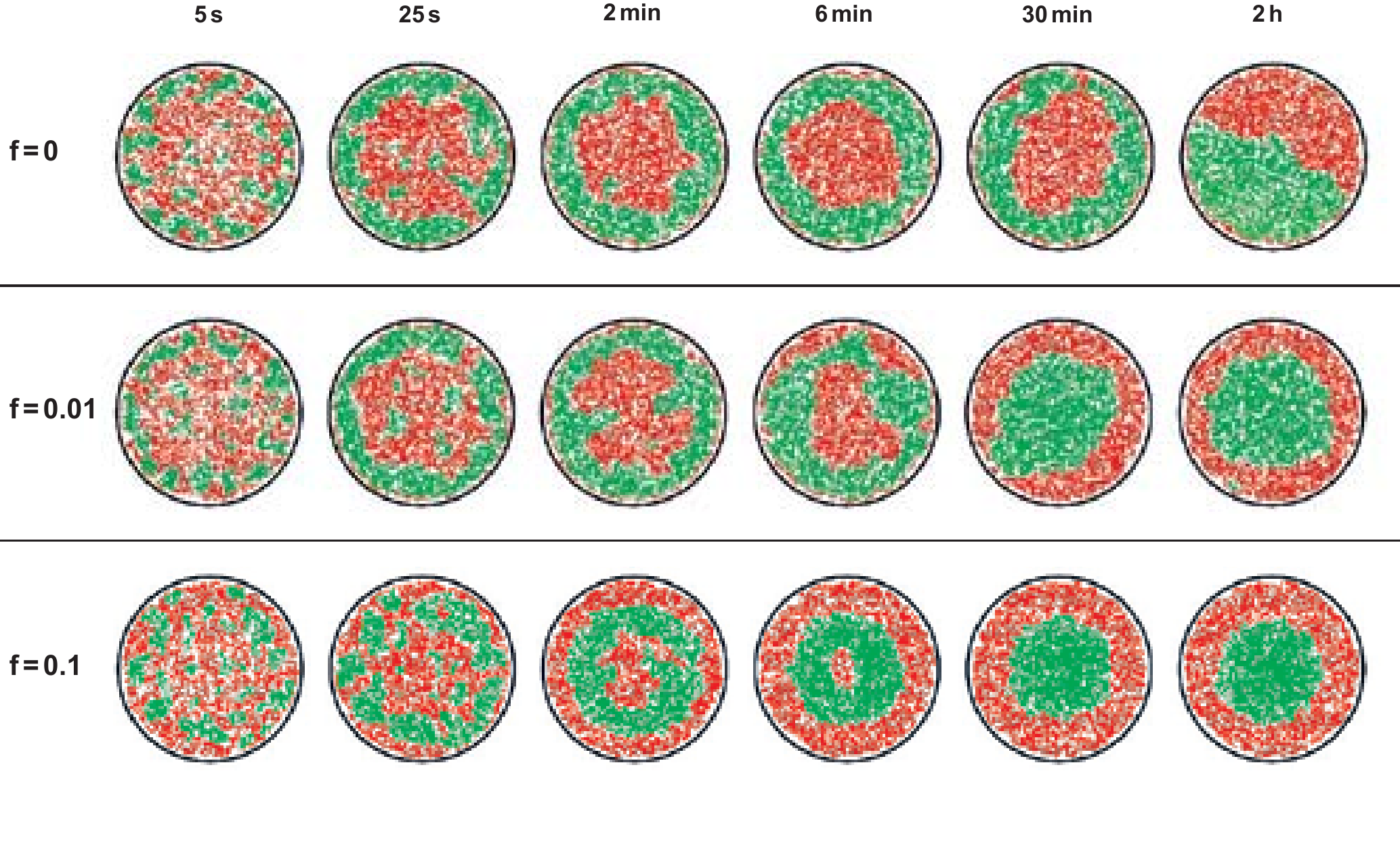}}
\vspace{-1cm}

\caption{Pattern evolution with active transport of TCRs towards the center of the contact zone. Membrane patches with bound TCR/MHCp complexes are shown in green, patches with LFA-1/ICAM-1 complexes in red. Molecular concentrations and binding energies are the same as in Fig.~\ref{figureTrings} (dynamic regime 1). (Top) At zero force, the intermediate TCR/MHCp pattern is stable for 30 minutes and more. In the final equilibrium pattern, both types of domains are in contact with the rim of the adhesion region, see section III. (Middle) At the force $F=0.01k_B T/a$, the final equilibrium state is the target-shaped mature synapse of T cells. This state is already established within 30 minutes. (Bottom) At the 10-fold stronger force  $F=0.1$ $k_B T/a$, the final target-shaped pattern already forms within 5 to 10 minutes. An intermediate pattern with a TCR/MHCp ring appears around 30 seconds after initial contact.
\label{figureTwithForce}}
\end{figure*}

Figure \ref{figureTwithForce} compares the pattern evolution at zero force with patterns at the forces $F=0.01k_B T/a\simeq 6\cdot 10^{-16}$~N and $F=0.1k_B T/a\simeq 6\cdot 10^{-15}$~N. The concentrations and binding energies are the same as in Fig.~\ref{figureTrings}. For these values, the force $F=0.01k_B T/a$ is close to the force threshold leading to a target-shaped final synapse with central TCR/MHCp cluster. Besides leading to a central TCR/MHCp cluster, the active forces speed up the pattern evolution. At the weaker force $F=0.01k_B T/a$, the final equilibrium state is reached after approximately 30 minutes, while the 10-fold stronger force $F=0.1k_B T/a$ leads to equilibrium within few minutes. The absolute times are based on the estimate that one Monte Carlo step roughly corresponds to 1 ms, see above. A TCR/MHCp ring at intermediate times is formed in all three cases shown in Fig.~\ref{figureTwithForce}.  The forces in the model are average forces acting on a single TCR. Since the transport of a TCR molecule over larger distances presumably involves several cytoskeletal binding and unbinding events, these average forces are significantly smaller than the local maximum forces around 1~pN$=10^{-12}$~N which can be exerted by a single molecular motor \cite{mehta99}. 

Experimentally, the mature synapse of T cells has been observed to form on timescales  between 5 and 30 minutes \cite{grakoui99,khlee02}. These timescales agree with the equilibration times for the force range $0.01k_B T/a \lesssim F \lesssim 0.1 k_B T/a$ with T-cell like pattern evolution, see Figs.~\ref{figureTwithForce}. In the absence of active forces ($F=0$), the intermediate peripheral TCR/MHCp ring seems to be metastable and appears in our simulations for times up to an hour. This metastability might explain the inverted NK cell synapse  which consists of a peripheral ring of short receptor/ligand complexes, and a central domain containing the longer integrins. The inverted synapse of NK cells seems to be formed by self-assembly, since it is not affected by ATP depletion or cytoskeletal inhibitors \cite{davis99,fassett01}. 

Some groups \cite{qi01,sjlee03,raychaudhuri03,coombs04} have proposed that the final T cell pattern can be obtained by self-assembly. In the model of Qi et al.\ \cite{qi01}, the central TCR/MHCp domain apparently results from the circular symmetry of the considered patterns. This symmetry prevents patterns with a single TCR/MHCp domain at the contact zone rim. Coombs et al.\ \cite{coombs04} investigate equilibrium aspects of T cell adhesion and focus on circular symmetric patterns similar to Qi et al. In the models of S.-J.\ Lee et al.\ \cite{sjlee03} and Raychaudhuri et al.\ \cite{raychaudhuri03}, the central TCR/MHCp domain seems to arise from the boundary condition that the membrane separation at the contact zone rim is close to the LFA-1/ICAM-1 length of 40 nm. This boundary condition favors LFA-1/ICAM-1 domains at the rim, and repels TCR/MHCp domains from the contact zone rim. However, directly adjacent to the contact zone of two cells, the membrane separation quickly attains values much larger than the lengths of the receptor/ligand complexes. Therefore, it seems more realistic to impose a membrane separation significantly larger than the lengths of LFA-1/ICAM-1 and TCR/MHCp complexes as boundary at the rim. In the discrete model presented here, the separation at the rim is 100 nm.

Active cytoskeletal processes may also play an important role for the multifocal patterns of adhering thymozytes  \cite{hailman02}. These patterns resemble the characteristic intermediate domain patterns in the dynamic regime 2, see Fig.~\ref{figureTmultifocal}. However, the intermediate patterns are only stable on the timescale of minutes.  After a few minutes, domain coalescence leads to a single TCR/MHCp domain in the model. In contrast, the multifocal synapse of thymozytes is stable for hours. One reason for the pattern stability might be the thymozyte cytoskeleton. Unlike the cytoskeleton of mature T cells, the cytoskeleton of thymozytes presumably remains in a mobile, nonpolarized state that still allows cell migration \cite{hailman02}. The few TCR/MHCp clusters of thymozytes may be coupled to the cytoskeleton, thus following its movements.  

\subsection*{Acknowledgements}

We would like to thank David Andelman, Jay Groves, Shige Komura, and Roland Netz for enjoyable and stimulating collaborations.

\newcounter{appendix}
\renewcommand{\theequation}{\mbox{\Alph{appendix}.\arabic{equation}}}

\section*{Appendices}
\addcontentsline{toc}{section}{Appendices}

\setcounter{appendix}{1}
\setcounter{equation}{0}
\appendix{{\bf \large A. Continuum model for homogeneous  membranes}}

\addcontentsline{toc}{subsection}{\protect\numberline{A} 
{Continuum model for homogeneous membranes} }

\vspace*{0.2cm}

The  membranes are regarded as  thin elastic sheets that exhibit an average 
orientation parallel to a reference plane for which we choose Cartesian coodinates 
$x \equiv  (x_1,x_2)$. First, let us consider a single membrane.  
 The shape  of this membrane can be parametrized by the local displacement field 
$ h(x)$  which describes   deformations  from the planar  reference state with 
$h(x) \equiv 0$. 
\footnote{For a deformable membrane interacting with a planar surface, 
the separation field $l$ as used in section \ref{S.EffectiveHamiltonian}
is identical with the displacement field $h$  considered here.}
 The associated elastic energy has the general form
\be
\cH_{\el}\{h \} = \cH_{\te}\{h \}  + \cH_{\ben}\{h \} 
\ee
where the first term  $\cH_{\te}$ represents 
the work against the membrane tension $ \sigma_1$, which is conjugate to the total 
membrane area,  and the second term $\cH_{\ben}$ corresponds to 
 the bending energy which is governed by the bending rigidity $ \kappa_1$. 

The work against the tension $\sigma_1$ is given by 
\be
\cH_{\te}\{h\} =  
 \sigma_1 \int d^2 x  \, \left[ \sqrt{1 + (\nabla h)^2} - 1 \right]
\approx \int d^2 x \, \, \half  \sigma_1 \,  (\nabla h)^2
\ee
where the asymptotic equality holds to leading order in the gradients of $h$. 
This term has the same form as for a fluctuating interface. \cite{lipo12}
The bending energy $\cH_{\ben} $  depends on the squared
mean  curvature \cite{helfrich73}. When 
expressed in terms of the displacement field $h$, the mean curvature
$M$ is given by
\be
2 M = \frac{
\nabla^2 h + (\p_2 l)^2 (\p_1\p_1 h)  - 2 (\p_1 h)  (\p_2 h) (\p_1 \p_2 h)  +
(\p_1 h)^2  (\p_2 \p_2  h) }{ [ 1 + (\nabla h)^2 ]^{3/2} }
\ee
where $\p_1$ and $\p_2$ represent partial derivatives with respect 
to $x_1$ and $x_2$, respectively. 
To leading order in the gradients of $h$, one has 
\be
2 M = [ \nabla^2 h + {\cal O} (\nabla^4 l^3) ] / [1 + (\nabla h)^2]^{3/2} 
\approx  \nabla^2 h. 
\ee
The bending energy is then given by 
\be
\cH_{\ben}\{h \} =  \kappa_1  \int d^2 x  \, \, \sqrt{1 + (\nabla h)^2 } \ \ 2 M^2
\approx \int d^2 x \, \, \half  \kappa_1 \,  (\nabla^2 h)^2 . 
\label{HBending}
\ee 
where the asymptotic equality corresponds again to the small gradient limit. 
Thus,  in this latter limit, the elastic  energy $\cH_{\el}$ behaves as    
\be
\cH_{\el}\{h \}  = \int d^2 x \, \, 
\left[  \half  \sigma_1  \left(\nabla h \right)^2 + 
\half  \kappa_1  \,  \left(  \nabla^2  h \right)^2 \right ] 
\ee
which is identical with the expression (\ref{HElastic})  if one sets
 $h \equiv l$, $\sigma_1 \equiv \sigma$,  and $\kappa_1 \equiv \kappa$.

Next, let us consider two membranes  which are, on average,  both oriented
 parallel to the $(x_1,x_2)$ plane. 
The two membranes are distinguished by the index $j$ with $j = 1,2$.  
The membrane with index
$j$ is described by the displacement field $h_j = h_j(x)$, which represents
the local distance of membrane $j$ from the reference plane, 
 has  bending rigidity $\kappa_j$, 
and  is subject to the tension $\sigma_j$. 
 To leading order in the gradients of $h_1$ and $h_2$,
 the elastic energy of the two membranes is  then  given by 
 \be
\cH_{2, \el}\{h_1, h_2 \}  = \int d^2 x \, \,  \sum_{j=1,2}
\left[  \half  \sigma_j  \left(\nabla h_j \right)^2 + 
\half  \kappa_j  \left(  \nabla^2  h_j \right)^2    \right ]  . 
\label{ShapeEnergy2Membranes}
\ee
In addition, the two membranes interact via the effective potential $V_\me$ which depends on 
the separation field  
\be
l(x) \equiv h_1(x)  - h_2(x) \ge 0
\ee
where we used the convention that membrane 1 is located above membrane 2. 
The inequality $l(x) \ge 0$ reflects the basic property that the two membranes cannot 
penetrate each other and, thus, exert a mutual hardwall potential. 
In general, the interaction  potential $V$ may depend both on the separation 
$l =  h_1 - h_2$ and on the  gradients of $h_1$ and $h_2$. As before, we keep 
only the leading term in a gradient expansion and, thus,  consider the simple interaction energy 
\be
\cH_{\rm 2, in} \{h_1, h_2 \}  = \int d^{2}x \,\ \ V(h_1 - h_2) = 
\cH_{\rm 2, in} \{h_1 - h_2 \}  . 
\label{PotentialEnergy}
\ee 
 The effective Hamiltonian for the two membranes is then given by 
\be
\cH_2\{h_1, h_2 \} =  \cH_{2, \el}\{h_1, h_2 \} + \cH_{\rm 2, in} \{h_1- h_2 \}  
\label{ConfigEnergy2Membranes}
\ee
with $ \cH_{2, \el}\{h_1, h_2 \} $ as in (\ref{ShapeEnergy2Membranes}). 

We now  change  variables from the displacement fields 
$h_1$ and $h_2$ to the separation field
 $l = h_1 - h_2$ and another new displacement field $m$. 
The new  field $m$ is most conveniently described  in 
Fourier space. Thus, let us define the wavenumber $q \equiv  (q_1, q_2)$ and 
the Fourier transform $\tilde f = \tilde f (q) $ of any function  $f = f(x)$ via 
\be
\tilde f (q)\equiv \int d^2 x \, \, e^{- i q \cdot x} f(x)  . 
\ee
The  elastic Hamiltonian as given by (\ref{ShapeEnergy2Membranes}) can now 
be expressed in terms of the Fourier transformed displacement fields 
$\tilde h_1 = \tilde h_1  (q) $ and $\tilde h_2 = \tilde h_2 (q) $ which leads to 
\be
\cH_{2, \el}\{ h_1,  h_2 \} = \int  \frac{d^2 q}{(2 \pi)^2} \, \, 
\left[ \half \chi_1 (q) | \tilde h_1 |^2 +  \half \chi_2 (q) | \tilde h_2 |^2 \right] 
\ee
with 
\be
\chi_j(q) \equiv \sigma_j q^2 + \kappa_j q^4  . 
\label{ChiDefined}
\ee

The Fourier transform of the new displacement field $l$ is simply given by 
$\tilde l = \tilde h_1 - \tilde h_2  $. The  new displacement 
field $m$, on the other hand, is now defined via its Fourier transform
\be
\tilde m  \equiv  \frac{\chi_1}{\chi_1 + \chi_2} \tilde h_1
+   \frac{\chi_2}{\chi_1 + \chi_2}  \tilde h_2 
\ee
with the $q$--dependent functions $\chi_1$ and $\chi_2$ as given by 
(\ref{ChiDefined}). 

When expressed in terms of the new displacement fields 
 $l$ and $m$, the effective Hamiltonian (\ref{ConfigEnergy2Membranes})
for two membranes is decomposed into two terms according to
\be
\cH_2\{h_1, h_2 \} = \cH \{l \} +  \cH^\prime \{m \} 
\ee
and, thus, does not contain any crossterms involving both $l$ and $m$.
The effective Hamiltonian $\cH^\prime \{m \}$ does not depend on the 
interactions of the membranes and describes the diffusive motion of the 
displacement field $m$ which is analogous to a 
`center-of-mass' coordinate. 
 
The effective Hamiltonian $\cH \{l \}$,  which governs the 
separation field $l = h_1 - h_2$,  is given by 
\be
\cH \{l \} = \int  \frac{d^2 q}{(2 \pi)^2} \, \,  \half  \chi (q) |\tilde l|^2
+ \int d^2 x \, \, V(l)
\label{H(l)}
\ee
with the `inverse propagator'
\be
\chi (q) \equiv \frac{(\sigma_1 q^2 + \kappa_1 q^4) (\sigma_2 q^2 + \kappa_2 q^4)}{
(\sigma_1 + \sigma_2) q^2 + (\kappa_1 + \kappa_2) q^4} . 
\ee
If the tensions and bending rigidities of the two membranes satisfy the linear relation 
\be
\sigma_1 / \sigma_2 = \kappa_1 / \kappa_2 ,
\label{ParameterRelation} 
\ee
the function $\chi(q)$ simplifies and becomes
\be
\chi(q) = \frac{\Sigma_1 \Sigma_2}{\Sigma_1 + \Sigma_2} \, q^2 + 
\frac{\kappa_1 \kappa_2}{\kappa_1 + \kappa_2}  \, q^4 .
\label{ChiSimple} 
\ee
After the inverse Fourier transform, the effective Hamiltonian
(\ref{H(l)}) then  has the simple form 
\cite{lipo50,lipo69}
\be
\cH \{l \} = \int d^2 x \, \, \left[
 \half \, \sigma \, (\nabla l)^2 
 + \half  \, \kappa \, (\nabla^2 l)^2
 + V(l) \right]  .
 \label{EffectiveH2}
\ee
with the effective tension 
\be
\sigma \equiv  \sigma_1 \sigma_2/ (\sigma_1 + \sigma_2)
\ee
and the effective rigidity
\be
\kappa \equiv  \kappa_1 \kappa_2/ (\kappa_1 + \kappa_2)
\ee
as in (\ref{EffectiveTension}) and (\ref{EffectiveRigidity}). 
This form also applies (i) to  tensionless membranes with $\sigma_1 = \sigma_2 = 0$, 
(ii) to very flexible membranes with $\kappa_1 = \kappa_2 = 0$, 
(iii) to  two identical membranes with $\sigma_1 = \sigma_2$
and $\kappa_1 = \kappa_2$, and (iv) to the limiting cases in which one of the membranes
becomes very tense and/or very rigid.
In all of these cases, the effective Hamiltonian for two interacting membranes as given by 
(\ref{EffectiveH2})  is 
identical with the effective Hamiltonian for one deformable membrane interacting
with another planar surface as described by(\ref{HHomogen}) -- (\ref{HInHomogen})
if  the effective tension 
$\sigma_1 \sigma_2 / (\sigma_1 + \sigma_2)$ is identified with $\sigma$ and 
the effective rigidity $\kappa_1 \kappa_2 / (\kappa_1 + \kappa_2) $ 
with $\kappa$ as in (\ref{EffectiveTension}) and (\ref{EffectiveRigidity}), respectively.

 If all tensions and rigidities are finite but do not satisfy 
the relation as given by (\ref{ParameterRelation}), the function $\chi(q)$ 
has  the asymptotic behavior
\be
\chi(q) \approx \frac{\sigma_1 \sigma_2}{\sigma_1 + \sigma_2} \, q^2 +
\frac{\kappa_1 \sigma_2^2 + \kappa_2 \sigma_1^2}{(\sigma_1 + \sigma_2)^2} \, q^4 
\ee
for small $q$ and 
\be
\chi(q) \approx \frac{\kappa_1 \kappa_2}{\kappa_1 + \kappa_2}  \, q^4 +
\frac{\sigma_1 \kappa_2^2 + \sigma_2 \kappa_1^2}{(\kappa_1 + \kappa_2)^2} \, q^2 
\ee
for small tensions $\sigma_1 \sim \sigma_2$. In both limits, the 
leading order term is again contained in (\ref{ChiSimple}).

The effective tension $\sigma$ and the effective rigidity $\kappa$ 
 define the length scale $\xi_* \equiv 
(\kappa / \sigma)^{1/2}$. On length scales that are large and compared
to $\xi_*$, the elastic energy is dominated by the tension and rigidity 
term, respectively. In the tension-dominated regime, the effective Hamiltonian 
(\ref{EffectiveH2}) for two interacting membranes is identical with the 
effective Hamiltonian for wetting transitions \cite{lipo12,lipo19,lipo39}.

\end{document}